**THE PRIMORDIAL MAGNETIC FIELD GENERATED AT LARGE FIELD**

**INFLATION, NATURAL INFLATION, AND $R^2$-INFLATION**

**BY $f^2FF$ MODEL**

APPROVED BY SUPERVISING COMMITTEE:

_________________________________________
Eric M. Schlegel, Ph.D., Co-Chair

_________________________________________
Rafael Lopez-Mobilia, Ph.D., Co-Chair

_________________________________________
Chris Packham, Ph.D.

_________________________________________
Marcelo Marucho, Ph.D.

_________________________________________
Lucio Tavernini, Ph.D.

Accepted: _________________________________________
Dean, Graduate School

## DEDICATION

*This dissertation is dedicated to my dear Mom, Dad, Wife, and my beloved children, Hassan, Fatimah, Muhammad, Zainab, Hussain, and Zahraa. Thank you all for providing me with constant support and inspiration.*

**THE PRIMORDIAL MAGNETIC FIELD GENERATED AT LARGE FIELD**

**INFLATION, NATURAL INFLATION, AND $R^2$-INFLATION**

**BY $f^2FF$ MODEL**

by

ANWAR SALEH ALMUHAMMAD, M.S.

THESIS
Presented to the Graduate Faculty of
The University of Texas at San Antonio
in Partial Fulfillment
of the Requirements
for the Degree of

DOCTOR OF PHILOSOPHY IN PHYSICS

THE UNIVERSITY OF TEXAS AT SAN ANTONIO
College of Sciences
Department of Physics and Astronomy
Dec 2015

# ACKNOWLEDGEMENTS

I would like to highly appreciate Dr. Rafael Lopez-Mobilia, who supervised this research and for very useful discussion, guidance, and correction. I would also like to appreciate his patience and cooperation during the period of this research. Further, I would like to appreciate the other dissertation committee members for cooperation and useful comments.

I would like to thank Saudi Arabian Cultural Mission (SACM) for the support they offered me during part of my Ph.D program. Likewise, I would like to thank the Department of Physics and Astronomy in the University of Texas at San Antonio (UTSA) for the teaching assistantship and fellowship they offered me during my Ph.D program.

Dec 2015

# THE PRIMORDIAL MAGNETIC FIELD GENERATED AT LARGE FIELD INFLATION, NATURAL INFLATION, AND $R^2$-INFLATION BY $f^2FF$ MODEL

Anwar Saleh AlMuhammad, Ph.D.
The University of Texas at San Antonio, 2015

Supervising Professors: Rafael Lopez-Mobilia, Ph.D. and Eric M. Schlegel, Ph.D.

Large scale magnetic fields are detected in almost all astrophysical systems and scales from planets to superclusters of galaxies. They have also been detected in very low density intergalactic media. The upper limit of primordial magnetic field (PMF) has been set by recent observations by the Planck satellite (2015) to be of the order of a few $n$G. The simple model $f^2FF$ used to generate the PMF during the inflation era, is attractive because it is a stable and leads to a scale invariant PMF. On March, 2014, a detection of the primordial tensor mode (B-mode) of the polarization of temperature anisotropy in CMB was initially announced by BICEP2, then put in doubt by Planck (Sep 2014) and finally disproved by a BICEP2/Keck Array and Planck joint analysis (Feb 2015). As a result, the attention on non-standard models of inflation such as Large Field Inflation (LFI) and Natural Inflation (NI) (originally favored by the BICEP2 results), has shifted to the old and more standard models, such as $R^2$ inflation. In this research, we compute magnetic and electric spectra generated by the $f^2FF$ model in both the LFI the NI for nearly all possible values of model parameters for de Sitter and power law expansion of inflation under the constraints of the first BICEP2 and for more general case. The necessary scale invariance property of PMF cannot be obtained in both LFI and NI under the first order of slow roll limits with the constraints implemented by the first BICEP2 results.

Furthermore, if the constraints were released, the scale invariant PMF can be achieved. In case of LFI, the associated electric field energy can fall below the energy density of inflation, $\rho_{\text{Inf}}$ for the ranges of comoving wavenumbers, $k > 8 \times 10^{-7} \text{Mpc}^{-1}$ and $k > 4 \times 10^{-6} \text{Mpc}^{-1}$ in de Sitter and power law (PL) expansion respectively. Further, it can drop below $\rho_{\text{Inf}}$ on the ranges, e-foldings $N > 51$, $p < 1.66$, $p > 2.03$, $l_0 > 3 \times 10^5 M_{\text{Pl}}^{-1} (H_i < 3.3 \times 10^{-6} M_{\text{Pl}})$, and $M > 2.8 \times 10^{-3} M_{\text{Pl}}$. All of the above ranges fit with the observational constraints. Also, they can be considered as upper/lower bounds of this model. Out of these ranges, generating PMF in LFI suffer from the backreaction problem. However, generating scale invariant PMF in NI model suffers from the backreaction problem for $k \lesssim 8.0 \times 10^{-7} \text{Mpc}^{-1}$ and Hubble parameter, $H_i \gtrsim 1.25 \times 10^{-3} M_{\text{Pl}}$. The former can be considered as a lower bound of $k$ and the later as an upper bound of $H_i$ for a model which is free from the backreaction problem. Further, there is a narrow range of the height of the NI potential, $\Lambda$, around $\Lambda_{\min} \approx 0.00874 M_{\text{Pl}}$ and of $k$ around $k_{\min} \sim 0.0173 \text{Mpc}^{-1}$, at which the energy of the electric field can fall below the energy of the magnetic field. These ranges also, lie within some observable scales. On the other hand, the generated PMF in the context of $R^2$-inflation can avoid the backreaction problem at $k\eta \ll 1$. The scale invariant PMF, in this model, is achieved at relatively high free power index of the coupling function, $|\alpha| \approx 7.44$. This model has no backreaction problem as long as, the rate of inflationary expansion, $H \lesssim 4.6 \times 10^{-5} M_{\text{Pl}}$, in both de Sitter and power law expansion, which both show similar results. We calculate the lower limit of the reheating parameter, $R_{\text{rad}} > 6.888$ in $R^2$-inflation. Based on the upper limit of inflationary energy density obtained from CMB, we find that the upper limits of magnetic field and reheating energy density as, $\left(\rho_{B_{\text{end}}}\right)_{\text{CMB}} < 1.184 \times 10^{-20} M_{\text{Pl}}^4$ and $(\rho_{\text{reh}})_{\text{CMB}} < 8.480 \times 10^{-22} M_{\text{Pl}}^4$. However, the limits derived from the inflationary model for the magnetic and reheating energy densities are $(\rho_{B_e})_{R^2-\text{inflation}} < 4.6788 \times 10^{-29} M_{\text{Pl}}^4$ and $(\rho_{\text{reh}})_{R^2-\text{inflation}} < 3.344 \times 10^{-30} M_{\text{Pl}}^4$. All of foregoing results are well more than the lower limit derived from WMAP7 for both large and small field inflation. By using the Planck inflationary constraints, 2015 in the context of $R^2$-inflation, the upper limit of reheating temperature and energy density for all possible values of, $\omega_{\text{reh}}$ are respectively constrained as, $T_{\text{reh}} < 4.32 \times 10^{13} \text{GeV}$ and $\rho_{\text{reh}} < 3.259 \times 10^{-18} M_{\text{Pl}}^4$ at $n_s \approx 0.9674$. This value of spectral index is well consistent with Planck, 2015 results. Adopting $T_{\text{reh}}$, enables us to constrain the reheating e-folds number, $N_{\text{reh}}$ on the range $1 < N_{\text{reh}} < 8.3$, for $-1/3 < \omega_{\text{reh}} < 1$. Finally and by using the scale invariant PMF generated by $f^2 FF$, we find that the upper limit of present magnetic field, $B_0 < 8.058 \times 10^{-9} \text{G}$. It is in the same order of magnitude of PMF, reported by Planck, 2015. Therefore, the $f^2 FF$ model may be viable model in the context of $R^2$-inflation to generate the PMF and may be considered as a possible avenue to solve the problem of backreaction.

# TABLE OF CONTENTS

# LIST OF FIGURES

# CHAPTER ONE: INTRODUCTION

Inflationary cosmology theory has solved many of the fundamental problems of the Big Bang model, such as the *flatness*, *horizon* problems. Also, it explains why we do not detect *magnetic monopole* in the universe. In order to solve these problems, the universe should follow a rapid way of expansion (exponential) and the expansion should last enough time (slow roll approximation).

As a result of these two conditions, the universe inflated enormously and becomes very flat afterward "*flatness problem*". In other words, the density of the universe is almost exactly equal the critical density $\Omega_0$ throughout all stages after inflation. Similarly, the universe was in thermal contact (equilibrium) at the onset of the inflation, so that the electromagnetic radiation CMB left over from inflation has the same temperature in all direction. Without proposing the stage of inflation, it was not clear how distant parts in different directions of the universe were in causal contact or in the same horizon "*horizon problem*". At that stage, the space-time (but not information) expanded in a speed greater than the speed of light, so the parts, which appear as unconnected, were in causal connection in the pre-inflation stage. Moreover, inflationary theory solved the problem of "*magnetic monopole*" in the same manner as it did for flatness. According to grand unification theory (GUT), there was a grand unification era at which, topological defects of space-time appeared as magnetic monopoles. Thus, their density should be very high in the current time, so it could be detected easily. But with inflation all point defects are removed and the density of magnetic monopole is negligible.

In the original form of inflation as proposed by A Guth (1981) [1], a fundamental scalar field (*inflaton*) is the physical generator of the inflationary era. That proposal is essentially based on a standard quantum field theory, which predicts the role of scalar fields at the very high energy limits, such as the energy of inflation. The scalar field inflationary universe model is stable under perturbations [2-11]. So, it is a physically viable model. It has also shown very accurate predictions of the cosmological properties of the universe like the fluctuations in both the temperature of the cosmic microwave background (CMB) and large scale structure (LSS) of the universe. The anisotropy of the CMB is also one of the natural predictions of that model. In order to get the observed amount of anisotropies, the initial conditions, potential, and number of e-foldings need to be fine-tuned. Based on that model, gravitational waves (GWs) have also been predicted as a result of space time tensor fluctuations.

Apart of these successes, there are some challenging problems associated with the scalar model. For example, there is no clear physical relation between the scalar field of early inflation and the dark energy that apparently is driving the current accelerated expansion of the universe. Also, over the last few years, where the precision of CMB detection is being increased, some levels of statistical anisotropies have been and are still being detected in the temperature fluctuation and the power spectrum of the CMB [12-17]. These levels of anisotropy are not easy to explain within scalar field model without fine-tuning to the initial conditions or the potential of the inflation. Also, within a scalar model only, one cannot explain the existing of primordial magnetic field (PMF) during the era of inflation. It is a necessary condition to explain the detection of PMF in almost all scales of the universe.

On the other hand, the existence of fundamental vector fields (e.g. electromagnetic fields) has been known since the nineteenth century. If the vector field is relevant to inflation, it may also produce anisotropy in an inflationary universe whatever small it is, which seems to contradict the perfect isotropy of the universe in the cosmic large scale "*the cosmic no-hair* *conjecture*" [18]. As a result, several massive vector field models have been proposed to replace the role of the scalar field model and to explain the anisotropies detected in the CMB. In some of these models, vector field is coupled to itself in the Lagrangian of the field [19-22] and in some other they are non-minimally coupled to gravity [23-26]. They show a consistency with the slow-roll approximation and the amount of detected anisotropies as much as the model of chaotic inflation with a scalar fields [27]. Also, they show a natural relation or transition between the early inflation and dark energy that is widely believed to drive the current expanding epoch of the universe [28-30]. But in these models, the square value of the mass of a vector turns out to be negative (ghost) which represents unstable growth of the linearized perturbations in the small wavelength (UV) regime. Moreover, in almost all massive fields, $U(1)$ gauge invariance of the vector field is broken [31-32].

In order to avoid these problems, models with a gauged non-massive vector field kinetically coupled to the scalar field are proposed, ($f^2FF$) [33-34]. Some researchers have shown that this scalar-vector model is stable under perturbations, see [32] and the references therein. According to these researches, the scalar field drives the inflationary expansion while the vector field causes the anisotropies. Moreover, it is more natural to assume the existing of both primordial vector and scalar field in the grand unified era than to assume one type of a field only. In addition, the primordial magnetic field (PMF) is a natural outcome of these models.

The magnetic fields are being observed in all kinds of galaxies and cluster of galaxies at wide range of redshift. Moreover, the PMF was revealed and its lower bound is constrained in a very low density intergalactic medium (void) [52]. As the common astrophysical model (galactic dynamo) of generating magnetic fields cannot explain the existence of PMF on the absence of the uniform rotating charged medium, such a detection and indication can add extra supportive point to this model [34, 40-43]. Finally, the latest constraints on the value of PMF were presented recently by Planck 2015 [95].

If the existence of a relatively high strength PMF during very early era of the universe is verified, there will be implications on some astrophysical and cosmological phenomena. As the magnetic field plays important role in magneto hydrodynamic, PMF may influence the process of gravitational collapse which in turns affects early stars and galaxies formation dynamics. The dynamics of galactic and intergalactic gases is affected as well. In addition to the foregoing astrophysical effects, some cosmological effects of PMF on most of the cosmological signals and spectra are expected. For example, the effects on; PGW, big bang nucleosynthesis (BBN), CMB, large scale structure (LSS), the thermal and chemical evolution during the dark ages of the universe, and reionization [53]. For these reasons, investigating models of generating PMF in the inflation period and problems associated with these models is itself a very important subject.

The main problems with this model are: the backreaction problem, where the scale of the electromagnetic field can exceed the scale of inflation itself [44-45], and the strong coupling between electromagnetic fields and charged matter at the beginning of inflation [40, 47]. It leads to a huge coupling between the electromagnetism and the charged particles. If the electromagnetic field couples to charged matter, the physical charge associated with, is so huge at the onset of inflation. For example, for the number of e-folds of inflation, $N = 60$, the physical electric charges, $q \propto e^{120}$ [40]. Such an incredibly huge charge makes this model un-trustable.

On March 2014, the first result of *Background Imaging of Cosmic Extragalactic Polarization* (BICEP2) was released [54]. They reported the detection of the tensor mode (B-mode) of the polarization of temperature anisotropy in CMB. The tensor to scalar ratio detected was, $r = 0.2^{+0.07}_{-0.05}$, with $r = 0$, disfavored at $7.0\sigma$, see Fig.1.1. Also, the scale of inflation energy is closed to *Grand Unified Theory* GUT scale ($\rho_{GUT}^{1/4} \sim 10^{16}$GeV). As a matter of fact, tensor perturbation directly indicates the energy scale of inflation [55]. According to Planck (2013) [56], the upper bound on the tensor-to-scalar ratio is $r < 0.11$ (95% CL) and the scalar spectral index was constrained by Planck, 2013 to $n_s = 0.9603 \pm 0.0073$ ($n_s = 0.9682 \pm 0.0062$, Planck, 2015 [96]).

As a result of BICEP2, many inflationary models including simple models endorsed by WMAP9 and Planck were in trouble [57-58]. Therefore, a non-standard models, such as Large Field Inflation (LFI) [59], Natural Inflation (NI) [60-61] models fit more with the new result of BICEP2 and got more interest.

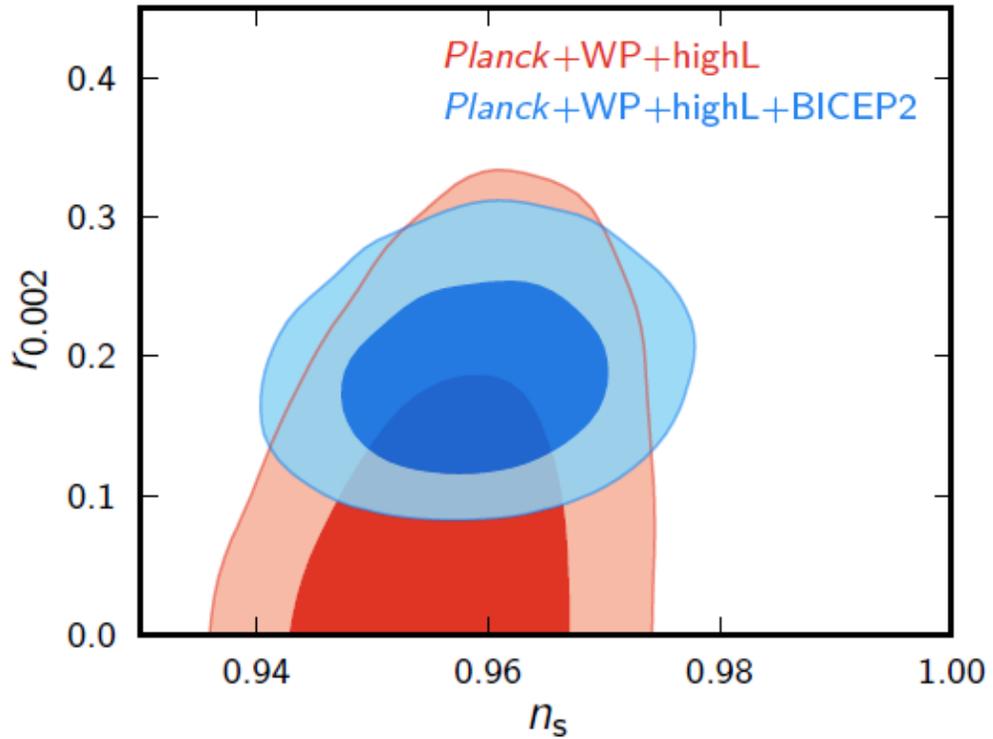

**Fig.1.1**. The indirect constraints of inflation from CMB spectrum, as shown in $r_{0.002} - n_s$ diagram. It is drawn from BICEP2, Planck, WMAP, highL. Courtesy, BICEP2 Collaboration [54].

In Sep 2014, Planck released the angular power spectrum of polarized dust emission at intermediate and high galactic latitudes [89]. The detection frequency of Planck (353 GHz) is different than that of BICEP2 (150 GHz). But in these results, they extrapolated the power spectrum to the same frequency. Also they observed the same patch of sky at high Galactic latitude, which was observed by BICEP2, at low multipoles $40 < l < 120$.

These above results indicated that, there is a significant contamination by dust over most of the high Galactic latitude sky in the same range of detecting a primordial B-mode by BICEP2. Consequently, there is a relatively high value of uncertainty that may raise the values of detected B-mode to the magnitude measured by BICEP2. This effect was previously proposed by [90]. These results of Planck created some serious challenges to BICEP2, which in turns urged more collaboration between the two probs.

Very recently, the joint analysis of BICEP2/Keck Array and Planck data was released on Feb 2015 [94]. The joint data of three probes eliminate the effect of dust contamination and show that the upper limit $r_{0.05} < 0.12$ at 95% CL, and the gravitational lensing B-modes (not the primordial tensor) are detected in $7\sigma$. By considering these inputs, the shape of the signals of the same patch of sky can be re-drawn as in Fig.1.2.

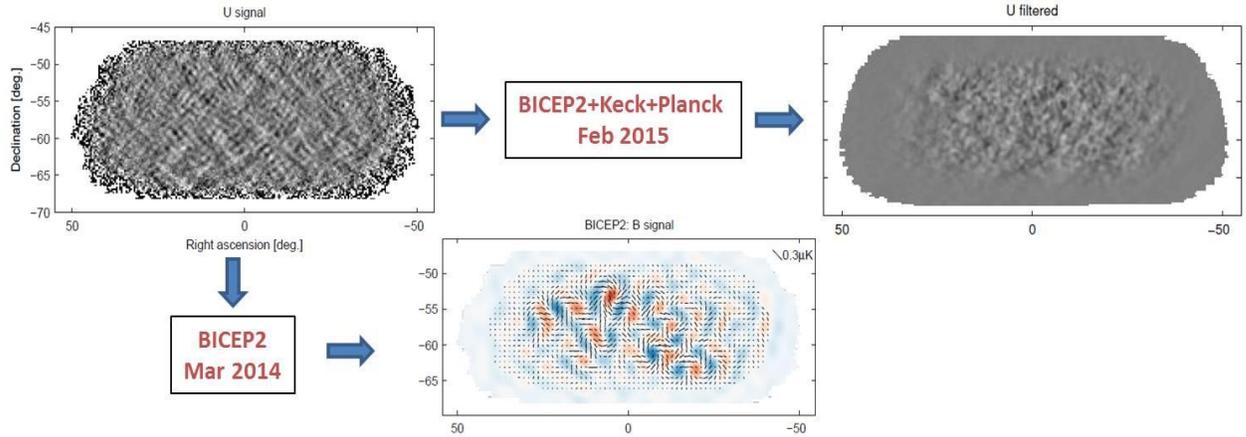

**Fig.1.2**. The signal map of the same patch of sky originally observed by BICEP2 which leads to the first result of BICEP2 (down). But after combing the data of the three probes, the primordial tensor B-modes will be smoothed out (right).

Few days later, Planck 2015 released new series of results that include the CMB anisotropies in both temperature and polarization based on the full Planck survey [95-96]. Both observed PMF and inflationary models are constrained. As a result, new comprehensive set of constraints on the upper limits of PMF have been presented. It will be introduced in detail in Chapter 4. Similarly, the inflationary models are constrained based on the new analysis of data. The more standard inflationary models, which result low value of $r$, are favored more by the new results. Among the best models favored by Planck is the earliest model proposed by Starobinsky ($R^2$-inflation) [2]. However, the chaotic inflationary models like LFI and NI are disfavored.

As we started part of this research months before Planck 2015 and applied both the BICEP2 limits and more general limits, it is still a good test to both the chaotic inflationary models and $f^2FF$ model. In fact, the results of this research can be considered in a one way against those models. Thus, goals of this research are to study the viability of PMF generated in inflation era by scalar fields coupled to vector field model, $f^2FF$, at flat universe for, LFI and NI models of inflation.

After Planck 2015, we started investigated $f^2FF$ model but in the context of $R^2$-inflation. We decide the condition under which, we have a scale invariant PMF and then calculate its magnitude during inflation. In this case we go beyond the inflation era to post-inflation phases, like reheating parameters. Thus, we constrain the reheating parameters both in $R^2$-inflation and based on the upper limits of PMF reported by Planck, 2015 [95]. The most important result of this research is the constraining of the present PMF. It is computed from the calculated magnitude of scale invariant PMF and the constrained reheating parameters.

The organization of this dissertation is as follows. In chapter 1, an introduction is given. In chapter 2, the fundamental of the scalar inflationary model is presented. In chapter 3, the vector field inflationary model is introduced. Chapter 4 shows the observational evidences of the PMF. Chapter 5 presents the generation of PMF by $f^2FF$ in exponential inflationary models as done in [44]. In chapter 6, we apply the same model with large field inflation LFI. In chapter 7, we use the same method but with natural inflation NI. In chapter 8, we investigate $f^2FF$ in the context of $R^2$-inflation and constrain both reheating parameters and present PMF on the light of Planck, 2015. Finally, in chapter 9, we recap the main results and conclusions of the dissertation.

# CHAPTER TWO: THE FUNDAMENTALS OF THE SCALAR FIELD INFLATION

Cosmic inflation is one of the most brilliant ideas which lead to understand the evolution of the Universe. It basically explains the reason why our Universe is so flat, and large, and homogeneous. Since the time it was proposed by Guth [1], it showed a successful explanation to the most cosmological puzzles. Also, one of the most important results of inflation is the generation of the curvature perturbation, which is essential to explain the degree of homogeneity in structure formation and the CMB anisotropy.

There have been two main constraints of the inflation since it has been proposed. First, the energy density of the universe during inflation was dominated by a slowly varying vacuum energy. This constraint is called *slow-roll approximation*. The second one, is that the scale factor of the universe $a(t)$ grew more-or-less exponentially during inflation. Without these two assumptions, one cannot solve the foregoing puzzles and explain many other aspects of the universe and astrophysical phenomena in a consistent way.

On the other hand, the proposed inflation was first thought [1] to be driven by a scalar field rolling dawn from local minimum slowly then it was stopped by quantum-mechanical barrier penetration which rolls to the global minimum at which the present universe lie. It is basically similar to a delayed first-order phase transition of a scalar field which was initially trapped in a local minimum of some potential, and then penetrates through the potential barrier and rolled toward a global minimum of the potential. Such inflation has fundamental problem because it leads to multi-bubbles inflation which in turn localize the energy on the boundaries of bubbles [3-7]. Consequently, the present universe will be highly inhomogeneous and anisotropic.

For these reasons, this model was replaced, soon later, by a new form of inflation [2-3] at which the potential $V(\phi)$ is adjusted to have zero second derivative at $\phi = 0$. In this point the potential is an unstable and has a local minimum at different point $\phi_0$. Such a model also leads to exponential expansion of the universe and slow rolling of the scalar field (*inflaton*), $\phi(t)$, and hence slow rolling of the initial large potential $V(\phi)$.

In this dissertation, we adopt the natural units, $\left[ c = \hbar = k_B = 1 \right]$, the signature $(-1, 1, 1, 1)$, and flat universe, where we use the reduced Planck mass, $M_{\text{Pl}} = \left( 8\pi G \right)^{-1/2}$, where $G$ is gravitational constant. However, the Planck mass will be taken, $M_{\text{Pl}} = 1$, in the computation parts.

One can drive the dynamical equation of the Universe from *Einstein field equation*,

$$R_{\mu\nu} - \frac{1}{2} g_{\mu\nu} R + g_{\mu\nu} \Lambda = -\frac{1}{M_{\text{Pl}}} T_{\mu\nu}, \tag{2.1}$$

where, $R_{\mu\nu}$ is Ricci tensor, $R$ is Ricci scalar, $g_{\mu\nu}$ is the space time metric, $\Lambda$ is the cosmological constant, and $T_{\mu\nu}$ is the energy-momentum tensor defined as,

$$T^{\mu\nu} = -g^{\mu\nu} \left[ \frac{1}{2} g^{\rho\sigma} \frac{\partial \phi}{\partial x^\rho} \frac{\partial \phi}{\partial x^\sigma} + V(\phi) \right] + g^{\mu\rho} g^{\nu\sigma} \frac{\partial \phi}{\partial x^\rho} \frac{\partial \phi}{\partial x^\sigma}, \tag{2.2}$$

where, $\phi(t)$ is a scalar field, and $V(\phi)$ is the inflation potential.

Simple models of inflation potential are based on a single scalar field, such as quadratic, $V(\phi) \sim \phi^2$, quartic, $V(\phi) \sim \phi^4$ [80], Higgs inflation, $V(\phi) \sim \left( 1 - \exp[-\sqrt{2/3}\, \phi / M_{\text{Pl}}] \right)^2$, and the exponential potential, $V(\phi) \sim \exp[-\sqrt{2\epsilon_1}\, (\phi - \phi_0)]$ [81]. The last one is used in [44-45] to find the magnetic and electric spectrum in the $f^2 FF$ model as we shall present in chapter 6. These models became more interesting after WMAP9 [17] and Planck 2013 [56]. As a result, the preferred potential class is the so called "*plateau inflation*", at which $V(0) \neq 0$.

On the other word Equation (2.1) shows the relation between the curvature of the space (left hand side) and its contents of energy and materials (right hand side). Adopting a homogenous and isotropic Universe, *Friedmann-Robertson-Walker FRW* metric $g_{\mu\nu}$, such that the invariant metric of such a Universe can be written in spherical coordinates as,

$$ds^2 = g_{\mu\nu}dx^\mu \, dx^\nu = -dt^2 + a^2(t)\left(\frac{dr^2}{1-kr^2} + r^2 d\theta^2 + r^2 sin^2\theta d\phi^2\right). \quad (2.3)$$

Where, $a(t)$, is scale factor. In the general Cartesian coordinates,

$$ds^2 = g_{\mu\nu}dx^\mu \, dx^\nu = -dt^2 + \frac{a^2(t)}{1+\frac{1}{4}Kx_ix^i}dx_jdx^j, \quad \begin{cases} K = 0, & \text{for a flat universe} \\ K = -1, & \text{for an open universe} \\ K = +1, & \text{for a closed universe} \end{cases} \quad (2.4)$$

In conformal time, $\eta$, FRW metric can be written as,

$$ds^2 = g_{\mu\nu}dx^\mu \, dx^\nu = a^2(\eta)\left(-d\eta^2 + \frac{1}{1+\frac{1}{4}Kx_ix^i}dx_jdx^j\right). \quad (2.5)$$

The solutions to (2.1) by adopting (2.3) are the well-known *Friedmann equations*

$$\dot{a}^{\,2} = \frac{\rho a^2}{3M_{Pl}^2} - k + \frac{\Lambda a^2}{3}, \quad (2.6)$$

$$\frac{\ddot{a}}{a} = -\frac{1}{6M_{Pl}^2}(\rho + 3p) + \frac{\Lambda}{3}, \quad (2.7)$$

And the conservation law,

$$\dot{\rho} = -\frac{3\dot{a}}{a}(\rho + p) = -3H(t)(\rho + p). \quad (2.8)$$

By using the general definition of energy-momentum tensor (2.2) and compare it with that for perfect fluid,

$$T_\phi^{\mu\nu} = pg^{\mu\nu} + (p + \rho)u^\mu u^\nu, \qquad g_{\mu\nu}u^\mu u^\nu = -1, \quad (2.9)$$

one can define the energy density $\rho$, pressure $p$, and velocity four vector $u^\mu$ as,

$$\rho = -\frac{1}{2}g^{\mu\nu}\frac{\partial\phi}{\partial x^\mu}\frac{\partial\phi}{\partial x^\nu} + V(\phi), \quad (2.10)$$

$$p = -\frac{1}{2}g^{\mu\nu}\frac{\partial\phi}{\partial x^\mu}\frac{\partial\phi}{\partial x^\nu} - V(\phi), \quad (2.11)$$

$$u^\mu = -\left[-g^{\rho\sigma}\frac{\partial\phi}{\partial x^\rho}\frac{\partial\phi}{\partial x^\sigma}\right]^{-1/2} + g^{\mu\tau}\frac{\partial\phi}{\partial x^\tau}. \tag{2.12}$$

Since the scalar field, $\phi(t)$ is only time dependent, then (2.10-2.12) can be written as,

$$\rho = \frac{1}{2}\dot{\phi}^2 + V(\phi), \quad p = \frac{1}{2}\dot{\phi}^2 - V(\phi), \quad u^\mu = [\dot{\phi}]^{1/2}. \tag{2.13}$$

Substituting of (2.13) into (2.6-2.7) yields,

$$H^2 = \frac{1}{3M_{\text{Pl}}^2}[\frac{1}{2}\dot{\phi}^2 + V(\phi)] - \frac{k}{a^2} + \frac{\Lambda}{3}. \tag{2.14}$$

If both $k$ and $\Lambda$ are neglected in the early universe, then (2.5) can be written as,

$$H^2 = \frac{1}{3M_{\text{Pl}}^2}[\frac{1}{2}\dot{\phi}^2 + V(\phi)]. \tag{2.15}$$

Combining (2.14) with (2.8) and (2.13) gives the inflation equation of motion,

$$\ddot{\phi} + 3H\dot{\phi} + \frac{dV}{d\phi} = 0. \tag{2.16}$$

In the slow roll approximation, one can neglect the second derivative in (2.16) which leads to the attractor condition,

$$\dot{\phi} = -\frac{1}{3H}\frac{dV}{d\phi}. \tag{2.17}$$

If the expansion during inflation period was exponential of the form,

$$\frac{a(t_2)}{a(t_1)} = exp\left[\int_{t_1}^{t_2} H dt\right], \tag{2.18}$$

where, $t_1, t_2$ are the cosmic time at the start of and during inflation respectively. Hence, the exponent is extremely high and,

$$\frac{|\dot{H}|}{H^2} \sim |\Delta H| \ll 1 \rightarrow |\dot{H}| \ll H^2, \tag{2.19}$$

which is combined with (2.15) to imply,

$$\dot{\phi}^2 \ll |V(\phi)|. \tag{2.20}$$

Both (2.15) and (2.19) imply that, during the inflation period,

$$H \sim \text{Const.} \tag{2.21}$$

If $H = \text{Constant}$, that is the case for de Sitter inflation at which the space time expand exponentially,

$$a(t) \propto exp[H\,t], \qquad (2.22)$$

where, $t$, is any time after inflation.

In fact, de Sitter model, is only zero order approximation which does not have graceful exit from inflation [91]. To have more realistic slow roll analysis that has a smooth exit from inflation, Hubble parameter should be taken as a function of scalar field, $\phi(t)$. If the field falls below certain value, it starts oscillates then converts to particles in the reheating era coming after inflation. Defining the conformal time, $\eta$, as,

$$dt = a(\eta)d\eta, \qquad (2.23)$$

Eq. (2.22) can be written in conformal time as,

$$a(\eta) = (H\,\eta)^{-1}. \qquad (2.24)$$

In more general power form [44, 55],

$$a(\eta) = l_0|\eta|^{1+\beta}, \qquad (2.25)$$

where, $-\infty < \beta \leq -2$ and $l_0$ is the integration constant which can be approximated as $l_0 \approx 1/H$. In the case of de Sitter model, we have a total scale invariance of CMB. Hence, $\beta$ is very close to $-2$.

From (2.13), the equation of state in that period is approximately similar to vacuum equation, $p \approx -\rho$, and (2.15) becomes,

$$H = \sqrt{\frac{V(\phi)}{3M_{\text{Pl}}^2}}\,, \qquad (2.26)$$

Combining (2.15), (2.20), and (2.26) yields what are called *flatness conditions*,

$$\left|\frac{V_\phi(\phi)}{V(\phi)}\right| \ll \frac{\sqrt{2}}{M_{\text{Pl}}}\,, \qquad (2.27)$$

$$\left|\frac{V_{\phi\phi}(\phi)}{V(\phi)}\right| \ll \frac{\sqrt{3}}{M_{Pl}}, \tag{2.28}$$

where, $V_\phi = \partial_\phi V$. The above two relations confirm the slow roll of $\phi$ and $\dot\phi$ and in turns confirm the exponential expansion for sufficient *e-folding*, $N$, which is needed to solve the main puzzles of the Big Bang cosmology. A typical values of $N \approx 60$, which may solve flatness, monopole, and horizon problem.

Hence, defining the slow roll parameters [55] of the single field inflation, which are assumed to be very small during inflation. They can be written in terms of inflationary potential,

$$\epsilon_{1V}(\phi) = \frac{1}{2} M_{Pl}^2 \left(\frac{V_\phi}{V}\right)^2, \tag{2.29}$$

$$\epsilon_{2V}(\phi) = M_{Pl}^2 \left(\frac{V_{\phi\phi}}{V}\right). \tag{2.30}$$

Also, they can be written in terms of Hubble parameter,

$$\epsilon_{1H}(\phi) = 2 M_{Pl}^2 \left(\frac{H_\phi}{H}\right)^2, \tag{2.31}$$

$$\epsilon_{2H}(\phi) = 2 M_{Pl}^2 \left(\frac{H_{\phi\phi}}{H}\right). \tag{2.32}$$

In terms of $\epsilon_{1V}(\phi)$, one can write the e-folding number $N$,

$$N \simeq \ln\left(\frac{a(t_2)}{a(t_1)}\right) = -\sqrt{\frac{1}{2M_{Pl}^2}} \int_\phi^{\phi_f} \frac{1}{\sqrt{\epsilon_{1V}}} d\phi \, . \tag{2.33}$$

In addition to the success, the inflation theory provides us with a very nice origin of almost equal tiny inhomogeneity observed in both the cosmic microwave background CMB radiation and the large scale structures in the Universe. That is based on the adiabatic perturbation at which the ratio of radiation and matter number densities is conserved. Other two important predictions are a scale-invariant spectrum of primordial Gaussian density, and prediction of PGW.

# CHAPTER THREE: VECTOR FIELD INFLATIONARY MODELS

In contrast to the success of inflationary theory, the nature of inflation field (*inflaton*) remains largely unknown. The link between the current dark energy which has been driving the acceleration of the universe, as confirmed by observing high $z$, Type Ia supernovae [121], and the early inflation is not yet understood. Furthermore, as a result of high precession cosmology, some deviation from homogeneity, isotropy, and scale invariant [17, 56], have been reported in the last few years. All of these reasons re-trigger the interest to explore some other conceivable types of inflationary models. Among these types is the vector field inflation. Physicists have dealt with vector (electromagnetic) field for over a century.

Another approach to the field of inflation is through field theory which is based on the Lagrangian density, $\mathcal{L}(\phi, \partial_\mu \phi, x^\mu)$, formulation [64]. Therefore, the equation of motion (2.16) can be derived either from field equation as we did in the first section, or by finding the minimum of the action, $S(\phi, \partial_\mu \phi)$, of the field $\phi$,

$$S(\phi, \partial_\mu \phi) = \int \mathcal{L}(\phi, \partial_\mu \phi, x^\mu) dx^\mu. \tag{3.1}$$

Hence, defining $\mathcal{L}(\phi, \partial_\mu \phi, x^\mu)$ is a key role in this method. The basic differences between inflation models can be referred to the definition of $\mathcal{L}(\phi, \partial_\mu \phi, x^\mu)$ which contains all the physics in the model.

For example the Lagrangian density for a free scalar field (particle) of mass $m$ can be written [65] as,

$$\mathcal{L}(\phi, \partial_\mu \phi, x^\mu) = \frac{1}{2} (\partial_\mu \phi)(\partial^\mu \phi) - \frac{1}{2} m^2 \phi^2. \tag{3.2}$$

However, the Lagrangian of the electromagnetic EM field, which is massless particles (photons) field of spin-1, is written as,

$$\mathcal{L}(A, \partial_\mu A, x^\mu) = -\tfrac{1}{4} F_{\mu\nu} F^{\mu\nu} - J^\mu A_\mu, \tag{3.3}$$

Where $F_{\mu\nu}$, is Maxwell's field tensor defined as,

$$F_{\mu\nu} = \partial_\mu A_\nu - \partial_\nu A_\mu. \tag{3.4}$$

$A_\mu$, is 4-vector potential, and $J^\mu$, is charge-current 4-vector. Hence, the Lagrangian density of free massive vector field can be written in correspondence to EM field as,

$$\mathcal{L}(A, \partial_\mu A, x^\mu) = -\tfrac{1}{4} F_{\mu\nu} F^{\mu\nu} + \tfrac{1}{2} m^2 A_\mu A^\mu. \tag{3.5}$$

Finding the minimum of (3.2) is similar to solving *Euler-Lagrange* (EL) equation in terms of field, $\phi$, which can be written as

$$\partial_\mu \frac{\partial \mathcal{L}}{\partial_\mu \phi} - \frac{\partial \mathcal{L}}{\partial \phi} = -\frac{\delta L}{\delta \phi} = 0 \ . \tag{3.6}$$

For scalar field $\mathcal{L}$ as in (3.2), solving (3.6) yields the inflation equation of motion (2.16). For example, solving EL for electromagnetic field Lagrangian (3.3) yields the compact form of Maxwell's equations.

$$\partial_\mu F^{\mu\nu} = J^\mu. \tag{3.7}$$

Similarly, substituting of (3.5) into (3.6) will result what is called *Proca Equation*,

$$\left(\partial_\mu \partial^\mu + m^2\right) A^\mu = 0. \tag{3.8}$$

The geometry of space time is assumed to be flat (Minkowski) in all of the above Lagrangian densities.

However, if the geometry of the background space time was not flat, then one has to use the Einstein-Hilbert Lagrangian, $\mathcal{L}_{EH}$,

$$\mathcal{L}_{EH} = \sqrt{-g} (\mathcal{L}_{\text{field}} + \mathcal{L}_R) \ . \tag{3.9}$$

Where $\mathcal{L}_{field}$ is the Lagrangian of a field, as defined above for different fields, and $\mathcal{L}_R$ is the geometric Lagrangian, which has the simplest form according to Einstein-Hilbert as,

$$\mathcal{L}_R = \frac{1}{16\pi} R, \tag{3.10}$$

where, $R$ is Ricci scalar which include the geometric curvature of the space time.

One main point is that, under a gauge transformation,

$$\hat{A}_\mu = A_\mu + \partial_\mu \lambda , \tag{3.11}$$

the physical Lagrangian should be invariant. If that is applied to the Lagrangian of the field as it is, then it is called minimally coupled field. However, if some other terms are added to the Lagrangian to satisfy gauge invariance, then it is called non-minimally coupled field. Applying this on the massive scalar field (3.2) and massless vector EM field Lagrangians shows that they are gauge invariant and minimally coupled.

In contrast to the massive scalar field, the Lagrangian of the massive vector field will not be invariant unless it is coupled with some matter to keep gauge invariant. Hence, it is a non-minimally coupled field.

There are two classes of vector field inflation, one at which the inflation is entirely driven by the vector field [23], and the other at which the vector field has a partial contribution to inflation [32]. Both of them end up either being entirely responsible for or only partially contributing to the curvature fluctuations and dark energy at the late times. The role of the scalar field in the second class is the dominant factor. Some models use one vector field and others use many fields.

# CHAPTER FOUR: THE OBSERVATIONAL EVEDENCES OF PMF

Magnetic fields of different strengths are measured or reported in all astrophysical scales from small planets to large galaxy clusters. The range of astrophysical magnetic fields varies from $10^{12}$G in neutron stars to $10^{-6}$G in galaxy clusters [67]. The existence of magnetic fields in all kinds of galaxies and galaxies clusters is well established. There are mounting evidences and hints for the existence of PMF in intermediate and high redshift clusters [53]. Moreover, the lower bound value of intergalactic magnetic field was reported as $B \geq 10^{-16}$G [52].

Recently, Planck 2015 presented new upper limits on the strength of PMF based on the CMB polarization [95]. The set of constraints imposed by Planck can be considered as comprehensive limits of PMF. In this chapter, we will summarize the main methods used to measure and study the astrophysical magnetic fields. We end this chapter by presenting the constraints of PMF imposed by Planck 2015.

## 4.1 Polarization of Optical Starlight

Polarized light from stars revealed the presence of large-scale magnetic fields in our Galaxy. The first unexpected observation of polarized starlight was made by Hiltner and Hall (1949) [66]. They expected to find time-variable polarization levels of 1-2% but they detected polarization levels as high as 10% for some stars. This observation led to the conjecture that a new property of the interstellar medium (ISM) had been discovered. In the same time Alfv´en (1949) and Fermi (1949) were proposing the existence of a galactic magnetic. This technique has many limitations to detect and trace the magnetic field accurately [66].

## 4.2 Zeeman Splitting

In the presence of magnetic fields, each spectral line of an atom splits into several lines according to its angular momentum number (spin plus orbital angular momentum). The most common spectral lines used in this method are the 21-cm line for natural hydrogen and 18-cm line for OH. Because of the difficulty of detection this effect in the far sources, it is used efficiently to measure magnetic fields in our galaxy only. The relation between the frequency of a line $\nu$, frequencies of the splitting lines $\nu_1, \nu_2$ and the magnetic field **B** is such that [66],

$$\frac{\nu_2 - \nu_1}{\nu} = 1.4g\left(\frac{\mathbf{B}}{\mu G}\right)\left(\frac{\text{Hz}}{\nu}\right), \tag{4.1}$$

Where, $g$ is Lande factor. That method was used to infer part of the magnetic field of the galaxy [68].

## 4.3 Synchrotron Radiation

It is the radiation emitted by relativistic electrons spiraling along magnetic field lines. It can be used to study magnetic fields in astrophysical sources ranging from pulsars to superclusters. The emissivity of the synchrotron radiation is proportional to the strength of magnetic fields whereas the degree of polarization is an indication of the field's uniformity and structure. Distinguishing this radiation from other types of radiation can be achieved by measuring the power spectral index, $\alpha$.

There is a definitive relation between $\alpha$ and the particle distribution index, $p$, at which the energy of a relativistic electron is distributed. That relation [69] is,

$$\alpha = \frac{p-1}{2}. \tag{4.2}$$

For a single electron moving in a magnetic field **B**, the emissivity $J(\nu, E)$ (energy emitted spontaneously per unit frequency per unit time per unit mass) as a function of frequency $\nu$ and electron energy $E$ is

$$J(\nu, E) \propto B_\perp \left(\frac{\nu}{\nu_c}\right)^{1/3} f\left(\frac{\nu}{\nu_c}\right),$$
(4.3)

where $B_\perp$ is the perpendicular component of **B** to the line of sight, $\nu_c$ is the critical frequency of the radiation, and $f(x)$ is a cut-off function. This method is used to detect and trace the magnetic field of galaxies in the order of $\sim$ (1-100 $\mu$G) /few Kpc [70].

## 4.4 Faraday Rotation

It occurs whenever a polarized electromagnetic waves, passes through a region of both magnetic field and free electrons. The initial waves of a left and right circular polarization states propagates with different phase velocities. However, the initially linear polarized waves, which are the most relevant case, will experience a rotation of the electric field vector. This rotation can be measured by the rotation measure (RM). The RM (in $rad/m^2$) for radiation of a source at $z_s$ redshift which passes through a region has the magnetic field **B** (in $\mu$G) and electron of $n_e$ (in $cm^{-3}$) density [71],

$$RM(z_s) \approx 8 \times 10^5 \int_0^{z_s} \frac{n_e B_\parallel(z)}{(1+z)^2} dL(z)(rad/m^2),$$
(4.4)

where, $B_\parallel(z)$ is the magnetic field along the line of sight. Using a multi-wavelength observation allow us to find the strength and the direction of $B_\parallel(z)$. This method is used to detect and trace the magnetic field of clusters of galaxies in the order of $\sim(1 - 10 \, \mu$G)/few 10 Kpc [72-73].

**4.5 Inverse-Compton Radiation**

This method based on exploiting the magnetic effects on the highly energetic photons emitted by a GeV or TeV distant sources like AGNs. The interaction between these photons and the low energy CMB photons can lead to electron pair production. The produced particles may in turns scatter with CMB photon via inverse-Compton process to produce $\gamma$-rays. If a very low density intergalactic medium has even a very weak magnetic field ($\sim 10^{-24}$G) [74], the shape of $\gamma$-rays spectra will be affected. Formation of a halo around the AGN source, which is confirmed, is an example of these effects.

According to Ref [71], "three independent groups have recently reported the detection of intergalactic magnetic fields with strengths close to $10^{-15}$G". Those groups mainly used the data of HESS Cherenkov Telescopes, Fermi satellite, and Large Area Telescope to constrain the lower bound of the intergalactic magnetic field. Some of them [52] assumed that the AGN source is isotropic and they got a lower bound of $10^{-16}$G in Mpc scales. However, the other [76] assumed it is highly beamed and they get a lower bound of $10^{-15}$ G in Mpc scales.

The most common astrophysical explanation to the large scale magnetic field is the galactic dynamo. It requires a charged material rotating in the presence of a magnetic field which in turns is amplified as the rotating rate increases in the process of collapsing or differential rotation of the galaxies. It was Larmor (1919) who first proposed the same principle to explain the magnetic field of the Sun and Earth. It was then equally applied on the magnetic fields of galaxies by Parker (1971). Although this theory was controversial in galactic scale, some researchers suggested it for cluster of galaxies case as well.

Regardless of all points raised against the galactic dynamo model, there are two challenging questions: What is the source of the first seed of magnetic fields? How to explain its presence in the very low density intergalactic regions (voids) [52]? Some researchers still argue these questions and they think it could have some astrophysical origin, like the transport of magnetic energy from the void galaxies and bordering AGNs into the voids by Cosmic Rays, See Ref [77] and the references therein. Apart from those investigations, the concept of the relic origin of the magnetic fields is gaining more interest. In the next chapter, the simple inflation model of the PMF will be presented.

## 4.6 Planck 2015 Constraints of PMF on CMB

Very recently (Feb 2015), Planck released the second series of cosmological results based on data from the entire Planck mission [95]. These results include both temperature and polarization of CMB. The constraints of PMF on CMB are among the released results. As a matter of fact, PMF imprints on the CMB in several ways. It includes the direct induced modes on CMB perturbation (scalar, vector, and tensor) and the ionization history of the Universe.

The constraints released were based on the initial conditions of PMF imprints. These includes; inflationary modes which are the most relevant to this research but not discussed in Planck 2015, passive modes which are generated if PMF contributes to the metric perturbation before neutrino decoupling ($T_\nu \gg$ MeV), and compensated modes which are generated if PMF contributes to the metric perturbation after neutrino decoupling ($T_\nu \sim$ MeV, $t \sim$ s).

The direct imprint of PMF on CMB can affect onto both temperature and polarization. The temperature correlation function of two points (power spectra) and higher order (bi-spectra, tri-spectra…) are presented in both passive and compensated modes. For example, the upper limit of passive modes of bi-spectra (non-Gaussianity analyses) at high $l$ of tensor perturbation is, $B_{1\text{Mpc}} < 2.8$nG. However, the upper limits of compensated modes of power spectra (scalar, vector, and tensor), are $B_{1\text{Mpc}} < 3.0\text{nG}$ for non-helical field, and $B_{h,1\text{Mpc}} < 5.6\text{nG}$, for helical field at 95% CL. Similarly the bi-spectra of high $l$ scalar modes is, $B_{1\text{Mpc}} < 2.97\text{nG}$ .

On the other hand, the upper limit determined by Faraday rotation induced by PMF on CMB polarization is, $B_{1\text{Mpc}} < 1380\text{nG}$. (at 70 GHz, low $l$<30). This calculation was based on the assumption that, the PMF was present in last scattering ($t \sim 10^5 yr$ after Big Bang). Also, on the assumption that, EE-spectrum transforms to BB-spectrum as a result of Faraday rotation. The upper limit of PMF is larger than the other limit, but it only apply at low frequency and angular scale, $l$. Therefore, it is a weak constraint and even weaker than the same constraint imposed by WMAP before.

Finally, the effect of PMF on ionization history of the Universe may modify CMB temperature and polarization power spectra, through energy dissipative effects. Considering this effect may result an upper limit for the scale invariant of PMF to be about $B_{1\text{Mpc}} < 0.67\text{nG}$. All of the forgoing constraints do not favor any particular model of generating PMF. In conclusion, the upper limit of PMF constrained by Planck 2015 is of the order of a few $n$G.

# CHAPTER FIVE: THE PMF GENERATED BY $f^2FF$ IN EXPONENTIAL INFLATION

After detecting magnetic fields in the intergalactic low density region, the early universe models have gained more attraction. One can divide the early universe models of PMF into two types of models; inflation and post-inflation models. The second type is mainly based on the phase transitions and preheating eras. It is believed that there were two main phase transitions processes took place in the early universe before the recombination era. Those are the electroweak (EW) phase transition, at $T_{EW} \sim 100\text{GeV}$, and quantum chromodynamical (QCD) phase transition, at $T_{QCD} \sim 200\text{MeV}$ [53, 66, 71]. However, the preheating era took place right after inflation and before the radiation dominated era. During the reheating, the field of inflation converted into particles. As a result of particles motion, the universe became very hot after super-cooling during inflation. This period can be divided into three stages: *preheating*, *heating*, and *thermalisation*.

The early universe models of PMF are not free from problems especially post-inflation ones. For example, phase transitions can produce strong magnetic fields but in small scale which is much less than what is observed. On the other hand, in the reheating models the amplification of the PMF is strongly suppressed by the electric conductivity. That in turns produced magnetic field much less than the required strength. In addition, many uncertainties and complexities are associated with these models, see [53, 66, 71] and the references therein. If the existence of PMF is confirmed in all directions and redshifts of the universe, then there will be a fundamental problem, similar to the horizon problem of CMB. Hence, inflation models are the best alternative way. They are the main objective of this research.

Several models were proposed to produce PMF in the inflation era [53, 66, 71]. Almost all of these models show some way to amplify the initial magnetic field. Unless the initial seed of magnetic fields is amplified, the exponential expansion of the space time will ultimately dilute the magnetic field to a level, at which it will not be sufficient to seed the observed values of PMF with or without galactic dynamo. Therefore, finding ways to amplify the original magnetic field during inflation era is the main idea of these models. To do so, one has to either, modify the classical electrodynamics theory in a flat universe, use non-flat universe, or use modified gravity. All of these avenues were richly investigated; see [53, 66, 71] and the references therein. Since the recent data collected by WMAP [17] and Planck [78] has been supporting the flat universe, the first way gets more attraction.

The model $f^2 FF$ at which a scalar field, like inflaton or dilaton, $\phi$, is coupled with a gauge vector field ,like electromagnetic vector potential, $A_\mu$, on a flat FRW universe is a one way to break the conformal symmetry. That in turns amplifies the PMF during inflation. Also, and as mentioned in chapter one, this model is stable under perturbation. For these reasons, we adopt this model in order to apply it on different inflationary models. We investigate its viability, the problems associate with, in the general homogenous and isotropic universe. In this chapter, this model will be investigated in the standard exponential inflationary potential. This is basically a reproducing of the results of Ref.[44]. In the next chapters, the same model will be investigated but in different inflationary models.

## 5.1 Scalar Fields Coupled to Vector Fields ($f^2 FF$) Inflationary Model of Generating PMF

The starting point is a Lagrangian of a scalar (so-called *inflaton*) field $\phi$ coupled to the electromagnetic (vector) field $A_\mu$ [43-44]. It can be written as

$$\mathcal{L} = -\sqrt{-g}\left(\frac{1}{2}(\partial_\mu \phi)(\partial^\mu \phi) + V(\phi) + \frac{1}{4}g^{\alpha\beta}g^{\mu\nu}f^2(\phi)F_{\mu\alpha}F_{\nu\beta}\right), \qquad (5.1)$$

where, $F_{\nu\beta} = \partial_\nu A_\beta - \partial_\beta A_\nu$ is the electromagnetic field tensor, and $g$ is the determinant of the spacetime metric $g_{\mu\nu}$. The first term in the Lagrangian is the standard kinetic part of the scalar field, and the second term, $V(\phi)$, is the potential, which decides the model of inflation. A Lagrangian of a pure electromagnetic field would be of the form $-\frac{1}{4}F_{\mu\nu}F^{\mu\nu}$, but here we are coupling it to the scalar field through the unspecified function $f(\phi)$. Thus, the action of the system, $S(\phi, A, \partial_\mu \phi, x^\mu)$, can be written as,

$$S(\phi, A, \partial_\mu \phi, x^\mu) = \int d^4x \, \mathcal{L}(\phi, A, \partial_\mu \phi, x^\mu) = -\int d^4x \sqrt{-g}\left(\frac{1}{2}(\partial_\mu \phi)(\partial^\mu \phi) + V(\phi) + \frac{1}{4}g^{\alpha\beta}g^{\mu\nu}f^2(\phi, t)F_{\mu\alpha}F_{\nu\beta}\right). \qquad (5.2)$$

Adopting the homogenous and isotropic space time, *Friedmann-Robertson-Walker FRW* metric, $g_{\mu\nu}$, (2.4), at which the cosmological distances change as a function of time only. Writing the Euler-Lagrange equations with respect to scalar and vector fields gives,

$$\partial_\tau \frac{\partial \mathcal{L}}{\partial_\tau \phi} - \frac{\partial \mathcal{L}}{\partial \phi} = -\frac{\delta L}{\delta \phi} = 0 \rightarrow \frac{1}{\sqrt{-g}}\frac{\partial}{\partial x^\mu}\left[\sqrt{-g}\,\partial^\mu \phi\right] - \frac{dV(\phi)}{d\phi} = \frac{1}{2}g^{\alpha\beta}g^{\mu\nu}f\frac{df}{d\phi}F_{\mu\alpha}F_{\nu\beta}, \quad (5.3)$$

$$\partial_\tau \frac{\partial \mathcal{L}}{\partial_\tau A_\tau} - \frac{\partial \mathcal{L}}{\partial A_\tau} = -\frac{\delta L}{\delta A_\tau} = 0 \rightarrow \partial_\mu\left[\sqrt{-g}\,g^{\mu\nu}g^{\alpha\beta}f^2(\phi, t)F_{\nu\beta}\right] = 0. \qquad (5.4)$$

We use the same method of [44, 45]. At spatially flat space time ($K = 0$), (2.4) can be written as,

$$ds^2 = g_{\mu\nu}dx^\mu \, dx^\nu = -dt^2 + a^2(t)dx^2 = g'_{\mu\nu}dx^\mu \, dx^\nu = -a^2(\eta)(d\eta^2 + dx^2), \qquad (5.5)$$

where, $\eta$ is the conformal time, $dt = a(\eta)d\eta$, $\sqrt{-g} = a^3(t)$, $\sqrt{-g'} = a^4(\eta)$, and $g^{\mu\nu}g_{\mu\beta} = \delta^\nu_\beta$. In inflation era, at which the universe is mainly driven by the scalar field in the exponentially expansion manner, the electromagnetic fields are negligible compared with inflation field. Thus, we can solve (5.3) by neglecting the right hand side. It yields the scalar field equation of motion which is the same as (2.16),

$$\ddot{\phi} + 3H\dot{\phi} + \frac{dV}{d\phi} = 0, \tag{5.6}$$

where, the single and double over dot respectively mean first and second derivative with respect to the cosmic time $t$, $H(t) = \frac{\dot{a}(t)}{a(t)}$, is the Hubble parameter or the rate at which the universe expands at arbitrary time. In the current time, as reported on nine year of WMAP, the Hubble constant, $H_0 \approx 69.32 \pm 0.80 \text{ kms}^{-1}\text{Mpc}^{-1}$ [17], and $H_0 \approx 67.3 \pm 1.2 \text{ kms}^{-1}\text{Mpc}^{-1}$ as reported by Planck [78].

In conformal time, (5.6) can be written as,

$$\phi'' + 2\mathcal{H}\phi' + a^2 \frac{dV}{d\phi} = 0, \tag{5.7}$$

where, $\phi' = \frac{\partial\phi}{\partial\eta} = a(\eta)\frac{\partial\phi}{\partial t} = a(\eta)\dot{\phi}$, and $\mathcal{H} = \frac{a'(\eta)}{a(\eta)}$. Adopting Coulomb (radiation) gauge, $\partial_i A^i(t,x) = 0$, and charge-free condition $A_0(t,x) = 0$, then (5.4) becomes,

$$\ddot{A}_i(t,x) + \left(H + \frac{2\dot{f}}{f}\right)\dot{A}_i(t,x) - \partial_j\partial^j A_i(t,x) = 0. \tag{5.8}$$

In conformal time,

$$A_i{}''(\eta,x) + 2\frac{f'}{f}A_i{}'(\eta,x) - a^2(\eta)\,\partial_j\partial^j A_i(\eta,x) = 0. \tag{5.9}$$

Define the function, $\overline{A}_i(\eta,x) = f(\eta)A_i(\eta,x)$, Eq (5.8) can be written as,

$$\overline{A}_i{}'' + \frac{f''}{f}\overline{A}_i - a^2\,\partial_j\partial^j\overline{A}_i(\eta,x) = 0. \tag{5.10}$$

If, $f = 1$, then (5.10) becomes,

$$\overline{A}_i{}'' - a^2\,\partial_j\partial^j\overline{A}_i = 0, \tag{5.11}$$

which is a simple harmonic oscillator.

The quantization of $\overline{A}_i$ can be written in terms of creation and annihilation operators, $b^\dagger{}_\lambda$ and $b_\lambda(k)$, as,

$$\overline{A}_i(\eta, x) = \int \frac{d^3 k}{(2\pi)^{3/2}} \sum_{\lambda=1}^{2} \varepsilon_{i\lambda}(k)[b_\lambda(k)\mathcal{A}(\eta, k)e^{ik.x} + b^\dagger{}_\lambda(k)\mathcal{A}^*(\eta, k)e^{-ik.x}], \quad (5.12)$$

where, $\varepsilon_{i\lambda}$ is the transverse polarization vector, and $k = \frac{2\pi}{\lambda}$, is the commoving wave number. A commoving frame (coordinates) is the frame that moves exactly with the cosmological flow of the universe (Hubble flow) without any peculiar velocity. According to the commoving observer in a *FRW* universe, the universe is perfectly isotropic.

The associate momentum conjugate to the gauge field, $A_j(t, x)$ can be defined as,

$$\pi^i(\eta, x) = \frac{\delta S}{\delta A'_i} = f^2(\phi) \, a^2(t) \, g^{ij} \, A'_j(t, x) \quad (5.13)$$

The canonical commutation relation between $A_i(t, x)$ and $\pi^i(t, x)$ is,

$$\left[ A_i(\eta, \boldsymbol{x}), \pi^i(\eta, \boldsymbol{y}) \right] = i \int \frac{d^3 \boldsymbol{k}}{(2\pi)^3} e^{i\boldsymbol{k}.(\boldsymbol{x}-\boldsymbol{y})} \left( \delta^i_j - \delta_{jl} \frac{k^i k^l}{k^2} \right) = i\delta_\perp^{(3)i}{}_j(\boldsymbol{x} - \boldsymbol{y}), \quad (5.14)$$

where, $\delta_\perp^{(3)}{}_{ij}$ is the transverse delta function which is consistent with Coulomb's gauge. Using (5.10-14), the Wronskian of $A_i(\eta, k)$ will be,

$$\overline{W}(\eta, k) = [\bar{A}\bar{A}'^* - \bar{A}^*\bar{A}'] = \frac{i}{f^2}. \quad (5.15)$$

Substituting of (5.12) into (5.10) gives,

$$\mathcal{A}''(\eta, k) + \left( k^2 - \frac{f''}{f} \right)\mathcal{A}(\eta, k) = 0, \quad (5.16)$$

where we have used $\partial^j = g^{jk}\partial_k = (\frac{\delta^{jk}}{a^2})\partial_k$.

Therefore, one has to solve (5.16) for $\mathcal{A}(\eta, k)$ in order to find the electromagnetic fields [46],

$$\mathrm{E}_\mu = u^\nu F_{\mu\nu}, \qquad \mathrm{B}_\mu = \frac{1}{2}\varepsilon_{\mu\nu\kappa\lambda}u^\lambda F^{\nu\kappa}, \quad (5.17)$$

where, $E_\mu$ and $B_\mu$ are the electric and magnetic fields respectively. For a commoving observer with 4-velocity $u^\nu = (1, 0)$ in cosmic time $t$, the spatial parts of electric and magnetic fields are respectively,

$$E_i = -\dot{A}_i, \qquad B_i = \frac{1}{a}\varepsilon_{ijk0}\partial_j A_k, \tag{5.18}$$

where, $\varepsilon_{\mu\nu\kappa\lambda}$ is a totally antisymmetric permutation tensor of space time. Similarly, the stress energy tensor can be written in terms of the action (5.2) [64],

$$T_{\mu\nu} = -\frac{2}{\sqrt{-g}}\frac{\delta S}{\delta g^{\mu\nu}} = (\partial_\mu\phi)(\partial_\nu\phi) - f^2(\phi,t)g^{\alpha\beta}F_{\mu\alpha}F_{\nu\beta} - \frac{1}{4}g_{\mu\nu}g^{\alpha\beta}g^{\gamma\delta}f^2(\phi,t)F_{\beta\delta}F_{\alpha\gamma}. \tag{5.19}$$

For the magnetic field, we ignore the kinetic part (first term) because it does not contribute to the electromagnetic field. Then,

$$T^B_{00} = \frac{1}{4}a^2 g^{ij}g^{kl}f^2(\phi,t)(\partial_j A_l - \partial_l A_j)(\partial_i A_k - \partial_k A_i). \tag{5.20}$$

The energy density of the magnetic field, $\rho_B$ is found by taking the average of the component of the stress energy tensor.

$$\rho_B(\eta) = -\langle T^B_{00}\rangle = \frac{1}{(2\pi)^3}\int d^3k \frac{1}{4}a^2 g^{ij}g^{kl}f^2(\phi,\eta)(\partial_j A_l - \partial_l A_j)(\partial_i A_k - \partial_k A_i). \tag{5.21}$$

Finally, the spectrum of the magnetic field can be found by $\frac{d\rho_B}{d\ln k}$, where $k$ is the commoving wave number,

$$\frac{d\rho_B}{d\ln k} = \frac{1}{2\pi^2}\left(\frac{k}{a}\right)^4 k|\mathcal{A}(\eta,k)|^2. \tag{5.22}$$

Similarly, the spectrum of the electric field can be calculated by,

$$\frac{d\rho_E}{d\ln k} = \frac{f^2}{2\pi^2}\frac{k^3}{a^4}\left|\left[\frac{\mathcal{A}(\eta,k)}{f}\right]'\right|^2, \tag{5.23}$$

where the prime is the derivative with respect to conformal time, $\eta$. Therefore, calculating $\mathcal{A}(\eta,k)$ is the main task in this research. Once it is obtained, then the rest of the results will be based on the form of $\mathcal{A}(\eta,k)$ under different circumstances.

## 5.2 Short Wavelength Regime, $\lambda \ll 1$ (inside Hubble radius)

At a very short wave length (subhorizon), $k^2 \gg 1$, inside the particle horizon (Hubble radius = c/H). Hence, $k^2$ dominates the second term of (5.16),

$$\mathcal{A}''(\eta, k) + k^2 \mathcal{A}(\eta, k) = 0, \tag{5.24}$$

The solution of (5.24) is the simple oscillator as the one in Minkowski space vacuum,

$$\mathcal{A}(\eta, k) = c_1(k) e^{ik\eta} + c_2(k) e^{-ik\eta}, \tag{5.25}$$

where, $c_1(k)$ and $c_2(k)$ are the integration constants. It is similar to the pure electromagnetic condition at which, $f(\phi) = 1$. Choosing the positive frequency as it is in Minkowski space vacuum state and from Wronskian (5.15), $c_2(k) = 1/\sqrt{2k}$. Thus, one can write the solution,

$$\mathcal{A}_{k \gg 1}(\eta, k) \to \frac{1}{\sqrt{2k}} e^{-ik\eta}, \tag{5.26}$$

Where, we used the limits ($t \to 0$ as $-k\eta \to \infty$).

During exponential inflationary model, one can solve Einstein field equation analytically to get the exact solution of scale factor in conformal time as [44, 121],

$$a(\eta) = a_0 \left| \frac{\eta}{\eta_0} \right|^{1+\beta}, \tag{5.27}$$

where, $a_0$ and $\eta_0$ are some initial values and, $\beta$, is the power of evolution. For example, for $= -2$ , one can achieve de Sitter space time evolution,

$$a(t) \propto \exp(Ht). \tag{5.28}$$

The inflation period in conformal time will lie in the period, $-\infty < \eta < 0$. Also, in this section and as done in Ref.[44], we adopt the power law form of coupling function,

$$f \propto a^\alpha \to f(\phi(\eta)) \propto \left( a_0 \left| \frac{\eta}{\eta_0} \right|^{1+\beta} \right)^\alpha = a_0 \left| \frac{\eta}{\eta_0} \right|^{(1+\beta)\alpha} = a_0 \left| \frac{\eta}{\eta_0} \right|^\gamma, \tag{5.29}$$

where, $\gamma = (1+\beta)\alpha$ in the exponential potential. In the next section, the more general form of coupling function will be derived. Hence from (5.22-23), the spectrum of magnetic and electric fields will respectively be,

$$\frac{d\rho_B}{d\ln k}(\eta, k) = \frac{k^5 \left|\frac{\eta}{\eta 0}\right|^{-4-4\beta}}{4a_0{}^4\pi^2|k|},$$ (5.30)

$$\frac{d\rho_E}{d\ln k}(\eta, k) = \frac{k^3\eta^{2\gamma} \left|-\frac{e^{-ik\eta}\gamma\eta^{-1-\gamma}}{\sqrt{2}\sqrt{k}} - \frac{ie^{-ik\eta}\sqrt{k}\eta^{-\gamma}}{\sqrt{2}}\right|^2 \left|\frac{\eta}{\eta 0}\right|^{-4-4\beta}}{2a_0{}^4\pi^2}.$$ (5.31)

As the short wavelength is not the relevant one, the electromagnetic spectra (5.30-31) are not derived explicitly in [44]. We compute them for completeness. The shape of electric and magnetic spectra for short wave length for, $a_0 = 1$, $\eta_0 = 1$, and $\beta = -2$ can be depicted in Fig.5.1.

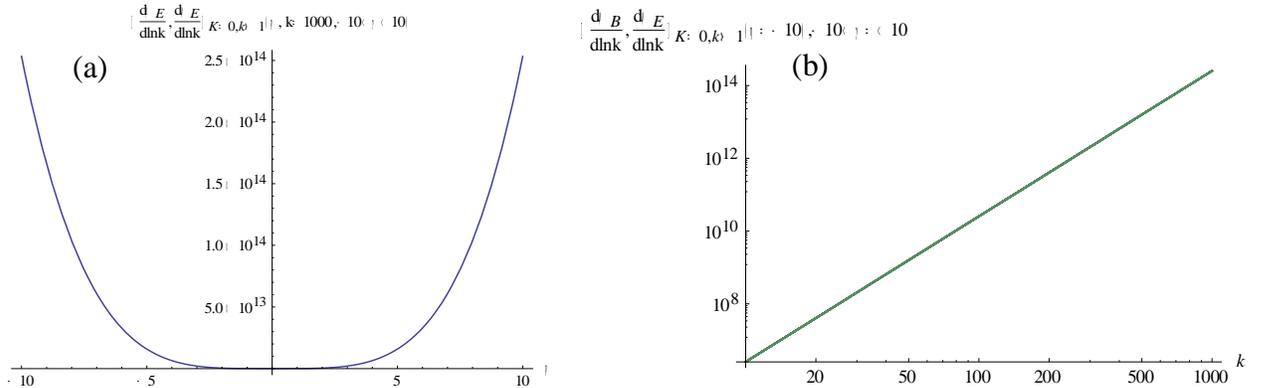

**Fig.5.1.** The magnetic and electric spectra, in the flat universe, $K = 0$, at the limit of $(k \gg 1, k = 1000)$, for, $-10 \leq \eta \leq 10$ and $-10 < \gamma < 10$ in electric field in (a) normal scale and (b) Log-Log scale. They have the same shape and magnitude, as we use the natural unit ($c$=1).

The shape of electromagnetic spectrum in Fig.5.1 is similar to the spectrum of conformal (normal) electromagnetic wave, as if $f = 1$. The magnetic and electric fields are of the same shape and magnitude, because of the unit we adopt ($c = 1$).

### 5.3 Long Wavelength Regime, $\lambda \gg 1$ (outside Hubble radius)

Long wavelength regime is the most relevant limit to PMF. As indicated in chapter 4, PMF is detected almost in all universal scales. Hence, the seed of PMF is most likely generated in the inflation era with a dimension larger than the Hubble horizon at that early time. So, the PMF can exit the horizon at some pivot point and re-enter the horizon later in radiation dominant or dark era of the universe. As a result, a scale invariant PMF can be detected in all scales of the Universe.

In this limit, $k^2 \ll 1$, one has to decide the form of coupling function $f(\phi)$, and then look for the general solution of (5.16). In this regime, the scale of wavenumber is much higher than Hubble radius, $\ll a H$. By assuming the de Sitter expansion, (2.24), one can re-write the previous limit as, $|k \eta| \ll 1$. On the other hand, the form of coupling function directly depends on the potential of inflation $V(\phi)$. In this chapter, the simple power law of the coupling function (5.29) is used. However, in the next section, the more general form of $f(\phi)$, will be derived.

By using the power law of scale factor, Eq.(5.27) , and adopting the form (5.29) for coupling function, Eq.(5.16) can be re-written as,

$$\mathcal{A}''(\eta, k) + \left(k^2 - \frac{\gamma(\gamma-1)}{\eta^2}\right)\mathcal{A}(\eta, k) = 0. \tag{5.32}$$

It is a Bessel differential equation of a form [79],

$$\frac{d^2y}{dx^2} - \frac{2\delta-1}{x}\frac{dy}{dx} + \left(\sigma^2\varpi^2 x^{2\varpi-2} + \frac{\delta^2-n^2\varpi^2}{x^2}\right)y = 0. \tag{5.33}$$

It has a general solution, which can be written as

$$y = \begin{cases} x^\delta[A\,J_n(\beta\,x^\varpi) + B\,Y_n(\beta\,x^\varpi)\,] & \text{for integer } n, \\ x^\delta[A\,J_n(\beta\,x^\varpi) + B\,J_{-n}(\beta\,x^\varpi)\,] & \text{for noninteger } n. \end{cases} \tag{5.34}$$

Comparing (5.32) and (5.33) implies $\delta = \frac{1}{2}$, $\varpi = 1$, $\sigma = k$, and $n = \gamma - \frac{1}{2}$. As $\gamma$ can be any real value, the general solution of (5.32) can be written as,

$$\mathcal{A}(\eta, k) = \eta^{1/2} \big[ C'_1(k) \, J_{\gamma-1/2}(k\eta) + C'_2(k) \, J_{-\gamma+1/2}(k\eta) \big], \qquad (5.35)$$

where the coefficients $C'_1(k)$ and $C'_2(k)$ are fixed by the initial conditions. Without losing the generality of the solution, (5.35) can be written as,

$$\mathcal{A}(\eta, k) = (k\eta)^{1/2} \big[ C_1(k) \, J_{\gamma-1/2}(k\eta) + C_2(k) \, J_{-\gamma+1/2}(k\eta) \big]. \qquad (5.36)$$

The shape of Bessel function is an oscillatory function, see Fig.5.2 which shows the shape of $J_\gamma$, for $\gamma = 0, 1/2, 1, \dots 5$.

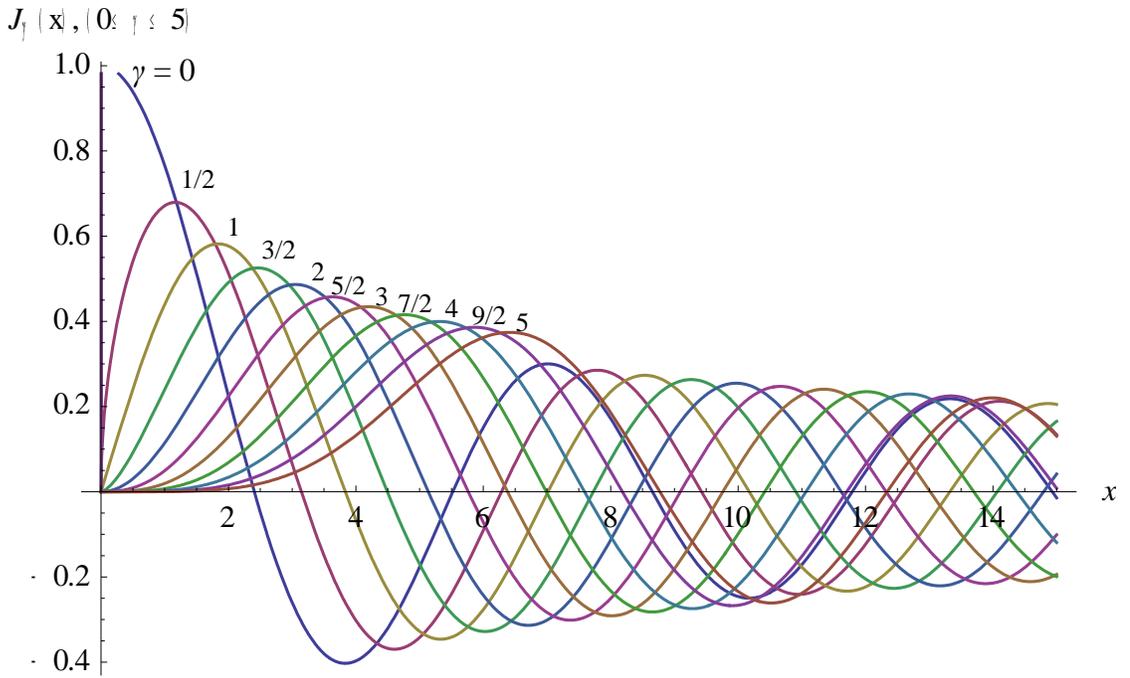

**Fig.5.2**. The Bessel function of the first kind, $J_\gamma(x)$, for $\gamma = 0, 1/2, 1, \dots 5$.

For example at ultraviolet regime $(k \gg 1)$, we can recover the solution (5.25) from (5.36) by using the asymptotic expansion of Bessel equation for $k\eta \to \infty$,

$$\lim_{x \to \infty} J_\nu(x) \; \to \; \sqrt{\frac{2}{\pi x}} \cos \Big[ x - (\nu + \tfrac{1}{2}) \tfrac{\pi}{2} \Big]. \qquad (5.37)$$

So, (5.36) becomes,

$$\mathcal{A}_{k \gg 1}(\eta, k) = (k\eta)^{1/2} \Big[ C_1(k) \sqrt{\frac{2}{\pi k\eta}} \cos \Big[ k\eta - \gamma \tfrac{\pi}{2} \Big] + C_2(k) \sqrt{\frac{2}{\pi k\eta}} \cos \Big[ k\eta - (1-\gamma) \tfrac{\pi}{2} \Big] \Big]. \, (5.38)$$

The two constants can be fixed by using (5.25),

$$C_1(k) = \sqrt{\frac{\pi}{4k}} \frac{\exp(\frac{-i\pi\gamma}{2})}{\cos(\pi\gamma)} \qquad ; \qquad C_2(k) = \sqrt{\frac{\pi}{4k}} \frac{\exp(\frac{i\pi(\gamma+1)}{2})}{\cos(\pi\gamma)}. \qquad (5.39)$$

On the other hand, at long wavelength regime ($k \ll 1$), using the asymptotic limit of Bessel,

$$\lim_{x \to 0} J_\nu(x) \to \frac{x^\nu}{2^\nu \Gamma(\nu+1)}, \qquad (5.40)$$

and substituting it into (5.36) gives,

$$\mathcal{A}_{k \ll 1}(\eta, k) = (k)^{-1/2} [c_1(\gamma) (-k\eta)^\gamma + c_2(\gamma)(-k\eta)^{1-\gamma}], \qquad (5.41)$$

where the constants of integration can be written as,

$$c_1(\gamma) = \frac{\sqrt{\pi}}{2^{\gamma+1/2}} \frac{e^{-i\pi\gamma/2}}{\Gamma\left(\gamma+\frac{1}{2}\right)\cos(\pi\gamma)},$$

$$c_2(\gamma) = \frac{\sqrt{\pi}}{2^{3/2-\gamma}} \frac{e^{-i\pi(\gamma+1)/2}}{\Gamma\left(\frac{3}{2}-\gamma\right)\cos(\pi\gamma)}. \qquad (5.42)$$

From the denominators of the constants, we can notice that, $c_1(\gamma)$ dominates for $\gamma \leq 1/2$, and $c_2(\gamma)$ dominates for $\gamma \geq 1/2$. See the shape of $\mathcal{A}_{k \ll 1}(\eta, k)$ in Fig.5.3.

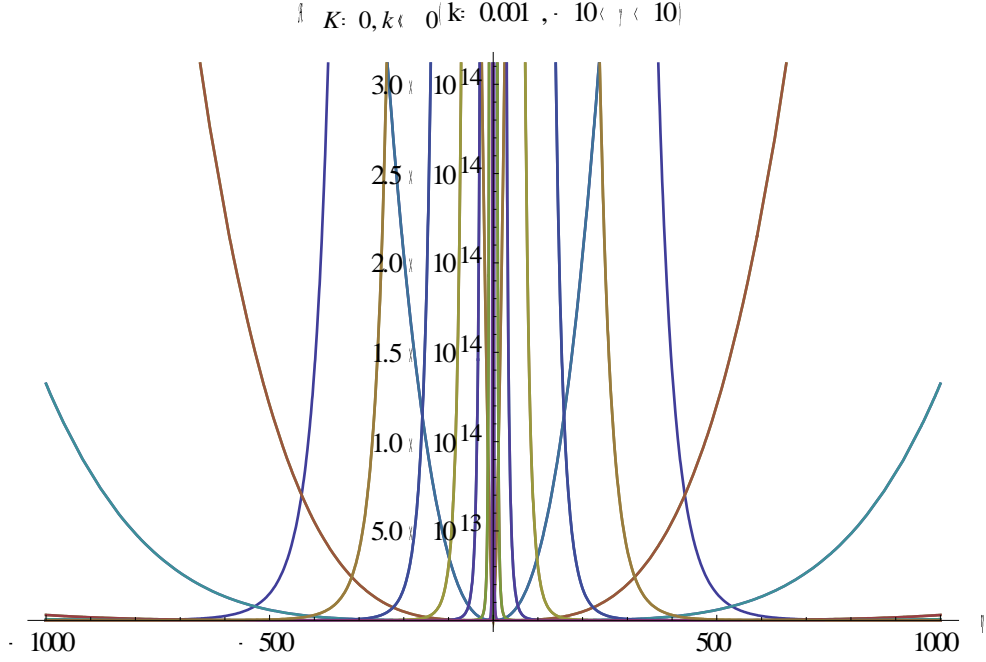

**Fig.5.3**. The shape of vector potential, $\mathcal{A}(\eta, k)$, in the flat universe, $K = 0$, at the limit of ($k \ll 1, k = 0.001$), for, $-10^3 \le \eta \le 10^3$ and $-10 < \gamma < 10$.

Another way to solve for $\mathcal{A}(\eta, k)$ at long wave length is by taking $k \to 0$ in the original differential equation, (5.16). A direct integration implies,

$$\mathcal{A}(\eta, k) = \bar{c_1} f(\eta) + \bar{c_2} f(\eta) \int \frac{d\eta}{f^2(\eta)}. \tag{5.43}$$

By using (5.29) and (5.27), one can get the same solution as (5.41).

Substituting of (5.41) into (5.22) gives the spectrum of magnetic field,

$$\frac{d\rho_B}{d\ln k}(\eta, k) = \frac{\mathcal{F}(n)}{2\pi^2} H^4 \left(\frac{k}{aH}\right)^{4+2n} \approx \frac{\mathcal{F}(n)}{2\pi^2} H^4 (-k\eta)^{4+2n}. \tag{5.44}$$

The dimensionless function $\mathcal{F}(n)$ is defined by,

$$\mathcal{F}(n) = \frac{\pi}{2^{2n+1}\Gamma^2\left(n+\frac{1}{2}\right)\cos^2(n\pi)}, \tag{5.45}$$

where, $n = \gamma$ for $\gamma \le 1/2$ and $n = 1 - \gamma$ for $\gamma \le 1/2$. In a more explicit form, one can compute the magnetic spectrum by substituting (5.24) and (5.41-42) into (5.22) to obtain,

$$\frac{d\rho_B}{d\ln k}(\eta, k) = \frac{k^5 \left|\frac{\eta}{\eta_0}\right|^{-4-4\beta} \left| \frac{2^{-\frac{3}{2}+\gamma} e^{\frac{1}{2}i\pi(1-\gamma)} \sqrt{\pi}(k\eta)^{1-\gamma} \operatorname{Sec}[\pi(1-\gamma)]}{\sqrt{k}\ \Gamma(\frac{3}{2}-\gamma)} + \frac{2^{-\frac{1}{2}-\gamma} e^{\frac{i\pi\gamma}{2}} \sqrt{\pi}(k\eta)^{\gamma} \operatorname{Sec}[\pi\gamma]}{\sqrt{k}\ \Gamma(\frac{1}{2}+\gamma)} \right|^2}{2 a_0^{\ 4} \pi^2}. \tag{5.46}$$

Similarly, by substituting of (5.41) into (5.23) and making use of the limit (5.40) and the relations,

$$J'_\nu(x) = \frac{1}{2}\left(J_{\nu-1}(x) - J_{\nu+1}(x)\right),$$

$$J_\nu(x) = \frac{x}{2\nu}\left(J_{\nu-1}(x) + J_{\nu+1}(x)\right), \tag{5.47}$$

one can write the spectrum of electric field in the similar form of (5.44) as,

$$\frac{d\rho_E}{d\ln k}(\eta, k) = \frac{\mathcal{G}(m)}{2\pi^2} H^4 \left(\frac{k}{aH}\right)^{4+2m} \approx \frac{\mathcal{G}(m)}{2\pi^2} H^4 (-k\eta)^{4+2m}. \tag{5.48}$$

The dimensionless function $\mathcal{G}(m)$ is defined by,

$$\mathcal{G}(m) = \frac{\pi}{2^{2m+3} \Gamma^2\left(m+\frac{3}{2}\right) \cos^2(m\pi)}, \tag{5.49}$$

where, $m = \gamma + 1$ for $\gamma \leq -1/2$ and $m = -\gamma$ for $\gamma \geq -1/2$. Similarly, in a more explicit form, one can compute the electric spectrum by substituting (5.24) and (5.41-42) into (5.23) to obtain,

$$\frac{d\rho_E}{d\ln k}(\eta, k) = \frac{1}{2 a_0^{\ 4} \pi^2} k^3 \eta^{2\gamma} \left|\frac{\eta}{\eta_0}\right|^{-4-4\beta} \left| \frac{1}{\sqrt{k}} \eta^{-\gamma} \frac{(2^{-\frac{3}{2}+\gamma} e^{\frac{1}{2}i\pi(1-\gamma)} k \sqrt{\pi}(1-\gamma)(k\eta)^{-\gamma} \operatorname{Sec}[\pi(1-\gamma)])}{\Gamma(\frac{3}{2}-\gamma)} + \right.$$

$$\frac{2^{-\frac{1}{2}-\gamma} e^{\frac{i\pi\gamma}{2}} k \sqrt{\pi}\gamma(k\eta)^{-1+\gamma} \operatorname{Sec}[\pi\gamma]}{\Gamma(\frac{1}{2}+\gamma)} ) - \frac{1}{\sqrt{k}}\gamma\eta^{-1-\gamma} \left(\frac{2^{-\frac{3}{2}+\gamma} e^{\frac{1}{2}i\pi(1-\gamma)} \sqrt{\pi}(k\eta)^{1-\gamma} \operatorname{Sec}[\pi(1-\gamma)]}{\Gamma(\frac{3}{2}-\gamma)} + \right.$$

$$\left. \frac{2^{-\frac{1}{2}-\gamma} e^{\frac{i\pi\gamma}{2}} \sqrt{\pi}(k\eta)^{\gamma} \operatorname{Sec}[\pi\gamma]}{\Gamma(\frac{1}{2}+\gamma)} ) \right|^2. \tag{5.50}$$

Assuming that the PMF is scale invariant, the magnetic spectrum should be constant, $\frac{d\rho_B}{d\ln k} = \text{const}$. Also, since $H$ is almost constant during inflation, then from (5.44) and (5.48) one should take, $4 + 2n = 0$. Therefore, power index $\gamma = -2, 3$. The magnetic fields spectra at $-5 < \gamma < 5$ in the increment of 0.1 values are shown in Fig.5.4 based on (5.46). The 3D plot of $(\frac{d\rho_B}{dlnk} - k - \gamma)$ shows that only $\gamma = -2, 3$ has flat spectra, see Fig.5.5.

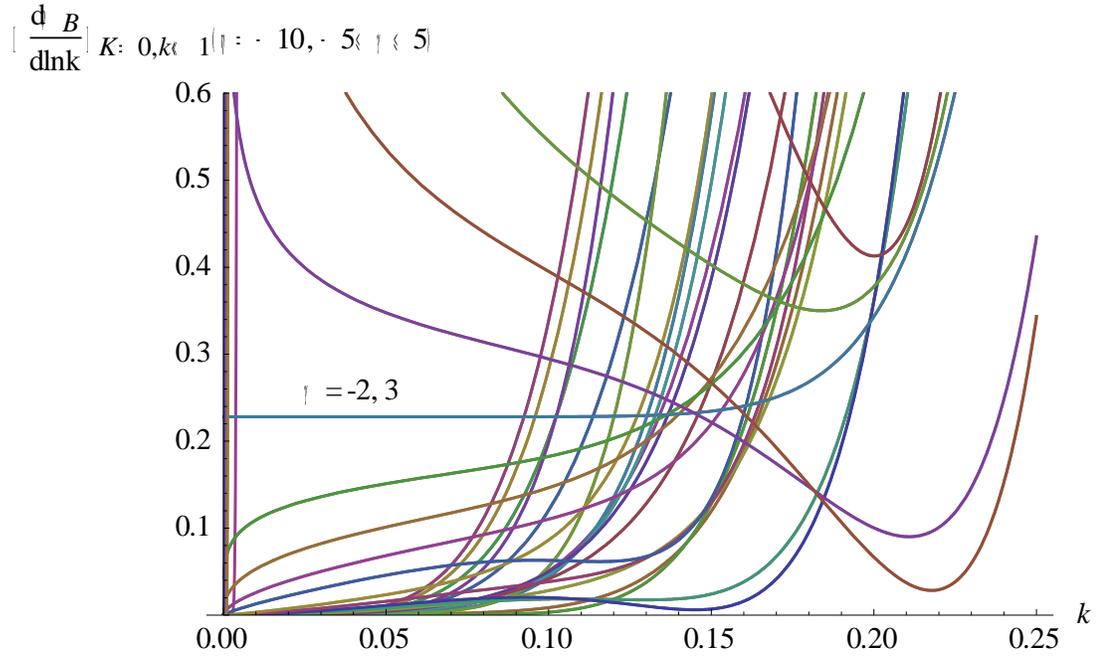

**Fig.5.4**. The magnetic spectra, in the flat universe, $K = 0$, at the limit of $(k \ll 1)$ for the values $-5 \leq \gamma \leq 5$, in the increment of 0.1. The spectrum is constant at $\gamma = -2, 3$.

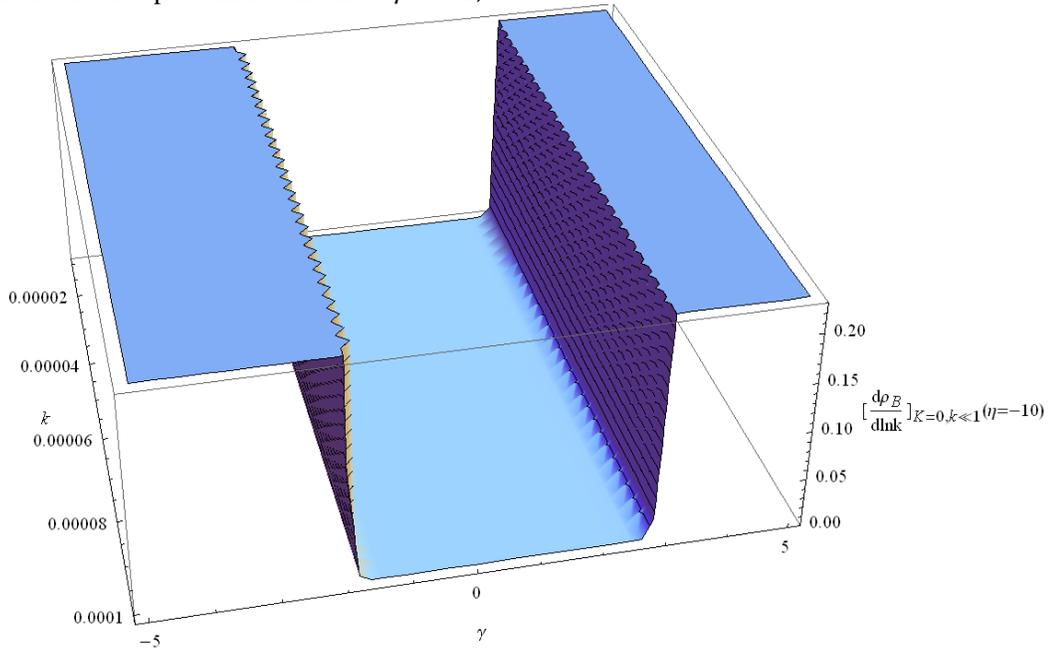

**Fig.5.5**. The 3D plot of magnetic field spectra, in the flat universe, $K = 0$, at the limit of $(k \ll 1)$, for the values $-5 \leq \gamma \leq 5$ where it is clearly flat at $\gamma = -2, 3$.

In contrast, the scale invariant of electric field spectrum leads to $4 + 2m = 0$, implying that $\gamma = -3, 2$. In fact, this condition is not necessary as the electric field is not detected everywhere. Also, the magnitude of electric field decays in the reheating era, because at that time the universe becomes electrically conducting. The electric fields spectra at $-5 < \gamma < 5$ in the increment of 0.1 are shown in Fig.5.6.

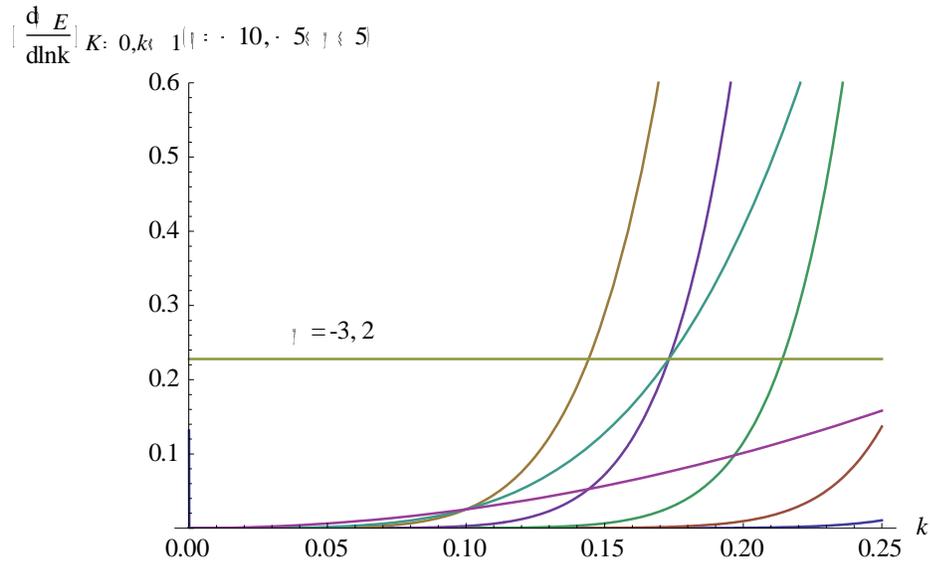

**Fig.5.6.** The electric field spectra, in the flat universe, $K = 0$, at the limit of $(k \ll 1)$ for the values $-5 \leq \gamma \leq 5$, in the increment of 0.1. The spectra are constant at $\gamma = -2, 3$.

Therefore, there is a mismatching between the values of $\gamma$ leading to the scale invariant condition in magnetic and electric field. This inconsistency in some cases leads to the divergence of electric fields which causes that the energy of the electric field density become more than the energy density of the universe. That may spoil the inflation and cause the *backreaction* problem [44, 45].

As a result, both electric and magnetic fields cannot be scale invariant in the same $\gamma$. As argued by [44, 45] and can be seen from Fig.5.7, at $\gamma = -2$, the problem of backreaction may be avoided for $k \ll 1$ as the spectrum of magnetic field is much higher than that of electric field.

On the other hand, some researchers [40] argued that, with this index, $\gamma = -2$, $f_i = f\left(\frac{a}{a_i}\right)^2$, where $f \to 1$, at the end of inflation. Hence, it may cause a strong coupling between electromagnetic fields and charged matter at the beginning of inflation. If the electromagnetic field couples to charged matter, the physical charge associated with, is so huge at the onset of inflation. For example, for the number of e-folds of inflation, $N = 60$, the physical electric charges, $q \propto e^{120}$ [40]. Such an incredibly huge charge creates another major problem to this model, the problem of strong coupling. This problem is an outstanding one associated with the $f^2FF$ model, regardless of the adopted inflationary model. *It is not addresssed in this research*. What we do is to investigate the problem of backreaction, which is most likely inflationary model dependent.

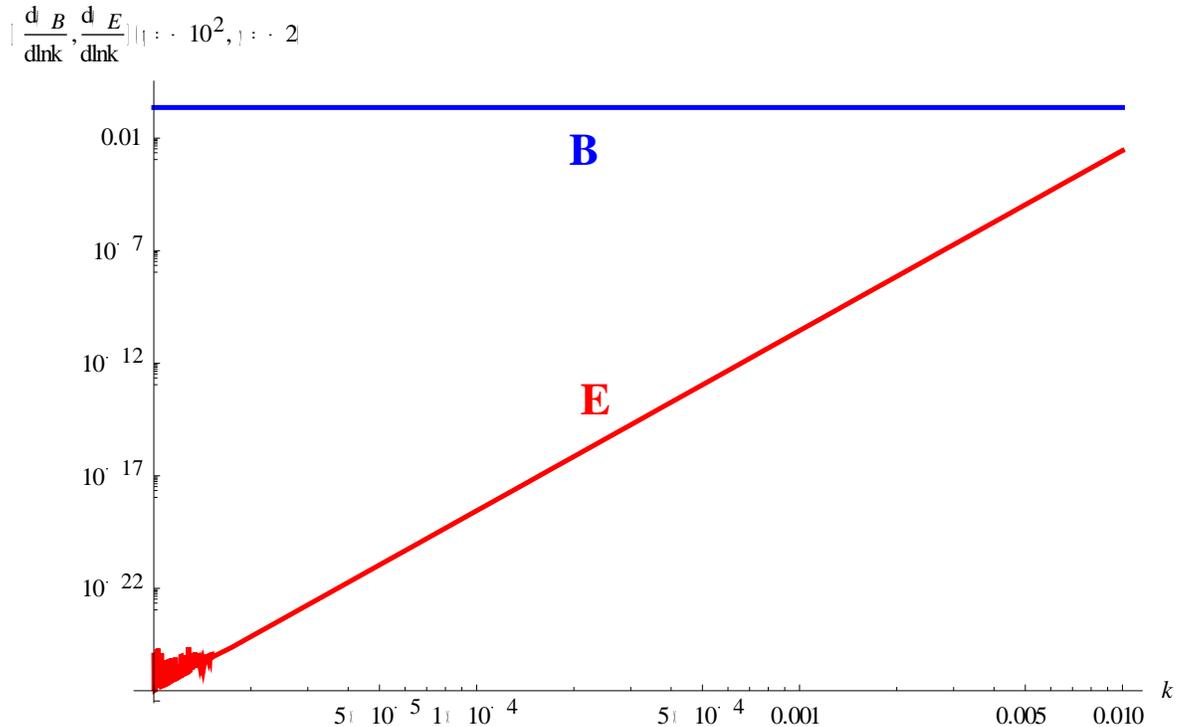

**Fig.5.7**. The magnetic and electric at the limit of ($k \ll 1$, $\gamma = -2$). It is clear that the spectrum of magnetic field is constant (scale invariant) in the small $k$ limit and much greater than the electric field spectrum.

However, for $\gamma = 3$, the electric field diverges for $k \ll 1$. As shown in Fig.5.8, in this case the backreaction problem cannot be avoided. In the next chapter, the generation of PMF in a flat universe will be discussed in the context of the large field inflation LFI, which is among models favored by the first BICEP2 results [54].

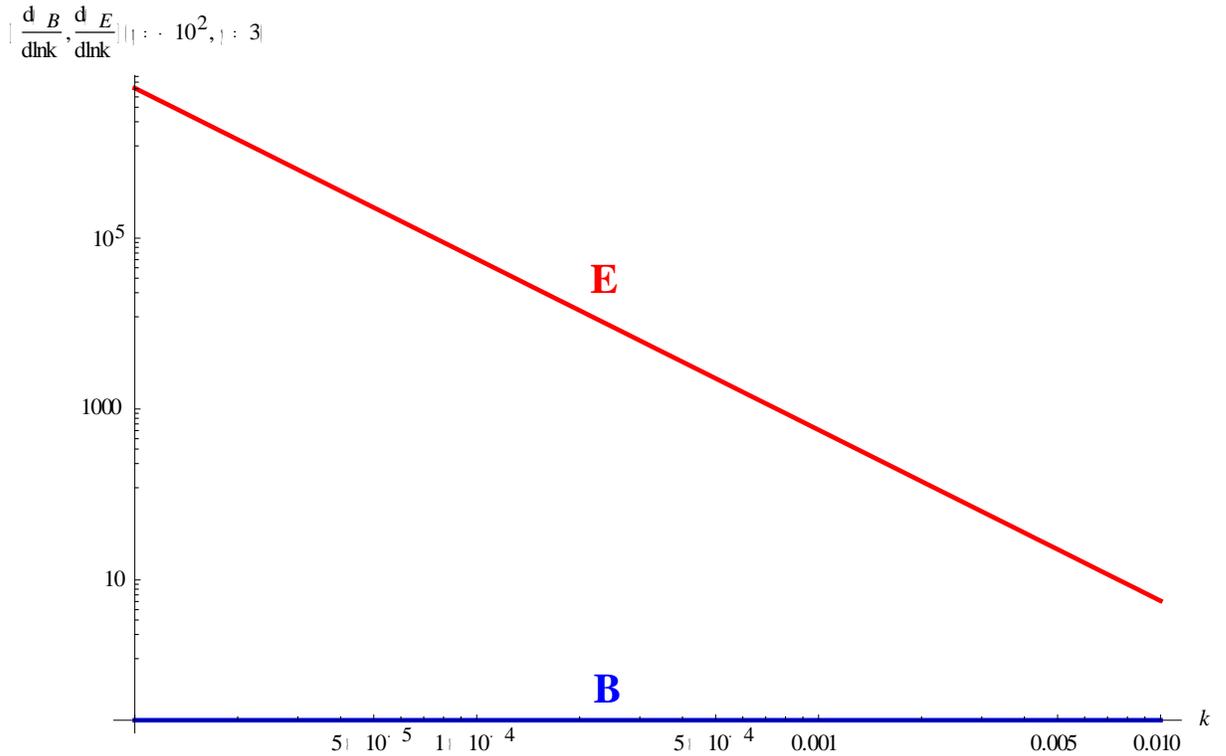

**Fig.5.8**. The magnetic and electric spectra in the limit of ($k \ll 1$, $\gamma = 3$). It is clear that the spectrum of electric field is much greater than the magnetic field spectrum.

## 5.4 The General Form of Coupling Function, $f(\phi)$

In this section, we derive the general relation between the coupling function, $f(\phi)$, and the form of an inflationary potential $V(\phi)$. Starting from the slow roll relations, (2.15) and (2.20) at which $\ddot{\phi} \ll \dot{\phi}$. Then they can be written as,

$$3H\dot{\phi} + \frac{dV}{d\phi} = 0, \qquad H = \frac{V(\phi)}{3M_{\mathrm{Pl}}^2} \qquad (5.51)$$

Using the definition of $H = \frac{\dot{a}}{a}$, one can combine (5.51) into

$$\frac{d\phi}{dt} = -\frac{dV}{3Hd\phi} = \frac{3M_{Pl}^2 \frac{\dot{a}}{a} dV}{V(\phi)d\phi}, \tag{5.52}$$

which can be re-written as,

$$\frac{da}{a} = -\frac{1}{3M_{Pl}^2} V(\phi) \left(\frac{dV}{d\phi}\right)^{-1} d\phi. \tag{5.53}$$

The scale factor, $a(\phi)$, can be written as a solution of (5.53) as

$$(\ln[a])_0^a = -\frac{8\pi G}{3} \int_0^\phi V(\phi) \left(\frac{dV}{d\phi}\right)^{-1} d\phi. \tag{5.54}$$

Hence, $a(\phi)$, will be,

$$a(\phi) = exp\left[-\frac{1}{3M_{Pl}^2} \int_0^\phi \frac{V(\phi)}{V'(\phi)} d\varphi\right]. \tag{5.55}$$

where, $V'(\phi) = \partial_\phi V$.

Assuming that the relation between couplings function and scale factor is in the power form, $f \propto a^\alpha$. Thus, by using (5.55), the coupling function can be written as,

$$f(\phi) \propto exp\left[-\alpha \frac{1}{3M_{Pl}^2} \int_0^\phi \frac{V(\phi)}{V'(\phi)} d\phi\right]. \tag{5.56}$$

The form (5.56) is more general and not necessary to be exactly of the same form as (5.29). Also, the index $\alpha$ is free for the general form of inflationary potential, $V(\phi)$. It can be constrained from observations, see section 3.2 of Ref.[44]. Furthermore, the relation, $\gamma = (1 + \beta)\alpha$, is no longer exactly the same as in the exponential inflationary model. Later on, whenever we refer to the value $\gamma$, it is only to decide the scale invariance property of PMF from Eq.(5.44). However, the dominant factor is the index of the Bessel solution, $n$, in (5.34). It will be denoted as, $\chi$ in the next chapters. Therefore, in the next three chapters we will use (5.56) to define the coupling function in the context of LFI, NI and $R^2$-inflation.

# CHAPTER SIX: THE PMF GENERATED BY $f^2 FF$ IN LARGE FIELD INFLATION

The interpretations as primordial gravitational waves of the detected signals of BICEP2 are most likely disproved by the latest results of Planck/Keck BICEP2 (PKB) 2015 [94]. Also, the constraints of inflationary models by Planck 2015 [96] favored the standard and old models that generally lead to low value of tensor to scalar ratio, $r$. However, we started this research before 2015, under the motivation of the first results of BICEP2 and fortunately we did not restrict the investigation on the constraints of BICEP2 only. Therefore, in this chapter an investigation of the PMF generated by $f^2 FF$ under LFI in general limits of inflation. Furthermore, the latest results announced in Feb 2015 are not needed to be employed and do not modify the main results of this research.

The tensor to scalar ratio reported by the first BICEP2 results is $r = 0.2^{+0.07}_{-0.05}$, with $r = 0$ disfavored at $7.0\sigma$. Also, the scale of inflation energy is close to the *Grand Unified Theory* GUT scale, $\rho_{\text{GUT}}^{1/4} \sim 10^{16} \text{GeV}$. As a matter of fact, tensor perturbation directly indicates the energy scale of inflation [21]. According to Planck (2013) [56], the upper bound on the tensor-to-scalar ratio is $r < 0.11$ at 95% CL ($r < 0.12$ at 95% CL, PKB 2015) and the scalar spectral index was constrained by Planck to be $n_s = 0.9603 \pm 0.0073$ ($n_s = 0.968 \pm 0.006$, Planck 2015). Many inflationary models, including simple models supported by WMAP9 and Planck were inconsistent with first interpretation of BICEP2 detection [57-58]. The diagram of $r_{0.002} - n_s$ relation drawn from BICEP2 and Planck 2013 in combination with other data sets compared to the theoretical predictions of selected inflationary models is shown in Fig.6.1 [56, 58].

Based on BICEP2 results, over 190 models of inflation [81] had been classified into 111 strongly disfavored models, 24 inconclusive models, and some good models [58]. Therefore, non-standard models, such as Large Field Inflation (LFI) [59], Natural Inflation (NI) [60-61] models had been favored by the BICEP2 results and for short period of time had gained more interest. We investigated the generating of PMF in natural inflation in [92]. As a result of raising the scale of inflation energy to GUT scale, the large field potentials, at which $V(0) \rightarrow 0$, are favored and the plateau inflation potentials at which $V(0) \neq 0$ are disfavored [58].

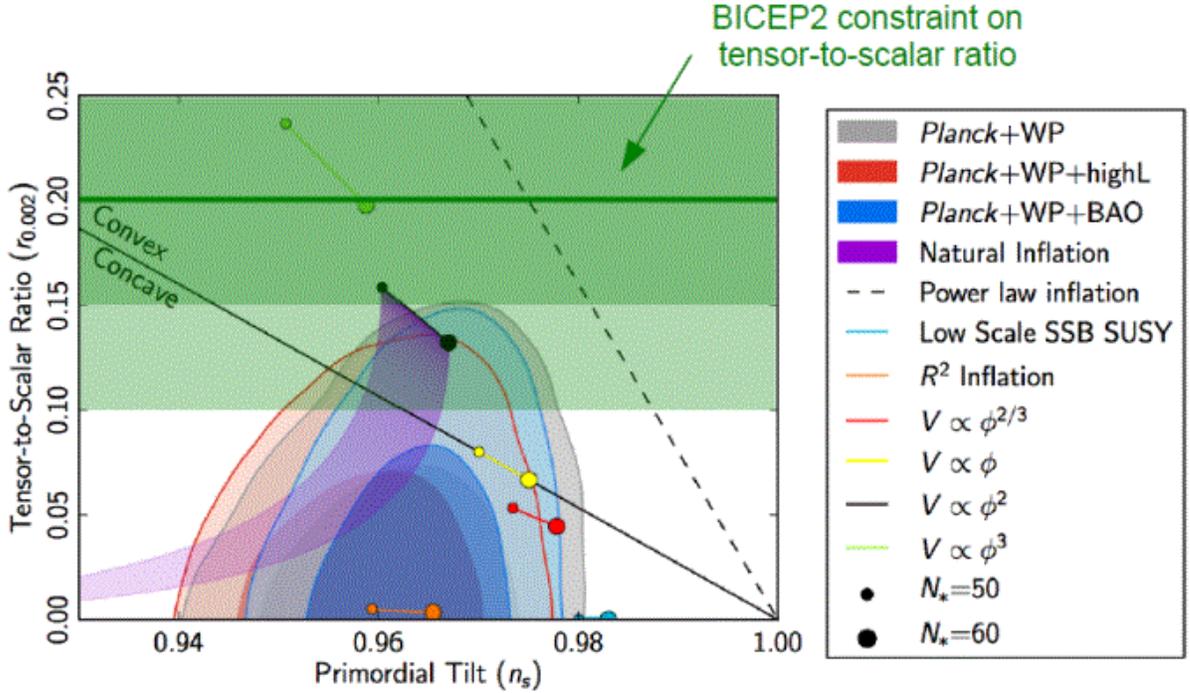

**Fig.6.1**: The constraints of inflation for different models as shown in $r_{0.002} - n_s$ diagram drawn from BICEP2 and Planck in combination with other data sets compared to the theoretical predictions of selected inflationary models. Courtesy, Planck 2013 Collaboration [56].

In Sep 2014, Planck released the angular power spectrum of polarized dust emission at intermediate and high galactic latitudes [89]. The detection frequency of Planck (353 GHz) is different than that of BICEP2 (150 GHz). But in these results, they extrapolated the power spectrum to the same frequency. Also they observed the same patch of sky at high Galactic latitude, which was observed by BICEP2, at low multipoles $40 < l < 120$.

These results of Planck indicate that there is a significant contamination by dust over most of the high Galactic latitude sky in the same region where BICEP2 detected B-mode polarization. Consequently, there is a good chance that the source of the observations reported by BICEP2 is all galactic dust and not of primordial origin. This problem was anticipated by [90]. The results of the present investigation can be considered, in a way, as conflicting with the BICEP2 results.

In this chapter, the simple inflation model, $f^2FF$, of PMF will be investigated in detail under the large field inflation LFI, for all possible values of the model parameter, $p$, in the same way as done in [44]. The potential of LFI can be written [59, 81] as,

$$V(\phi) = M^4 \left( \frac{\phi}{M_{Pl}} \right)^p,$$ (6.1)

where, $p$ is the model parameter and $M$ is the normalization of the potential. The value of $M$ is constrained from the amplitude of CMB anisotropies to be, $M \simeq 3 \times 10^{-3} M_{Pl}$ [81].

The order of this chapter will be as follows, in section 6.1, the slow roll inflation formulation will be presented for both simple de Sitter model of expansion and the more general power expansion. In section 6.2, the PMF and associated electric fields spectra will be computed for LFI at some interesting values of $p$. In section 6.3, a summary and discussion of the results will be presented.

## 6.1 Sow Roll Analysis of LFI

The slow roll parameters of inflation in terms of the potential of a single field inflation for LFI can be written [55] as

$$\epsilon_{1V}(\phi) = \frac{1}{2} M_{\text{Pl}}{}^2 \left( \frac{V_\phi}{V} \right)^2 = \frac{1}{2} M_{\text{Pl}}{}^2 \left( \frac{p}{\phi} \right)^2, \tag{6.2}$$

$$\epsilon_{2V}(\phi) = M_{\text{Pl}}{}^2 \left( \frac{V_{\phi\phi}}{V} \right) = M_{\text{Pl}}{}^2 \frac{p(p-1)}{\phi^2}. \tag{6.3}$$

They also, can be written in terms of the Hubble parameter,

$$\epsilon_{1H}(\phi) = 2 M_{\text{Pl}}{}^2 \left( \frac{H_\phi}{H} \right)^2, \quad \epsilon_{2H}(\phi) = 2 M_{\text{Pl}}{}^2 \left( \frac{H_{\phi\phi}}{H} \right). \tag{6.4}$$

The relation between the two formalisms [55] can be written as

$$\epsilon_{1V} = \epsilon_{1H} \left( \frac{3 - \epsilon_{2H}}{3 - \epsilon_{1H}} \right)^2. \tag{6.5}$$

All of the above parameters are assumed to be very small during the slow roll inflation, $(\epsilon_{1V}, \epsilon_{2V}, \epsilon_{1H}, \epsilon_{2H}) \ll 1$. Further, inflation ends when, the values of $(\epsilon_{1V}, \epsilon_{1H}) \to 1$. In the first order of approximation, one can neglect $\epsilon_{1H}$ and $\epsilon_{2H}$ comparing with the number 3 in (6.5), obtaining $\epsilon_{1V} \simeq \epsilon_{1H}$. Therefore, using (6.4) and the relation between the cosmic, $t$, and the conformal time, $\eta$, $dt = a(\eta) d\eta$, one can write the relation between conformal time and slow roll parameter, $\epsilon_{1H}$, as [55],

$$\eta = -\frac{1}{aH} + \int \frac{\epsilon_{1H}}{a^2 H} da. \tag{6.6}$$

Recall that $H(\eta) = a'(\eta) / a^2(\eta)$ in (6.6), where $a'(\eta) = da / d\eta$. Assuming, $\epsilon_{1H} \approx const$, and then integrating (6.6) yields the power law expansion of the universe during inflation,

$$a(\eta) = l_0 |\eta|^{-1 - \epsilon_{1H}}. \tag{6.7}$$

where, $l_0$ is integration constant.

On the other hand, in the simplest form of inflationary expansion (de Sitter), the universe expands exponentially during inflation at a very high but constant rate, $H_i$,

$$H_i = \frac{\dot{a}}{a} \simeq \text{const} , \qquad (6.8)$$

$$a(t) = a(t_1) \exp[H_i t], \qquad (6.9)$$

where, $t_1$, is the start time of inflation. We use $H_i$ with subscript $i$ to denote the constant expansion rate in the case of de Sitter expansion. Thus in a conformal time, Eq.(6.9) can be written as,

$$a(\eta) = -\frac{1}{H_i \eta + c_1} , \qquad (6.10)$$

where, $c_1$, is the integration constant. Plugging (6.10) into the relation between cosmic and conformal time and integrating implies that $\eta \rightarrow (-\infty, \, 0^-)$ as $t \rightarrow (0, \, \infty)$. Thus, if $\eta \rightarrow 0^-$, then $a(\eta) \rightarrow \infty$ and $c_1 \rightarrow 0$. Therefore, we can write $a(\eta) = -1 / (H_i \eta)$.

Also, the relation between slow roll parameters and the scalar power spectrum amplitude, $A_s$, the tensor power spectrum amplitude, $A_t$, the scalar spectral index, $n_s$, and tensor-to-scalar ratio, $r$, can be written as follows [55],

$$A_s = \frac{V}{24\pi^2 M_{\text{Pl}}{}^4 \epsilon_{1V}} , \qquad (6.11)$$

$$A_t = \frac{2V}{3\pi^2 M_{\text{Pl}}{}^4} , \qquad (6.12)$$

$$n_s = 1 - 6\epsilon_{1V} + 2\epsilon_{2V}, \tag{6.13}$$

$$r = \frac{A_t}{A_s} = 16\epsilon_{1V}. \tag{6.14}$$

Using LFI potential (6.1) into (6.2)-(6.3) yields

$$n_s = 1 - M_{\text{Pl}}^2 \frac{p(2p+1)}{\phi^2}, \tag{6.15}$$

$$r = 8M_{\text{Pl}}^2 \frac{p^2}{\phi^2}, \tag{6.16}$$

One can find the relation between $r$ and $n_s$ which depends on the number of e-folds of inflation, $N$. The first order of approximation for $N$ can be written as,

$$N \simeq -\sqrt{\frac{1}{2M_{\text{Pl}}^2}} \int_{\phi}^{\phi_f} \frac{1}{\sqrt{\epsilon_{1V}}} d\phi \sim \frac{\phi^2}{2M_{\text{Pl}}^2 p}, \tag{6.17}$$

where, $\phi$, is the initial field and $\phi_f$ is the field at the end of inflation. In (6.17), we have adopted BICEP2 constraint, (6.23), at which $\phi_f \ll \phi$. Solving for $\phi$ from (6.17) and plugging it into (6.15)-(6.16) yields

$$r \simeq \frac{4p}{N}\left(n_s + \frac{2p+1}{2N}\right). \tag{6.18}$$

The relation (6.18) fits Fig.6.1, in a good precision. For BICEP2 constraints, $0.15 < r < 0.27$, at $n_s = 0.960$, we have, $2.62 < p < 4.57$ for $N = 70$, $2.24 < p < 3.92$ for $N = 60$, and $1.86 < p < 3.26$ for $N = 50$. However, for BK/Planck, we have $r < 0.12$ see Fig.6.2.

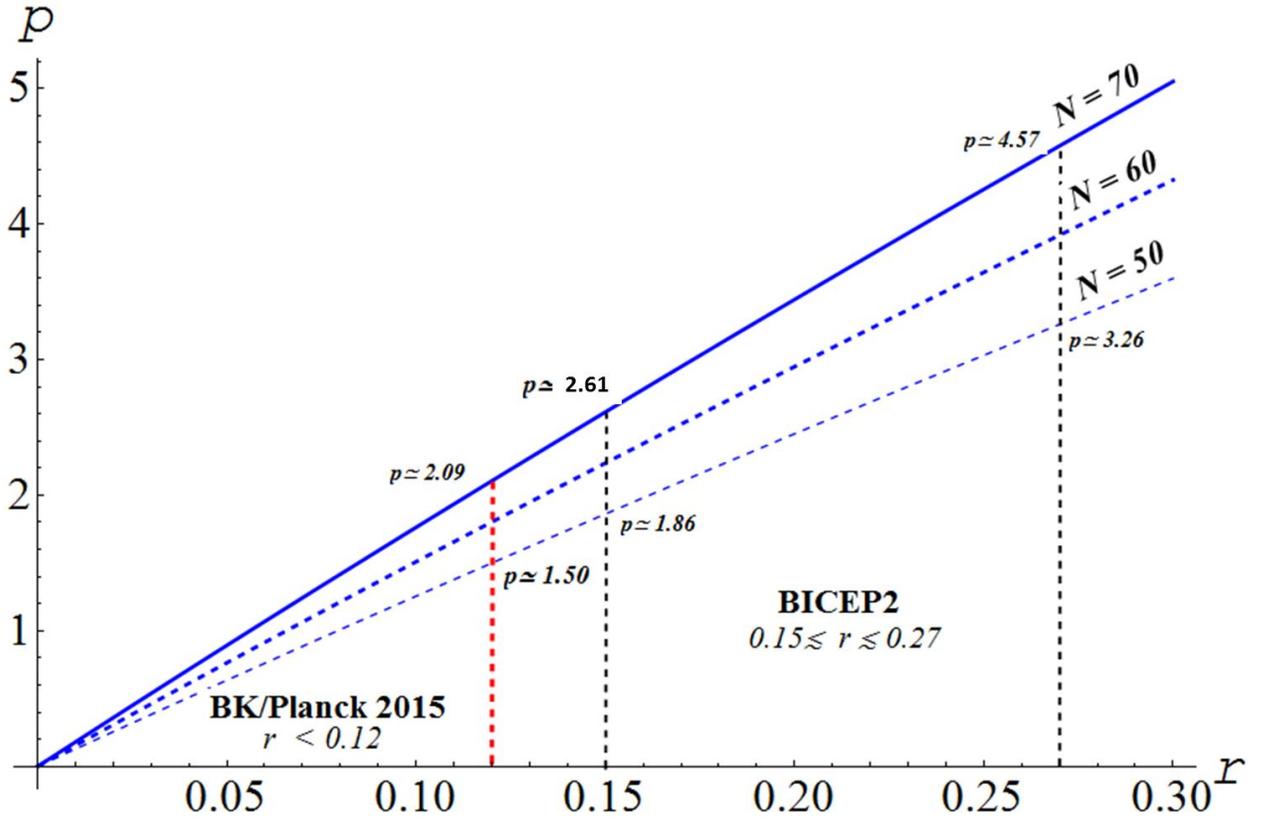

**Fig.6.2**. The limits of LFI parameter, $p$, based on the constraint of BICEP2, $0.15 < r < 0.27$, and BKP 2015, $r < 0.12$ as shown in, $p - r$ at $n_s = 0.960$, which is favored by Planck.

Shortly after the onset of inflation the value of $H$ becomes very high and is approximately constant, but later on it decreases as the value of the field changes. We use the subscript $i$ in the Hubble parameter during inflation, $H_i$. For the zeroth approximation, we can consider $H_i$ as a constant, after the first few e-foldings. That is basically the de Sitter expansion, which is exactly exponential expansion as described by (6.9). Also, after few $N_*$, the spacetime (pivot scale, $k_*$) exits from the Hubble horizon. In this research, we adopt Planck, 2015 pivot scale, $k_* = 0.05 \text{Mpc}^{-1}$, and in most of the cases we use the upper limit of Hubble parameter in the inflationary era, $H_i \lesssim 3.6 \times 10^{-5} M_{\mathrm{Pl}}$, [96]. So, it is worthwhile to investigate both cases; the de Sitter, and the more realistic power law model described by (6.7).

### 6.1.1 LFI on the de Sitter Expansion

In fact, de Sitter model is only zeroth order approximation which does not have graceful exit from inflation [91]. But it can be assumed as a valid way of expansion during most of the inflationary era. In conformal time, $\eta$, Eq. (2.17) can be written as,

$$\frac{1}{a(\eta)}\phi' \simeq -\frac{V_\phi}{3H_i},$$ (6.19)

where, $\phi' = d\phi / d\eta$. Substituting of (6.1) and (6.10) into (6.19) and integrating both sides yields,

$$\frac{d\phi}{p\,\phi^{p-1}} = \frac{M}{M_{\mathrm{Pl}}{}^p}\frac{d\eta}{H_i}.$$ (6.20)

By solving (6.20) for $\phi(\eta)$, we have two different solutions,

$$\phi(\eta) = (H_i \eta)^{\frac{2M^4}{3H_i^2 M_{\mathrm{Pl}}^2}} . \exp\!\left(-\frac{2M^4 c_2}{3H_i^2 M_{\mathrm{Pl}}^2}\right), \quad p = 2,$$ (6.21)

$$\phi(\eta) = \left(\frac{2(2-p)pM^4}{3H_i^2 M_{\mathrm{Pl}}{}^p}\Big[\ln(H_i \eta) + c_2\Big]\right)^{1/(2-p)}, \quad p \neq 2,$$ (6.22)

where, $c_2$, is the integration constant. Adopting BICEP2 favored model, at which the inflation potential vanishes at the end of inflation, $\eta_f$,

$$V\big(\phi(\eta_f)\big) \propto \Big[\phi(\eta_f)\Big]^p \ll 1.$$ (6.23)

Hence, for $p = 2$, we have the limit,

$$c_2 \gg \frac{3H_i{}^2 M_{\text{Pl}}{}^2}{2M^4} \quad . \tag{6.24}$$

However, for $p < 2$, we have the limit,

$$c_2 \to -\ln\left(H_i \eta_f\right), \tag{6.25}$$

and for $p > 2$, we have the limit,

$$c_2 \gg \frac{3H_i{}^2 M_{\text{Pl}}{}^p}{2(2-p)pM^4} . \tag{6.26}$$

On the other hand, if we do not adopt (6.23), then we can solve for $c_2$ that leads to a scale invariance condition of PMF. Both cases, at which BICEP2 constraint is adopted and not adopted, will be studied in section 6.2 to derive the coupling function $f(\phi)$ and then to solve equation of motion for the electromagnetic field, $A_\mu$.

### 6.1.2 LFI on the Power Law Expansion

To have a more optimal slow roll analysis that has a smooth exit from inflation, the Hubble parameter can be written as a function of $\phi(\eta)$, $H(\phi)$. If the field falls below a certain value, it starts to oscillate and then converts to particles in the reheating era, right after inflation. The expansion of space-time during inflation can be described by the power law. Thus, plugging (6.2) into (6.7) yields

$$a(\eta) = l_0 \left|\eta\right|^{-1 - \frac{1}{2}M_{\text{Pl}}{}^2\left(\frac{p}{\phi}\right)^2} \quad . \tag{6.27}$$

Solving for $\phi$ from (6.17) and then substituting it into (6.27) gives,

$$a(\eta) = l_0 \left| \eta \right|^{-1 - \frac{p}{4N}}, \qquad H(\eta) = \frac{a'(\eta)}{a^2(\eta)} = -\frac{1}{l_0}\left(1 + \frac{p}{4N}\right)\left| \eta \right|^{\frac{p}{4N}}. \qquad (6.28)$$

Substituting of (6.28) into (2.17) yields,

$$\frac{d\phi}{\phi^{p-1}} = \frac{pM^4 l_0^2 \eta^{-\left(1 + \frac{p}{2N}\right)}}{3M_{\mathrm{Pl}}^p \left(1 + \frac{p}{4N}\right)} d\eta. \qquad (6.29)$$

Again, the solution of (6.29) will be $\phi(\eta)$ and depends on the model parameter, $p$. We have two different cases; $p = 2$, and $p \neq 2$.

In the case of $p = 2$, one can write the solution of (6.29) as,

$$\phi(\eta) = c_2 \exp\left(-\frac{2}{3}\frac{M^4 l_0^2 \eta^{-\frac{1}{N}}}{M_{\mathrm{Pl}}^2 \frac{1}{N}\left(1 + \frac{1}{2N}\right)}\right), \qquad (6.30)$$

where, $c_2$ is the integration constant. Adopting (6.23) yields $\phi(\eta_{end}) \ll 1$, and $c_2 \ll 1$, where $\eta_{end} \ll -1$.

However, for $p \neq 2$, the solution of (6.29) will be,

$$\phi(\eta) = \left(-\frac{2}{3}\frac{NM^4 l_0^2 \eta^{-\frac{p}{2N}}(2-p)}{M_{\mathrm{Pl}}^p \left(1 + \frac{p}{4N}\right)} + c_2\right)^{\frac{1}{2-p}}. \qquad (6.31)$$

By adopting (6.23), for $p < 2$, the integration constant, $c_2 \ll 1$. However, for $p > 2$, the integration constant will be,

$$c_2 \gg \frac{2}{3} \frac{NM^4 l_0^2 \eta_{end}^{-\frac{p}{2N}} (2-p)}{M_{\text{Pl}}^p \left(1 + \frac{p}{4N}\right)}, \qquad (6.32)$$

Relaxing (6.23) and choosing the values of $c_2$, that implies the scale invariance of PMF, will be investigated in the next section.

## 6.2 The PMF Generated in LFI Model

This subject was investigated in [44], their result is that the LFI do not lead to sensible model building. We use the same method used in the last chapter, but with LFI potential, to investigate the generation of PMF for all possible values of $p$. Substituting of (6.1) into (5.56) gives,

$$f\big(\phi(\eta)\big) = D \, exp\left[ -\frac{\alpha}{6 M_{\text{Pl}}^2} \frac{\phi^2}{p} \right], \qquad (6.33)$$

where, $D$, is a coupling constant. Substituting (6.33) into (5.16) gives,

$$\mathcal{A}''(\eta, k) + \big(k^2 - Y(\eta)\big)\mathcal{A}(\eta, k) = 0, \qquad (6.34)$$

where the function $Y(\eta) = \frac{f''}{f}$. We solve (6.34) in two models of inflationary expansion. In simple de Sitter model of expansion (6.9), and in the power law expansion (6.7).

### 6.2.1   The PMF Generated in LFI in a Simple de Sitter Model

Applying de Sitter approximation was used by Ref.[40] to investigate PMF. One can investigate PMF under de Sitter model by substituting (6.21)-(6.22) into (6.34), and apply the limits (6.24)-(6.26) for the selected values of model parameter, $p$.

For $p = 2$, which is the most interesting value, because it fits well with both spectrum index, $n_s$ detected by Planck [56], and $r$ value reported by BICEP2 [54]. Also, it is the closest case to the standard inflationary models. Substituting of (6.21) into (6.33) and (6.34) yields,

$$Y(\eta) = \frac{e^{-\frac{8c_2 M^4}{3M_{Pl}^2 H_i^2}} M^8 \alpha^2 (\eta H_i)^{-2 - \frac{8M^4}{3M_{Pl}^2 H_i^2}}}{81 M_{Pl}^8 H_i^2} \ . \tag{6.35}$$

Since $M \simeq 3 \times 10^{-3} M_{Pl}$ and $H_i \simeq 3.6 \times 10^{-5} M_{Pl}$, hence,

$$\frac{8M^4}{3M_{Pl}^2 H_i^2} \ll 1 \ . \tag{6.36}$$

Therefore, (6.34) becomes,

$$\mathcal{A}''(\eta, k) + \left( k^2 - \frac{e^{-\frac{8c_2 M^4}{3M_{Pl}^2 H_i^2}} M^8 \alpha^2 (\eta H_i)^{-2}}{81 M_{Pl}^8 H_i^2} \right) \mathcal{A}(\eta, k) = 0 \ , \tag{6.37}$$

The solution of (6.37) is a Bessel function [79]. Its solution with the initial condition, $\mathcal{A}(0^-, k) \to 0$, will be similar to (5.36). This initial condition insures that the electromagnetic field will ultimately goes to zero, as $t \to \infty$. Hence, $\mathcal{A}(\eta, k)$, can be written as,

$$\mathcal{A}(\eta, k) = (k\eta)^{1/2} \left[ C_1(k) \mathbf{J}_\chi(k\eta) + C_2(k) \mathbf{J}_{-\chi}(k\eta) \right], \tag{6.38}$$

where, $\chi$ can be written as,

$$\chi = \frac{1}{18} \sqrt{81 + \frac{4 e^{\frac{8c_2 M^4}{3M_{Pl}^2 H_i^2}} M^8 \alpha^2}{M_{Pl}^8 H_i^4}}, \tag{6.39}$$

Using both (6.24) and (6.36), one can write, $\chi \approx 1/2$. Compare it with (5.36), implies that it is similar to the case of $\gamma = 0$. Hence, it cannot generate a scale invariant PMF.

In order to find the electric field spectra, one can substitute from (6.21) into (6.33), to get,

$$f(\eta) = D \exp\left[ -\frac{\alpha}{6 M_{\text{Pl}}^2} \frac{\left( (H_i \eta)^{-\frac{2M^4}{3H_i^2 M_{\text{Pl}}^2}} \cdot \exp\left( -\frac{2M^4 c_2}{3 H_i^2 M_{\text{Pl}}^2} \right) \right)^2}{2} \right]. \qquad (6.40)$$

Employing (6.24), the exponent of (6.40) will be so small, hence, $f(\eta) \to D$. Hence, the plot of the spectra of both PMF and electrical field shows that they are of the same order at low value of $k$, and diverge at relatively high $k$, see Fig.6.3.

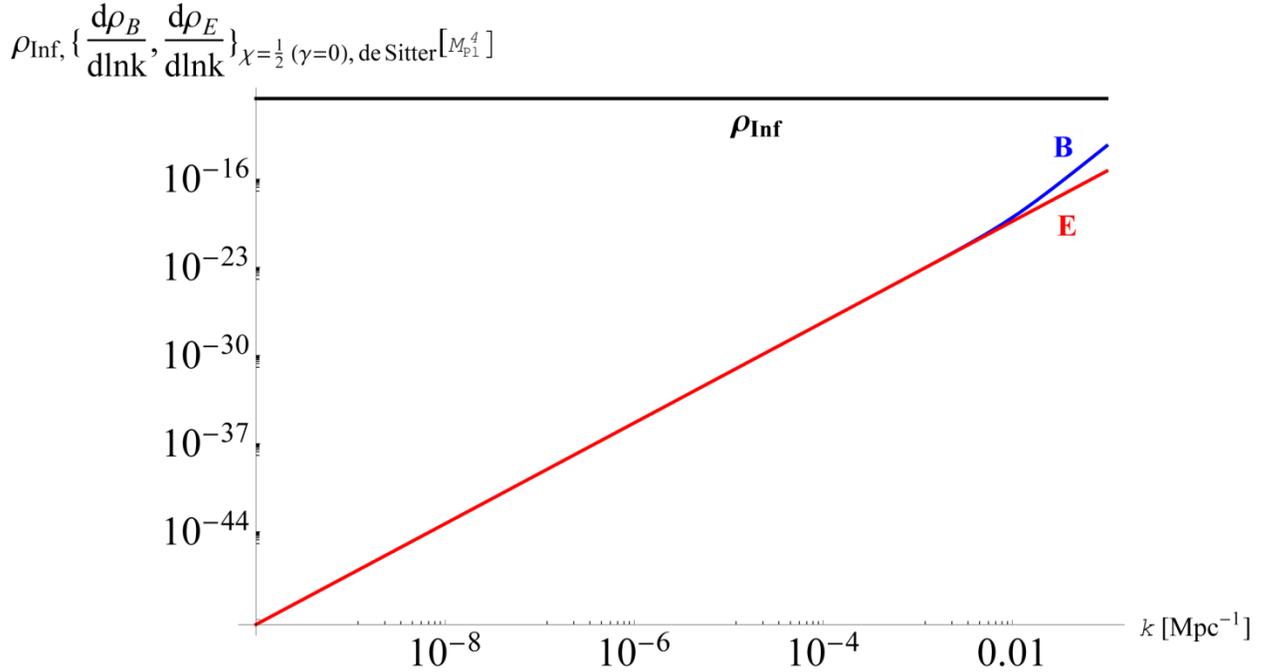

**Fig.6.3**. The magnetic and electric spectra, generated under LFI model in the simple de Sitter expansion, and by considering the constraint, $V(0) \to 0$, where, $p = 2$, $k\eta \ll 1$, and $\chi = 1/2$ which is similar to the case of $\gamma = 0$ in the exponential inflationary potential. The electromagnetic spectra are of the same order at low value of $k$. In the plot, we assume, $H_i = 3.6 \times 10^{-5} M_{\text{Pl}}$ and use Planck pivot scale, $\eta = -20$. In this case the spectra are not scale invariant and for the observable scale, $k$, they are less than the energy of inflation, $\rho_{\text{Inf}}$.

If we relax the limits (6.24) and (6.36), and choose $c_2$ to enforce $\chi = 5/2$, which is the case at which PMF is scale invariant. We have,

$$c_2 = -\frac{3M_{\text{Pl}}^2 H_i^2}{8M^4} \ln\left[\frac{486 M_{\text{Pl}}^8 H_i^4}{M^8 \alpha^2}\right]. \qquad (6.41)$$

Substituting from (6.41) into (6.40) to find the coupling function, and use it to plot the electromagnetic spectra on the same method, see Fig.6.4. It shows that a scale invariant PMF can be achieved without a backreaction problem as long as $k \gtrsim 8 \times 10^{-7} \text{Mpc}^{-1}$. However, the electric spectra can go over the scale of the inflation, $\rho_{\text{Inf}}$, for $k \lesssim 8 \times 10^{-7} \text{Mpc}^{-1}$. Hence, the backreaction problem still exists at extremely very low values of $k$. In the above calculation, we use $\alpha = 2$, $H_i = 3.6 \times 10^{-5} M_{\text{Pl}}$, $M = 3 \times 10^{-3} M_{\text{Pl}}$ and $(M_{\text{Pl}}, D) = 1$.

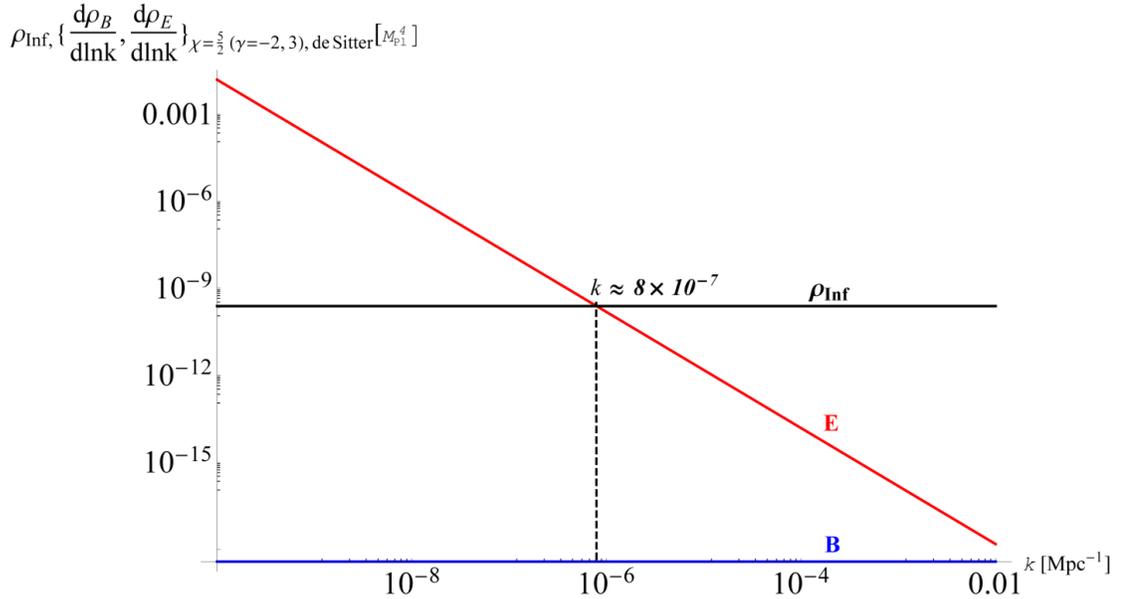

**Fig.6.4**. The magnetic and electric spectra, generated under LFI model, in the simple de Sitter expansion, with, $p = 2$, $k \ll 1$, and $\chi = 5/2$ which is similar to the case of $(\gamma = -2, 3)$ in the exponential inflationary potential. The electric spectra can go over the scale of the inflation, $\rho_{\text{Inf}}$, for $k \lesssim 8 \times 10^{-7} \text{Mpc}^{-1}$. However, the backreaction problem can be avoided for $k \gtrsim 8 \times 10^{-7} \text{Mpc}^{-1}$. In the plot, we use, $\eta = -20$, $\alpha = 2$, $H_i = 3.6 \times 10^{-5} M_{\text{Pl}}$, $M = 3 \times 10^{-3} M_{\text{Pl}}$ and $(M_{\text{Pl}}, D) = 1$.

On the other hand, plotting the spectra versus the Hubble rate $H_i$ shows that changing that rate will change both magnetic and electric field almost in the same manner, see Fig.6.5. However, for $H_i \gtrsim 1.3 \times 10^{-3} M_{Pl}$, the electric energy can go over the $\rho_{Inf}$ which causes the backreaction problem. The value $H_i \sim 1.3 \times 10^{-3} M_{Pl}$ is well above the upper bound of $H_i$ reported by Planck ($\lesssim 3.6 \times 10^{-5} M_{Pl}$).

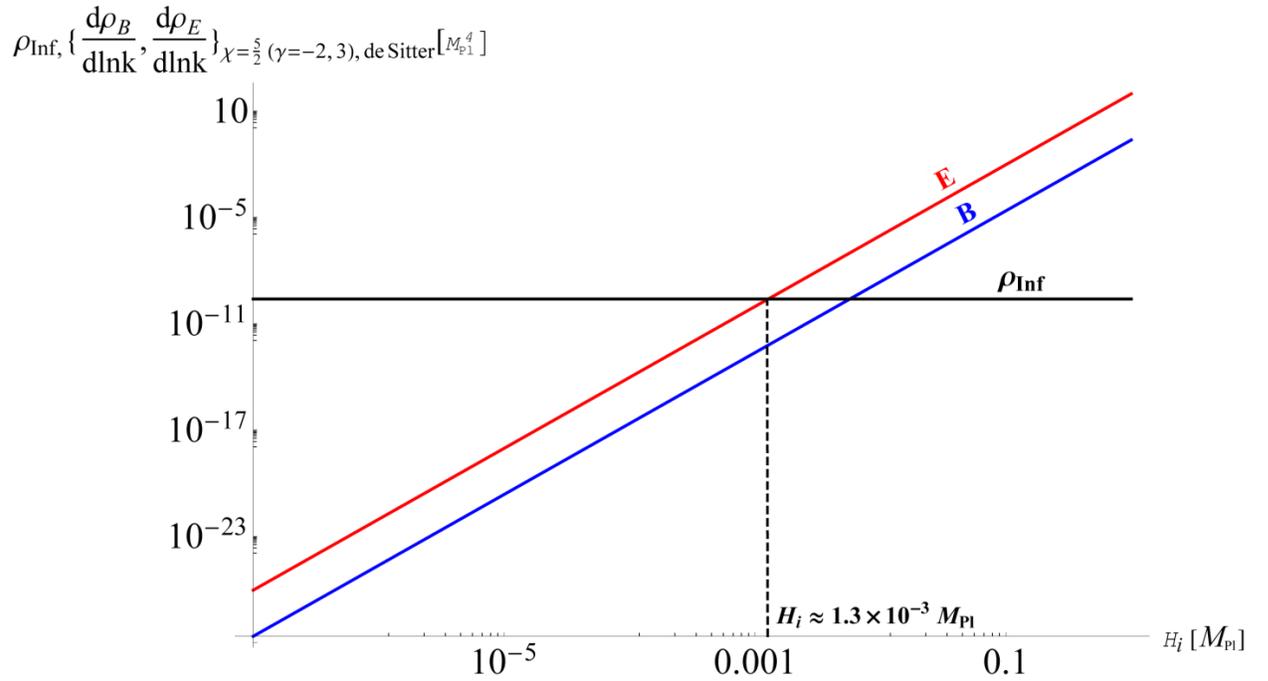

**Fig.6.5**. The magnetic and electric spectra, generated under LFI model, in the simple de Sitter expansion, as a function of $H_i$, with $p = 2$, $k = 1$, and $\chi = 5/2$ which is similar to the case of $(\gamma = -2,3)$ in the exponential inflationary potential. They both change in the same manner. For $H_i > 1.3 \times 10^{-3} M_{Pl}$, the electric energy can go over the $\rho_{Inf}$ which causes the backreaction problem. But the value $H_i \sim 1.3 \times 10^{-3} M_{Pl}$ is well above the upper bound of $H_i$ reported by Planck ($3.6 \times 10^{-5} M_{Pl}$). In the plot, we use, $k = 10^{-3} Mpc^{-1}$, $\eta = -20$, $\alpha = 2$, $H_i = 3.6 \times 10^{-5} M_{Pl}$, $M = 3 \times 10^{-3} M_{Pl}$ and $(M_{Pl}, D) = 1$.

Similarly, plotting electromagnetic spectra as function of the parameter, $M$, shows that the backreaction problem can be avoided if $M \gtrsim 8.5 \times 10^{-5} M_{\text{Pl}}$, see Fig.6.6. It is well below $3 \times 10^{-3} M_{\text{Pl}}$, the value calculated by the amplitude of CMB.

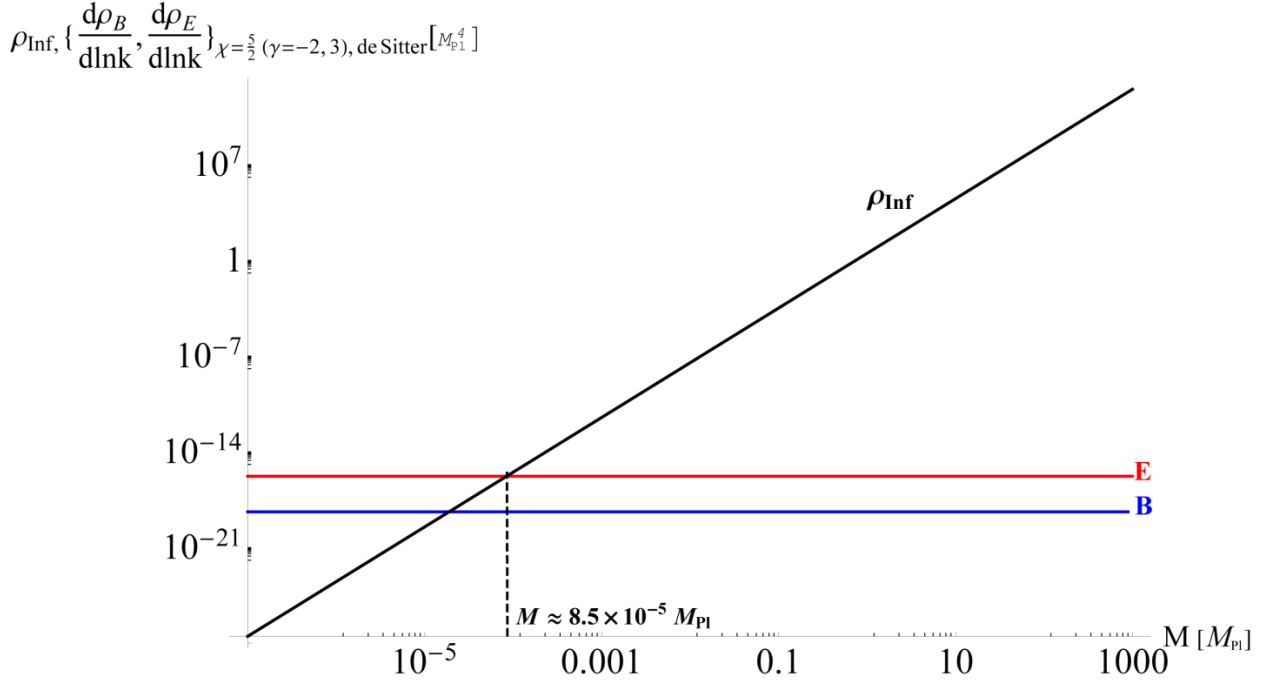

**Fig.6.6.** The magnetic and electric spectra and inflationary density of energy $\rho_{\text{Inf}}$, generated under LFI model, in the simple de Sitter expansion, with, $p = 2$, $k \ll 1$, and $\chi = 5/2$ ,which is similar to the case of ($\gamma = -2, 3$) in the exponential inflationary potential, as a function of $M$. The electromagnetic spectra are independent of $M$. For $M \gtrsim 8.5 \times 10^{-5} M_{\text{Pl}}$, the electric energy can go below the $\rho_{\text{Inf}}$ which avoid the backreaction problem. But the value $M = 8.5 \times 10^{-5} M_{\text{Pl}}$ is well below the one calculated from the amplitude of CMB, $M = 3 \times 10^{-3} M_{\text{Pl}}$. In the plot, we use, $k = 10^{-3} Mpc^{-1}$, $\eta = -20$, $\alpha = 2$, $H_i = 3.6 \times 10^{-5} M_{\text{Pl}}$, and $(M_{\text{Pl}}, D) = 1$.

As a result of the foregoing discussion, one can conclude that a scale invariant PMF cannot be generated in LFI, for $p = 2$, in the limit, $V(0) \approx 0$, which was the now-discredited BICEP2 favored shape of inflationary potential. However, a scale invariant PMF can be generated if we relax that limit. Also, the backreaction problem can be avoided under some conditions which fit with some observable scales of $k$.

**For $p < 2$,** there are some interesting cases, such as, $p = 1, 2/3$. These are shown in Fig.6.1. Thus, following the same way as done in the previous subsection, one has to substitute (6.22) into (6.33) and (6.34). In this case,

$$Y(\eta) = \frac{2^{\left(\frac{4}{2-p}\right)} 3^{\left(-2-\frac{4}{2-p}\right)} M^8 M_{\text{Pl}}{}^{-4-2p} \alpha^2 \left(\frac{M^4 M_{\text{Pl}}{}^{-p}(2-p) p(c_2 + \ln[H_i \eta])}{H_i^2}\right)^{-2+\frac{4}{2-p}}}{H_i^4 \eta^2} . \quad (6.42)$$

Using limits, (6.25), (6.36), and the fact that, $\ln(\eta/\eta_f) \simeq C$, since both, $\eta_f, \eta \ll -1$. Hence, substituting (6.42) into (6.34), yields that,

$$\mathcal{A}(\eta, k) = (k\eta)^{1/2} \left[ C_1(k) \mathbf{J}_\chi(k\eta) + C_2(k) \mathbf{J}_{-\chi}(k\eta) \right], \quad (6.43)$$

where, $\chi$ can be written as,

$$\chi = \frac{3^{-1+\frac{2}{-2+p}} \sqrt{3^{\frac{2(-4+p)}{-2+p}} C^2 M_{\text{Pl}}{}^4 (-2+p)^2 p^2 + (-1)^{-\frac{4}{-2+p}} 2^{\frac{2(-4+p)}{-2+p}} C^{-\frac{4}{-2+p}} H_i^{\frac{8}{-2+p}} M^{-\frac{16}{-2+p}} M_{\text{Pl}}{}^{\frac{4p}{-2+p}} (-2+p)^{-\frac{4}{-2+p}} p^{-\frac{4}{-2+p}} \alpha^2}}{2 C M_{\text{Pl}}{}^2 (-2+p) p} . \quad (6.44)$$

Employing the limit (6.36), the second term under the square root will vanish and the value of (6.44), reduces to $\chi \simeq 1/2$. Therefore, PMF cannot be scale invariant when it is generated in LFI for $p < 2$, in $V(0) = 0$, limit. In order to calculate the electric spectrum, one has to fix, $f(\eta)$, from (6.33),

$$f(\eta) \propto \exp\left\{-\frac{2^{-1+\frac{2}{2-p}} 3^{-\frac{2}{2-p}} \left(\frac{C M^4 M_{\text{Pl}}{}^{-p}(2-p) p}{H_i^2}\right)^{\frac{2}{2-p}} \alpha}{p}\right\}, \quad (6.45)$$

which is approximately constant. Therefore, one expects to get the same magnetic and electric spectra as shown in Fig.6.3. Similarly, if we relax the BICEP2 limits and enforce, $\chi = 5/2$ to get scale invariant PMF, then we expect to have a backreaction problem similar to the case of $p = 2$.

**For $p > 2$,** there are some interesting cases, like, $p = 3, \ 4$. In the case of $p = 3$, Eq.(6.18) gives the tensor to scalar ratio, $r \simeq 0.204$ at $N = 60$ and $n_s \simeq 0.960$, which fits very well with BICEP2 results, as shown in Fig.6.1. However, for $p = 4$, the tensor to scalar ratio $r \simeq 0.276$ at $N = 60$ and $n_s \simeq 0.960$, which is on the upper limit of the BICEP2 range, See Fig.6.2. The main problem with higher order model is that they are not bounded from below, so their expansion does not converge [56]. Also, there is a constraint imposed by WMAP7, at which $p < 2.2$ at 95% of confidence [17].

In order to investigate PMF for, $p > 2$, one has to substitute (6.22) into (6.33) and (6.34) and employing limit (6.26). In this case, we can neglect, $\ln[H_i \eta]$ comparison with $c_2$. Hence, we end up with the same equation as (6.42). Therefore, we expect to have the same magnetic and electric spectra manner as in the case of $p < 2$. As a result, a scale invariant PMF is not expected to be generated in LFI by the simple inflation model $f^2 FF$ in a de Sitter model of expansion unless, we choose the integration constant which enforces the scale invariance condition for PMF. Thus, the problems of backreaction is expected at extremely low values of $k$ and $M$ and very high values of $H_i$. However, it can be avoided for some values of $k$, $M$ and $H_i$ that fit with observations.

Finally, plotting the electromagnetic spectra as a function of $p$, in the case of $\chi = 5/2$, for the de Sitter way of inflationary expansion, at the adopted values of model parameters shows that the electromagnetic energy always much less than that of inflation, $\rho_{\text{Inf}}$, see Fig.6.7.

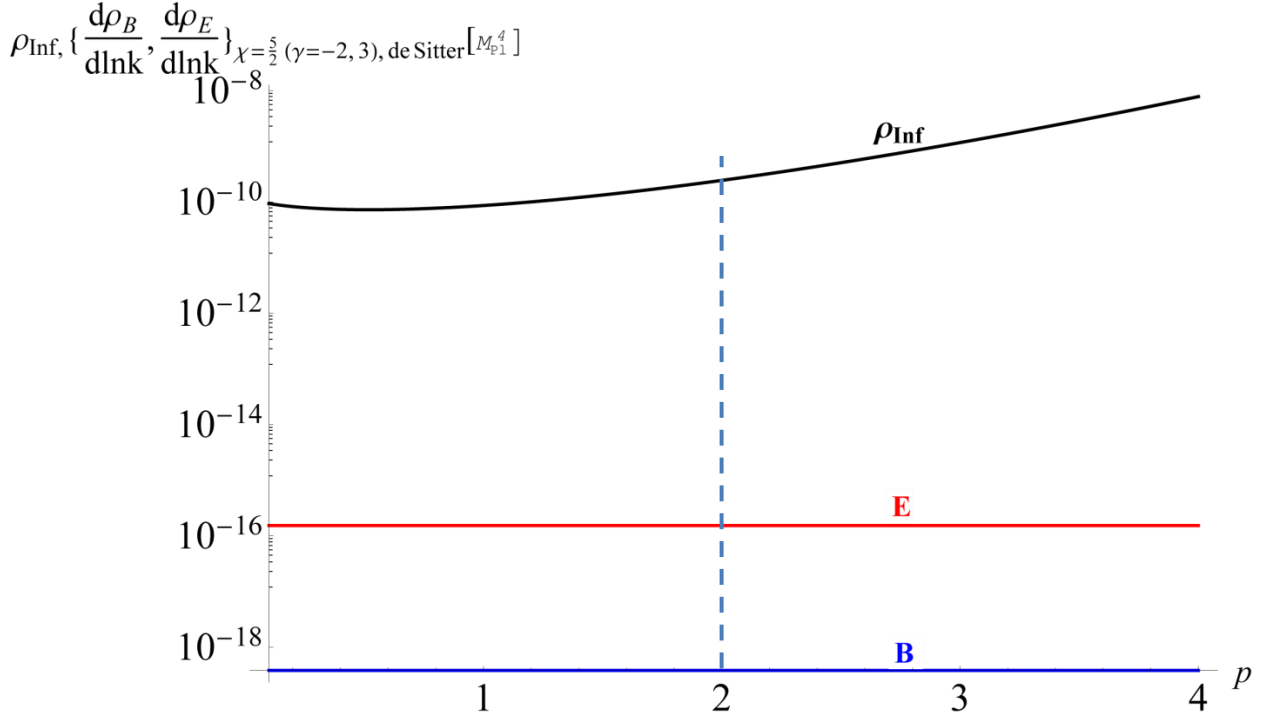

**Fig.6.7.** The magnetic and electric spectra, generated under LFI model, in the simple de Sitter expansion, as a function of $p$, but ($p \neq 2$). The electromagnetic spectra always less than the scale of inflation, $\rho_{\text{Inf}}$. In the plot, we use, $k = 10^{-3} Mpc^{-1}$, $\eta = -20$, $\alpha = 2$, $H_i = 3.6 \times 10^{-5} M_{\text{pl}}$, $M = 3.0 \times 10^{-3} M_{\text{pl}}$, and $(M_{\text{pl}}, D) = 1$.

## 6.2.2  The PMF Generated in LFI in the Power Law Model of Expansion

The power law model is more optimal and realistic description of the expansion of space time during inflation. It leads to a graceful exit from inflation. As done in [44] but in more general form, (6.27), for LFI, one can either adopt the limit, (6.32), or solve for $c_2$ to have a scale invariant PMF for a selected interesting values of model parameter, $p$.

**For**, $p = 2$, if we substitute (6.30) into (6.33) and (6.34), for $\eta \ll -1$, we have,

$$Y(\eta) = \frac{c_2^{\,4} l_0^{\,4} M^8 \alpha^2 \eta^{-2(1+\frac{1}{N})}}{81 M_{\text{Pl}}^{\,8} \left(1 + \frac{1}{2N}\right)^2} \; . \tag{6.46}$$

For a relatively large, $N \geq 50$, one can assume, $1 + \frac{1}{N} \approx 1$. Without this approximation, the solution of (6.34) is not generally in a closed form. With this assumption, the solution will be Bessel function with,

$$\chi = \frac{\sqrt{81 M_{\text{Pl}}^{\,8} + 324 M_{\text{Pl}}^{\,8} N + 324 M_{\text{Pl}}^{\,8} N^2 + 16 c_2^{\,4} l_0^{\,4} M^8 N^2 \alpha^2}}{18 M_{\text{Pl}}^{\,4} \left(1 + 2N\right)} \; . \tag{6.47}$$

If we adopt the limit, (6.23), then $c_2 \ll 1$ and $\chi \simeq 1/2$. Therefore, PMF is not scale invariant in this condition. It is similar condition to de Sitter case when we adopt BICEP2 result. The magnetic and electric spectra will be similar to Fig.6.3.

However, if we choose $c_2$ to have the scale invariance condition, $\chi = 5/2$, the coupling function can be written as,

$$f(\eta) = D \exp \left\{ -\frac{3\sqrt{\frac{3}{2}} e^{-\frac{4 l_0^{\,2} M^4 N \eta^{-1/N}}{3 M_{\text{Pl}}^{\,2} \left(1 + \frac{1}{2N}\right)}} M_{\text{Pl}}^{\,2} \left(1 + 2N\right)}{4 l_0^{\,2} M^4 N} \right\} , \tag{6.48}$$

where, $D$, is a coupling constant. As we can see from (5.22-23), the electromagnetic spectra do not depend on $D$. In this case the magnetic and electric spectra are shown in Fig.6.8. This is similar to the de Sitter case, Fig.6.4, but with one order of magnitude higher value of comoving wavenumber, ( $k \sim 4 \times 10^{-6} Mpc^{-1}$ ) under which the backreaction problem may occur.

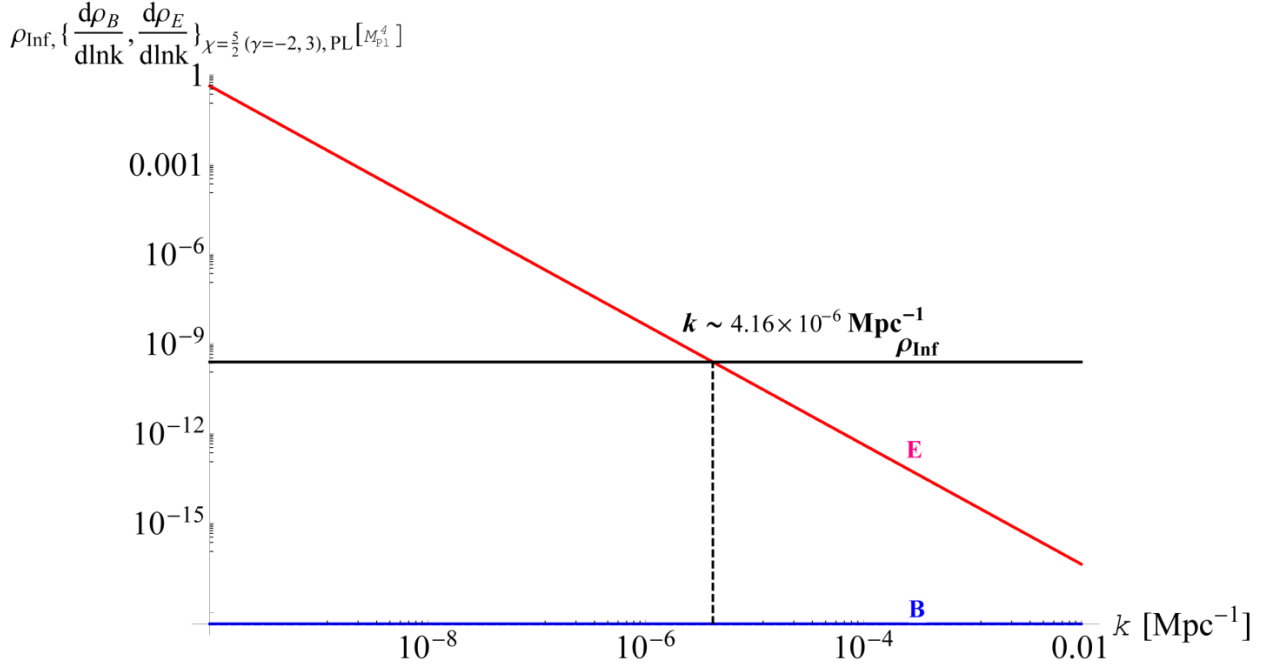

**Fig.6.8**. The magnetic and electric spectra, generated under LFI model, in the power law (PL) expansion, with, $p \approx 2$, $k \ll 1$ and $\chi \approx 5/2$ ,which is similar to the case of $(\gamma = -2,3)$ in the exponential inflationary potential. The electric spectra can go over the scale of the inflation, $\rho_{Inf}$ , for $k \lesssim 4 \times 10^{-6} Mpc^{-1}$ . However, the backreaction problem can be avoided for $k \gtrsim 4 \times 10^{-6} Mpc^{-1}$ . In the plot, we use, $\eta = -20$ , $\alpha = 2$ , $l_0 = 1/(3.6 \times 10^{-5} M_{Pl})$ , $M = 3 \times 10^{-3} M_{Pl}$ and $(M_{Pl}, D) = 1$ .

Calculating the spectra versus the e-folding number $N$ , shows that the electric field can drop less than $\rho_{Inf}$ for $N > 51$ , see Fig.6.9. It fits with the reported values of $N$ to end of inflation, by Planck, 2015 [96].

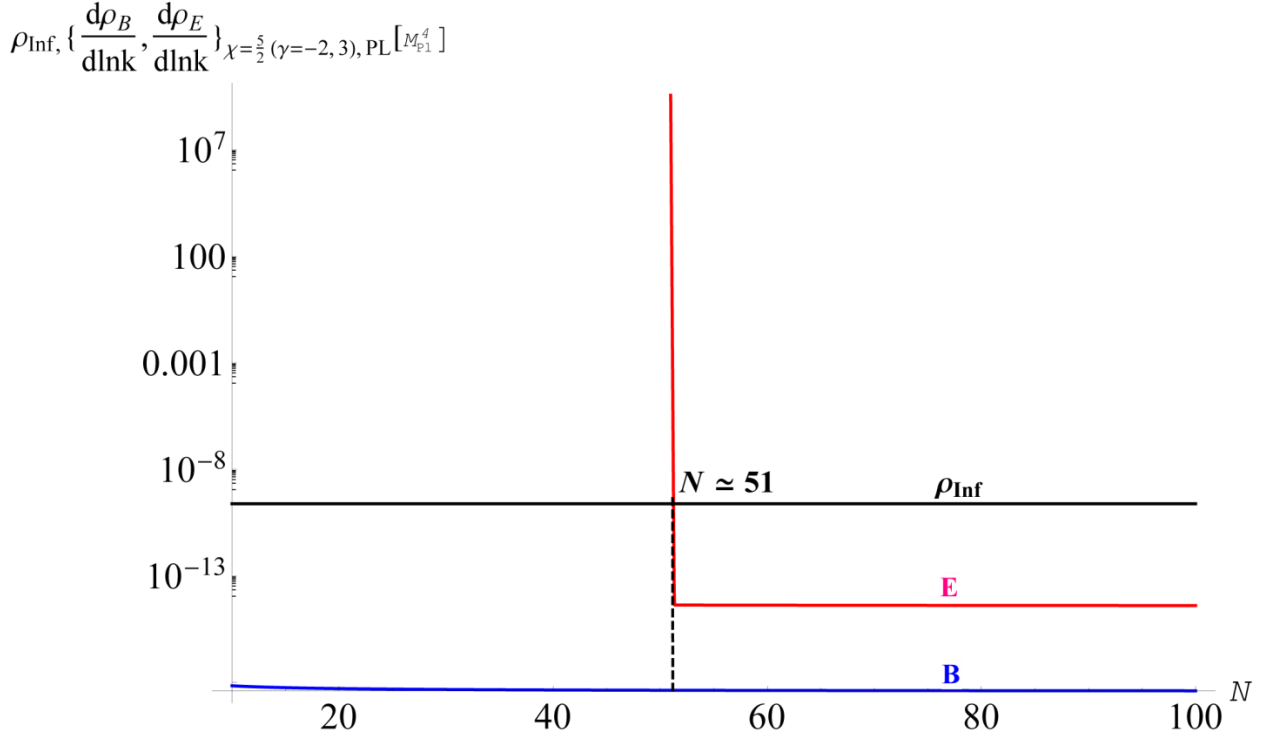

**Fig.6.9.** The magnetic and electric spectra and inflationary density of energy $\rho_{\text{Inf}}$, generated under <u>LFI</u> model, in the PL expansion, with, $p \simeq 2$, $k \ll 1$ and $\chi = 5/2$, which is similar to the case of $(\gamma = -2, 3)$ in the exponential inflationary potential, as a function of $N$. For $N \gtrsim 51$, the electric energy can go below the $\rho_{\text{Inf}}$ which avoids the backreaction problem. It also fits with the reported rang of $N$ by Planck. In this plot, we use, $k = 10^{-3} Mpc^{-1}$, $\eta = -20$, $\alpha = 2$, $l_0 = 1/(3.6 \times 10^{-5} M_{\text{Pl}})$, $M = 3 \times 10^{-3} M_{\text{Pl}}$, and $(M_{\text{Pl}}, D) = 1$.

The same is true for the electromagnetic spectra as a function of the parameter $l_0$, the electric field energy falls below $\rho_{\text{Inf}}$ for $l_0 > 3 \times 10^5 M_{\text{Pl}}^{-1}$, see Fig.6.10. Since $l_0 \propto 1/H_i$, then that value is corresponding to $H_i \sim 3.3 \times 10^{-6} M_{\text{Pl}}$ which is less than the upper bound, $H_i \simeq 3.6 \times 10^{-5} M_{\text{Pl}}$.

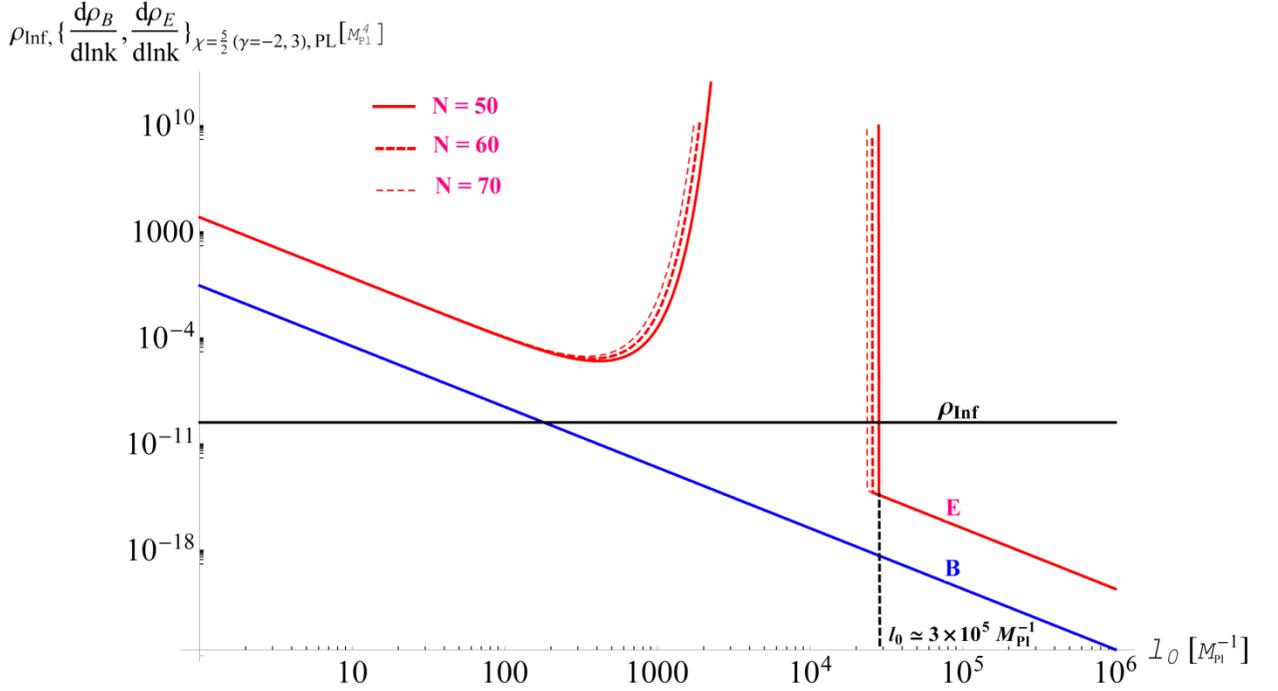

**Fig.6.10**. The magnetic and electric spectra and inflationary density of energy $\rho_{\text{Inf}}$, generated under LFI model, in the PL expansion, with, $p = 2$, $k \ll 1$, and $\chi = 5/2$, which is similar to the case of ($\gamma = -2, 3$) in the exponential inflationary potential, as a function of $l_0$. For $l_0 \gtrsim 3 \times 10^5 M_{\text{Pl}}^{-1}$ ($H_i \lesssim 3.3 \times 10^{-6} M_{\text{Pl}}$), the electric energy can go below the $\rho_{\text{Inf}}$ which avoid the backreaction problem. It also fits with the upper bound of $H_i$ reported by Planck. In this plot, we use, $k = 10^{-3} \text{Mpc}^{-1}$, $\eta = -20$, $\alpha = 2$, $M = 3 \times 10^{-3} M_{\text{Pl}}$, and $(M_{\text{Pl}}, D) = 1$.

**For $p < 2$**, in this case we substitute (6.31) into (6.33) and (6.34), and use the fact that $|\eta| \gg 1$ to yield,

$$Y(\eta) = \frac{l_0^4 M^8 p^2 \alpha^2 \eta^{-2-\frac{p}{N}} c_2^{-2+\frac{4}{2-p}}}{324 M_{\text{Pl}}^{4+2p} \left(1 + \frac{p}{4N}\right)^2} \ . \tag{6.49}$$

Adopting, (6.23) and implies that the integration constant, $c_2 \ll 1$. Assuming that, $-2 - \frac{p}{N} \approx -2$, for $0 < p \leq 3$, and $N \geq 50$, hence from (6.43),

$$\chi = \frac{\sqrt{81 M_{\text{Pl}}^{4+2p} (4N+p)^2 + 16\, c_2^{-\frac{2p}{-2+p}} l_0^4 M^8 N^2 p^2 \alpha^2}}{18 (4N+p) M_{\text{Pl}}^{2+p}}. \tag{6.50}$$

Since, $c_2$, $l_0^4 M^8 \ll 1$, then $\chi \simeq 1/2$. Thus, a scale invariant PMF cannot be generated under the limit (6.23). The magnetic and electric field spectra will be similar to Fig.6.3.

**For $p > 2$**, we substitute (6.31) into (6.33) and (6.34), and use the fact that $l_0^4 M^8 \ll 1$, and $|\eta| > |\eta_{end}| \gg 1$. By means of (6.32), we have (6.49). Also, since $p > 2$, we end up with, $\chi \simeq 1/2$. Again, a scale invariant PMF cannot be generated under the BICEP2 favored inflationary model.

Similarly, for $p \neq 2$, if we enforce the scale invariance condition of PMF at which, $\chi = 5/2$, and substitute it into (6.31) to find the coupling function, $f(\eta)$, and in turn the magnetic and electric fields spectra. They all show the backreaction problem at some ranges. Plotting the electromagnetic spectra as a function of $p$, for $p \neq 2$, is shown in Fig.6.11. As a result, the backreaction problem can be avoided in the range $p \in [1.66, 2.03]$, for the range, $50 \leq N \leq 70$. In fact, it is much easier in this case to solve for $\alpha$ rather than $c_2$, and then substitute into (6.33) to find the coupling function.

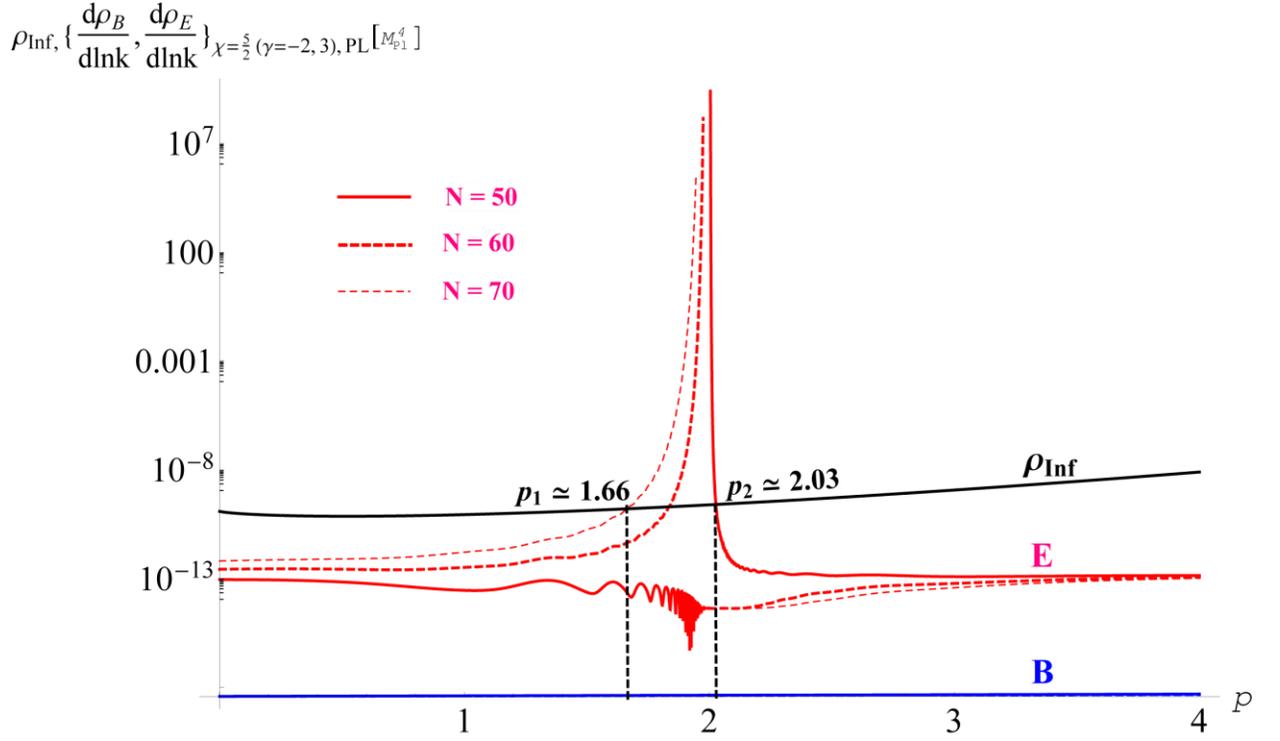

**Fig.6.11**. The electromagnetic spectra and inflationary density of energy $\rho_{\text{inf}}$, generated under LFI model, in the PL expansion, with, $k\eta \ll 1$, and $\chi \simeq 5/2$, which is similar to the case of $(\gamma = -2,3)$ in the exponential inflationary potential, as a function of $p$, $p \neq 2$. The electromagnetic spectra falls below $\rho_{\text{inf}}$ on the range, $1.66 \lesssim p \lesssim 2.03\ (p \neq 2)$, for the interesting values of $N$. Hence, the backreaction problem can be avoided on this range. In this plot, we use, $k = 10^{-3}\,Mpc^{-1}$, $\eta = -20$, $\alpha = 2$, $M = 3 \times 10^{-3} M_{\text{Pl}}$, $c_2 = 1$, and $(M_{\text{Pl}}, D) = 1$.

Plotting the electromagnetic spectra as a function of $M$ shows that the backreaction problem can be avoided for, $M \gtrsim 2.8 \times 10^{-3} M_{\text{Pl}}$, see Fig.6.12. This value is consistent with observational value calculated from the amplitude of CMB anisotropies, $M \approx 3.0 \times 10^{-3} M_{\text{Pl}}$ [81].

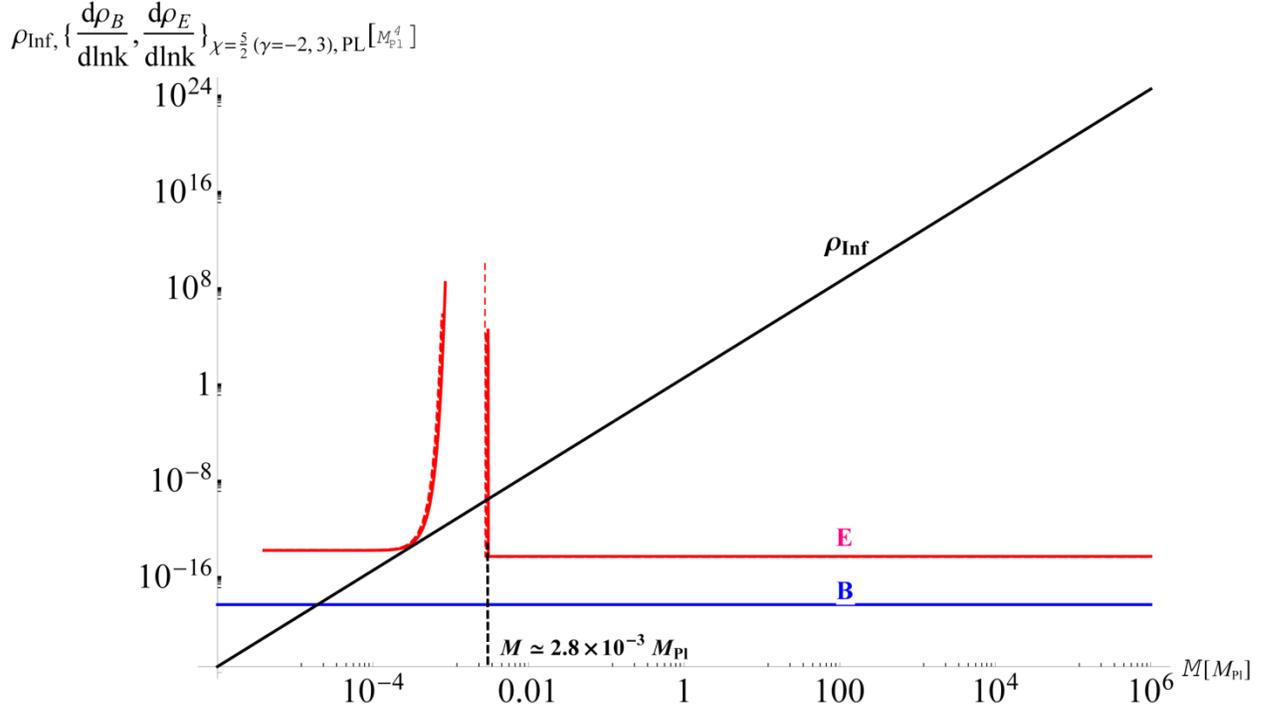

**Fig.6.12.** The electromagnetic spectra and inflationary density of energy $\rho_{\text{Inf}}$, generated under LFI model, in the PL expansion, with, $k\eta \ll 1$, and $\chi \simeq 5/2$, which is similar to the case of $(\gamma = -2,3)$ in the exponential inflationary potential, as a function of $M$. The electromagnetic spectra falls below $\rho_{\text{Inf}}$, for, $M \gtrsim 2.8 \times 10^{-3} M_{\text{Pl}}$, for the interesting values of $N$. Hence, the backreaction problem can be avoided on this range. In this plot, we use, $k = 10^{-3} Mpc^{-1}$, $\eta = -20$, $\alpha = 2$, $p = 2$, $c_2 = 1$, and $(M_{\text{Pl}}, D) = 1$.

Finally, plotting the electromagnetic spectra as a function of the integration constant, $c_2$ shows that the backreaction problem can be avoided for $c_2 \gtrsim 1$, see Fig.6.13, for the expansion of $N > 50$.

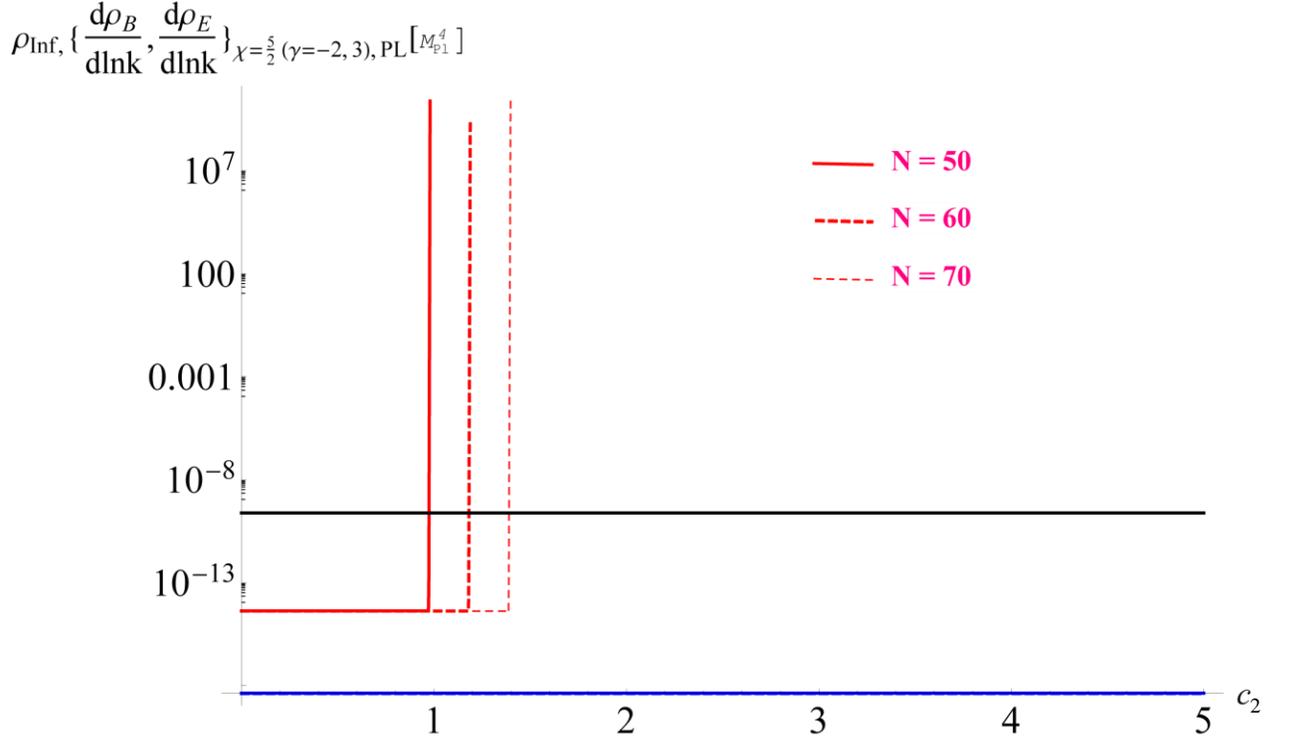

**Fig.6.13**. The electromagnetic spectra and inflationary density of energy $\rho_{\text{Inf}}$, generated under <u>LFI</u> model, in the PL expansion, with, $k\eta \ll 1$, and $\chi \simeq 5/2$, which is similar to the case of $(\gamma = -2,3)$ in the exponential inflationary potential, as a function of integration constant $c_2$. The electromagnetic spectra falls below $\rho_{\text{Inf}}$ for, $c_2 > 1$, and $N > 50$. Hence, the backreaction problem can be avoided on this range. In this plot, we use, $k = 10^{-3} Mpc^{-1}$, $\eta = -20$, $\alpha = 2$, $p \simeq 2$, $M = 3 \times 10^{-3} M_{\text{Pl}}$, and $(M_{\text{Pl}}, D) = 1$.

## 6.3 Summary and Discussion of the PMF Generated in LFI

PMFs can be generated in the simple model with gauge invariant coupling $f^2 FF$, in the standard models of inflation. It requires a breaking of the conformal symmetry of the electromagnetism. In this chapter, we used the same method of [44] to investigate the PMF in the context of LFI model. In [44], it was shown that LFI is not a good model to generate a scale invariant PMF, because the coupling function needed, is too complicated to be justified.

In this chapter, we do an analysis for all reasonable values of $p$ , $H_i$ , $N$ , $l_0$ , $M$ , and $c_2$. Also, we solve for the complicated coupling function. We first present the slow roll analysis of the LFI, and derive the, $r - n_s$ relation. We then investigate the PMF generated in the context of LFI by both de Sitter expansion, Eq.(6.10), and the power law expansion, Eq.(6.7).

The de Sitter expansion is used as an approximation at the onset of inflation [40]. After investigating the PMF in LFI, outside Hubble radius, ( $k\eta \ll 1$ ), in de Sitter expansion, the main results of this research are that the PMF can in principle be generated in the LFI model in all values of the argument, $p$. However, for the shape of the inflationary model favored by the BICEP2 results (6.23), the scale invariant PMF cannot be achieved by LFI. Furthermore, in this case, the generated electric field spectrum is in the same order of the PMF's, see Fig.6.3.

On the other hand, for the general limit, one can enforce the condition, $\chi = 5/2$ , to generate a scale invariant PMF. Although, the energy of the electric field increases excessively and becomes much greater than the energy of PMF at ( $k\eta \ll 1$ ), the electromagnetic energy falls below the energy of inflation, $\rho_{\text{Inf}}$ at some observable scales, $k$ . For example, the backreaction problem can be avoided for $k \gtrsim 8 \times 10^{-7}\text{Mpc}^{-1}$ and $k \gtrsim 4 \times 10^{-6}\text{Mpc}^{-1}$ in de Sitter and power law expansion respectively. Thus, one can avoid the backreaction problem under some circumstances of LFI model.

Similarly, computing the electromagnetic spectra as a function of $p$ , $H_i$ , $N$ , $l_0$ , $M$ , and $c_2$, shows that the electromagnetic spectra can fall below $\rho_{\text{Inf}}$ at certain ranges. Under de Sitter expansion, the backreaction problem can be avoided on the ranges, $H_i < 1.3 \times 10^{-3} M_{\text{Pl}}$ , and $M > 8.5 \times 10^{-5} M_{\text{Pl}}$ . However, under the power law of expansion, it can be avoided on the ranges, $N \gtrsim 51$, $p < 1.66$, $p > 2.03$, $l_0 \gtrsim 3 \times 10^5 M_{\text{Pl}}^{-1} (H_i \lesssim 3.3 \times 10^{-6} M_{\text{Pl}})$, $M \gtrsim 2.8 \times 10^{-3} M_{\text{Pl}}$ , and $c_2 > 1$. Interestingly enough, all of the above ranges fit with the observational constraints. Beyond these ranges, the backreaction problem is more likely to occur. In these cases, the results of this research provide more arguments against the simple gauge invariant coupling $f^2 FF$, as way of generating PMF.

Therefore, the results of this chapter is consistent with investigating PMF in the natural inflation NI [93], and $R^2$-inflation [97], at which, the backreaction problem can be avoided under certain parameters of the models.

# CHAPTER SEVEN: THE PMF GENERATED BY $f^2FF$ IN NATURAL INFLATION

In this chapter, the simple inflation model, $f^2FF$, of PMF will be investigated under the simplest version of natural inflation NI, in the same way as done under large field inflation in the last chapter. The potential of Natural Iinflation NI can be written [61, 81] as,

$$V(\phi) = \Lambda^4 \left( 1 + \cos(\frac{\phi}{\sigma}) \right),$$  (7.1)

where $\Lambda$ is fixed by CMB normalization of the potential, and $\sigma$ is the mass scale of the model. Since Eq.(7.1) is a periodic, even function of $\phi$, one can study the potential in the interval $\phi \in [0, \pi\sigma]$ [81]. In order to reach the GUT scale of inflation ( $10^{15} - 10^{16}$ GeV ), (as would follow from the BICEP2 results), the mass scale has to be of the order of $\sigma \sim M_{Pl}$ and $\Lambda \sim M_{GUT} \sim 10^{-3} M_{Pl}$ [61].

The order of this chapter will be as follows. In section 7.1, the slow roll inflation formulation will be presented for both the simple de Sitter model of expansion and the more general, power law expansion in the context of NI. In section 7.2, the PMF and associated electric fields are computed for NI at different values of the parameters. In the last section, we summarize and discuss the results.

## 7.1 Slow Roll Analysis of NI

Defining the slow roll parameters of inflation in terms of the potential [55], of a single inflation field for NI,

$$\epsilon_{1V}(\phi) = \frac{1}{2} M_{\text{Pl}}^2 \left(\frac{V_\phi}{V}\right)^2 = \frac{1}{2\zeta^2} \left(\frac{\sin(\frac{\phi}{\sigma})}{1 + \cos(\frac{\phi}{\sigma})}\right)^2, \tag{7.2}$$

$$\epsilon_{2V}(\phi) \simeq 2 M_{\text{Pl}}^2 \left(\left(\frac{V_\phi}{V}\right)^2 - \frac{V_{\phi\phi}}{V}\right) = \frac{2}{\zeta^2} \frac{1}{\cos(\frac{\phi}{\sigma}) + 1}, \tag{7.3}$$

where $\zeta \equiv \sigma / M_{\text{Pl}}$. If $V = 0$ ($\phi = \pi\sigma$), then $(\epsilon_{1V}, \epsilon_{2V}) \to \infty$. Hence, one needs to avoid this value of the field in this model. Also, the slow roll parameters can be written in terms of the Hubble parameter as Eq.(6.4) and the relation between the two formalisms, as Eq.(6.5). Inflation ends when $\epsilon_{1V}(\phi) \approx 1$, then from Eq.(7.2),

$$\phi_{end} \approx \sigma \arccos\left(\frac{1 - 2\zeta^2}{1 + 2\zeta^2}\right). \tag{7.4}$$

If $\phi_{end} \ll 1$, then $\sigma_{end} \ll 1$, or $\sigma_{end} \to \frac{M_{\text{Pl}}}{2}\left(1 - \frac{(2n+1)}{2}\pi\right)$, where, $n$ = integer.

Also, from the relations between the slow roll parameters and the scalar spectral index, $n_s$, (6.13) and tensor-to-scalar ratio, $r$, (6.16), one can write the relations,

$$n_s = 1 + \frac{3}{\zeta^2} - \frac{1}{\zeta^2} \sec^2\left(\frac{\phi}{2\sigma}\right) \tag{7.5}$$

$$r = \frac{8}{\zeta^2}\left(\frac{\sin(\frac{\phi}{\sigma})}{1+\cos(\frac{\phi}{\sigma})}\right)^2. \tag{7.6}$$

Similarly, the first order of approximation for $N$, can be written [55] as,

$$N = \ln\frac{a(t_{end})}{a(t)} \simeq -\sqrt{\frac{1}{2M_{\text{Pl}}^2}}\int_\phi^{\phi_{end}}\frac{1}{\sqrt{\epsilon_1}}d\phi = \zeta^2\ln\left[\frac{1-\cos(\frac{\phi_{end}}{\sigma})}{1-\cos(\frac{\phi}{\sigma})}\right], \tag{7.7}$$

where $N$ is the difference between the final e-fold and the e-fold at $t$. Solving for $\phi$ in (7.7),

$$\phi = \sigma\arccos\left(1-[1-\cos(\frac{\phi_{end}}{\sigma})]\exp(-N/\zeta^2)\right). \tag{7.8}$$

Substituting $\phi_{end}$ from (7.4) into (7.8),

$$\phi \simeq \sigma\arccos\left[1-\left(\frac{4\zeta^2}{1+2\zeta^2}\right)\exp(-N/\zeta^2)\right]. \tag{7.9}$$

From (7.9), $\phi$ changes as a function of $N$. Hence, both of them are functions of time. Now, combine (7.5) and (7.6),

$$r = \frac{8}{3}\left\{n_s - 1 + \frac{1}{\zeta^2}\sec^2\left(\frac{\phi}{2\sigma}\right)\right\}\left(\sin(\frac{\phi}{\sigma})/\left\{1+\cos(\frac{\phi}{\sigma})\right\}\right)^2. \tag{7.10}$$

Substituting (7.9) into (7.10), one can plot $r-n_s$ for different values of $\zeta$ and $N$ at the end of inflation, see Fig.7.1. [81]. In this figure, $f$ stands for $\sigma$ in the text.

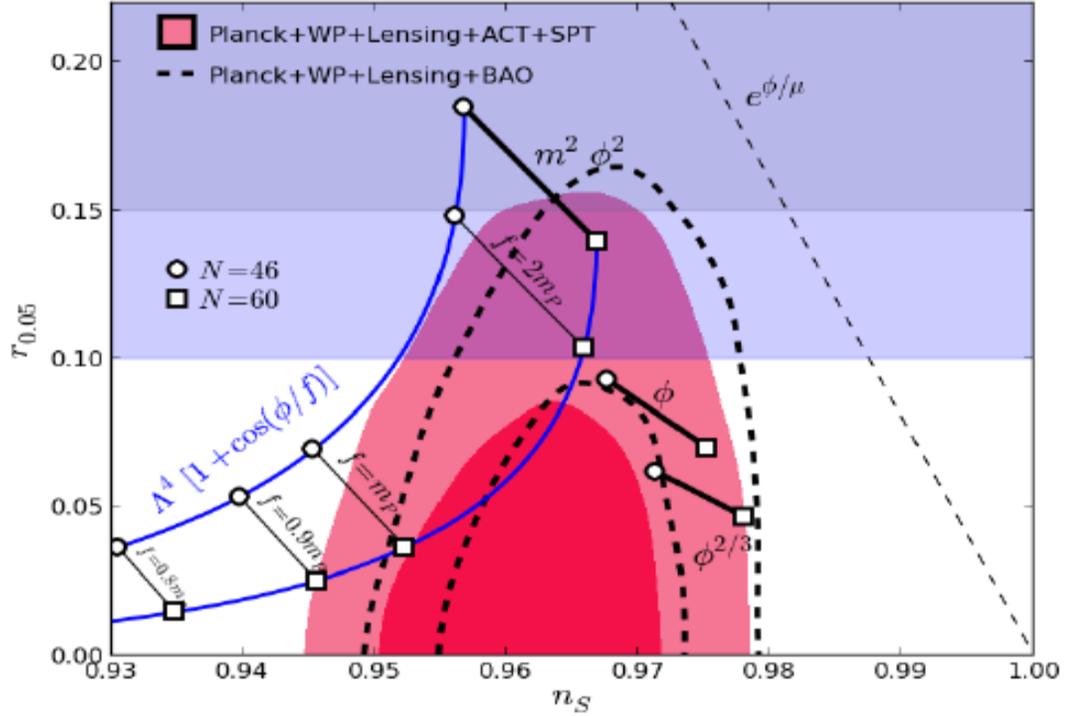

**Fig.7**.1. The natural inflation NI, the solid blue curves for, $\zeta = 0.8,\ 0.9,\ 1,$ and $2$ , and $N = 46,\ 60$ as shown in $r_{0.05} - n_s$ diagram, which is drawn from Planck+WMAP+ACT+SPT. Here $f$ stands for $\sigma$ in the text. Courtesy, [81].

### 7.1.1    NI on de Sitter Expansion

Again, the de Sitter model can be used as an approximation at the early stages of inflation. Thus, substituting of (7.1) and (6.10) into (6.19) and then integrating both sides yields,

$$\sigma \csc(\frac{\phi}{\sigma})\, d\phi = \frac{\Lambda^4}{3H_i^2} \frac{d\eta}{\eta}. \tag{7.11}$$

Solving for $\phi(\eta)$,

$$\phi(\eta) = 2\sigma \arctan[c_2 \, |\eta|^{-\frac{\Lambda^4}{3H_i^2\sigma^2}}], \tag{7.12}$$

where $c_2$ is the integration constant and $H_i$ stand for the rate of expansion during inflation. Adopting the BICEP2 preferred model, where the inflation potential is very small at the end of inflation, $\eta_{end}$,

$$V(\phi_{end}) = \Lambda^4 \left( 1 + \cos(2\arctan[c_2 \left| \eta_{end} \right|^{-\frac{\Lambda^4}{3H_i^2\sigma^2}}]) \right) \ll 1 \,, \tag{7.13}$$

Hence, the argument of cos should be $\pi$. Thus, the integration constant has to be

$$c_2 \gg \left| \eta_{end} \right|^{\frac{\Lambda^4}{3H_i^2\sigma^2}} \,. \tag{7.14}$$

The form (7.12) and the limit (7.14) will be used to derive the coupling function, $f(\eta)$, and then calculate the electromagnetic spectra in the de Sitter model approximation, in the next section.

### 7.1.2   NI on a Power Law Expansion

For more optimal slow roll analysis that has a smooth exit from inflation, inflation can be described by the power law expansion. Thus, plugging (7.2) and (7.9) into (6.7), yields

$$a(\eta) = l_0 \left| \eta \right|^{-1-\frac{1}{2\zeta^2}\left(\frac{\sin(\varpi)}{1+\cos(\varpi)}\right)^2} \,, \tag{7.15}$$

where, $\varpi = \arccos\left[ 1 - \left( \frac{4\zeta^2}{1+2\zeta^2} \right) \exp(-N/\zeta^2) \right]$. The Hubble parameter is then

$$H(\eta) = \frac{a'(\eta)}{a^2(\eta)} = -\frac{(1+\frac{1}{2\zeta^2}\left(\frac{\sin(\varpi)}{1+\cos(\varpi)}\right)^2)}{l_0} \left| \eta \right|^{\frac{1}{2\zeta^2}\left(\frac{\sin(\varpi)}{1+\cos(\varpi)}\right)^2} \,. \tag{7.16}$$

For $N \geq 50$, and $\zeta \leq 2$, we have $\varpi \approx 0$. Then we end up with $a(\eta) \approx l_0 |\eta|^{-1}$, and $H \approx const$, which is the same as the de Sitter model. In fact, that is a good approximation for most of the inflationary era. However, to end inflation, $(\epsilon_{1V}, \epsilon_{1H}) \to 1$, hence, $H \neq const$ at very last e-foldings, before the end of inflation.

However, in more general solution, one can substitute (7.15) and (7.16) into (7.17) and solve for $\phi$, to yield

$$\phi(\eta) = 2\sigma \arctan[c_2 \exp\left\{-\frac{l_0^{\,2}\left(e^{\frac{N}{\zeta^2}} - 2\zeta^2 + 2e^{\frac{N}{\zeta^2}}\zeta^2\right)^2 |\eta|^{\frac{2}{e^{\frac{N}{\zeta^2}} - 2\zeta^2 + 2e^{\frac{N}{\zeta^2}}\zeta^2}} \Lambda^4}{6\left(1 + e^{\frac{N}{\zeta^2}} - 2\zeta^2 + 2e^{\frac{N}{\zeta^2}}\zeta^2\right)\sigma}\right\}], \qquad (7.18)$$

where, $c_2$ is the integration constant. Adopting BICEP2 constraint, $V(\phi(\eta_{end})) \ll 1$, implies that,

$$c_2 \gg \exp\left\{\frac{l_0^{\,2}\left(e^{\frac{N}{\zeta^2}} - 2\zeta^2 + 2e^{\frac{N}{\zeta^2}}\zeta^2\right)^2 |\eta|^{\frac{2}{e^{\frac{N}{\zeta^2}} - 2\zeta^2 + 2e^{\frac{N}{\zeta^2}}\zeta^2}} \Lambda^4}{6\left(1 + e^{\frac{N}{\zeta^2}} - 2\zeta^2 + 2e^{\frac{N}{\zeta^2}}\zeta^2\right)\sigma}\right\}. \qquad (7.19)$$

Using (7.18) and (7.19) we can derive the coupling function, $f(\eta)$, but it is very complicated and actually not needed, as the simple de Sitter approximation is sufficient to investigate PMF under NI.

## 7.2 The PMF Generated in the NI Model

The electromagnetic spectra can be calculated by using the same method we used in the previous chapter but with NI potential. Therefore, we need first to define the coupling function, $f(\eta)$, in order to solve for the electromagnetic vector field, $A_\mu$. Substituting of (7.1) into (5.56) gives,

$$f(\phi(\eta)) = D \, \sin\left[\frac{\phi(\eta)}{2\sigma}\right]^{\frac{2\alpha\sigma^2}{3M_{\text{Pl}}^2}},$$
(7.20)

where, $D$, is a coupling constant. Substituting into (5.16) gives,

$$\mathcal{A}''(\eta,k) + \left(k^2 - Y(\eta)\right)\mathcal{A}(\eta,k) = 0,$$
(7.21)

where, $Y(\eta) = \dfrac{f''}{f}$, which can be calculated, by fixing $\phi(\eta)$ in Eq.(7.20).

Since the power law expansion approaches de Sitter for relatively high $N$ ($>50$), as shown from (7.15) and (7.16), in this section, we will solve (7.21) in a simple de Sitter model of expansion. However, using (7.18) to find $f(\eta)$ explicitly yields a very complicated $Y(\eta)$, and an analytical solution of (7.21) cannot be found. However, we can safely use de Sitter expansion.

The de Sitter approximation was used by [40] to investigate PMF. One can investigate PMF under de Sitter model by substituting (7.12) into (7.20) and (7.21). Hence,

$$Y(\eta) = \frac{f''}{f} = \frac{2\alpha\eta^{-2+\frac{2\Lambda^4}{3\sigma^2 H_i^2}}\Lambda^4\left(3c_2^2 M_{\text{Pl}}^2\left(3\sigma^2 H_i^2 - 2\Lambda^4\right) + \sigma^2\eta^{\frac{2\Lambda^4}{3\sigma^2 H_i^2}}\left(9H_i^2 M_{\text{Pl}}^2 + 2\alpha\Lambda^4\right)\right)}{81\sigma^2 H_i^4 M_{\text{Pl}}^4\left(c_2^2 + \eta^{\frac{2\Lambda^4}{3\sigma^2 H_i^2}}\right)^2}.$$
(7.22)

By using the limit (7.14) and the facts that, $(\eta,\ \eta_{end}) \ll -1$, $\sigma \sim M_{Pl}$, $\Lambda \sim M_{GUT} \sim 10^{-3} M_{Pl}$ [61] and the upper limit of $H_i < 3.7 \times 10^{-5} M_{Pl}$ at 95% CL [56], then $\dfrac{2\Lambda^4}{3\sigma^2 H_i^2} \ll 1$. Thus, $Y(\eta)$ can be written as,

$$Y(\eta) \approx \frac{6\alpha\eta^{-2}\Lambda^4\left(3\sigma^2 H_i^2 - 2\Lambda^4\right)}{81\sigma^2 H_i^4 M_{Pl}^2 c_2^2}. \qquad (7.23)$$

Substituting (7.23) into (7.21),

$$\mathcal{A}''(\eta,k) + \left(k^2 - \frac{6\alpha\eta^{-2}\Lambda^4\left(3\sigma^2 H_i^2 - 2\Lambda^4\right)}{81\sigma^2 H_i^4 M_{Pl}^2 c_2^2}\right)\mathcal{A}(\eta,k) = 0. \qquad (7.24)$$

Eq.(7.24) is a Bessel differential equation. Its solution with the initial condition, $\mathcal{A}\left(0^-,\ k\right) \to 0$, will be similar to (5.36). Hence, $\mathcal{A}(\eta,k)$, can be written [79] as

$$\mathcal{A}(\eta,k) = (k\eta)^{1/2}\left[C_1(k)\boldsymbol{J}_\chi(k\eta) + C_2(k)\boldsymbol{J}_{-\chi}(k\eta)\right]$$

where, $\chi$ is given by

$$\chi = \frac{\sqrt{27 + \dfrac{8\alpha\left(3\sigma^2 H_i^2\Lambda^4 - 2\Lambda^8\right)}{c_2^2\sigma^2 H_i^4 M_{Pl}^2}}}{6\sqrt{3}}. \qquad (7.25)$$

Adopting BICEP2 limit (7.14), implies that $\chi = 1/2$, which is comparable with, $\gamma = 0$, in exponential potential. As the form of $f(\eta)$ in (7.20) is different than (5.29), the implication of the value $\chi$ on the expansion in NI is different than the implication of $\gamma$ in exponential potential. Therefore, a scale invariant PMF cannot be generated in NI under simple de Sitter model of inflation, if the BICEP2 favored model of potential is adopted. Calculating the electromagnetic spectra shows that, they are almost of the same order of magnitude, at $k \ll 1$, see Fig.7.2.

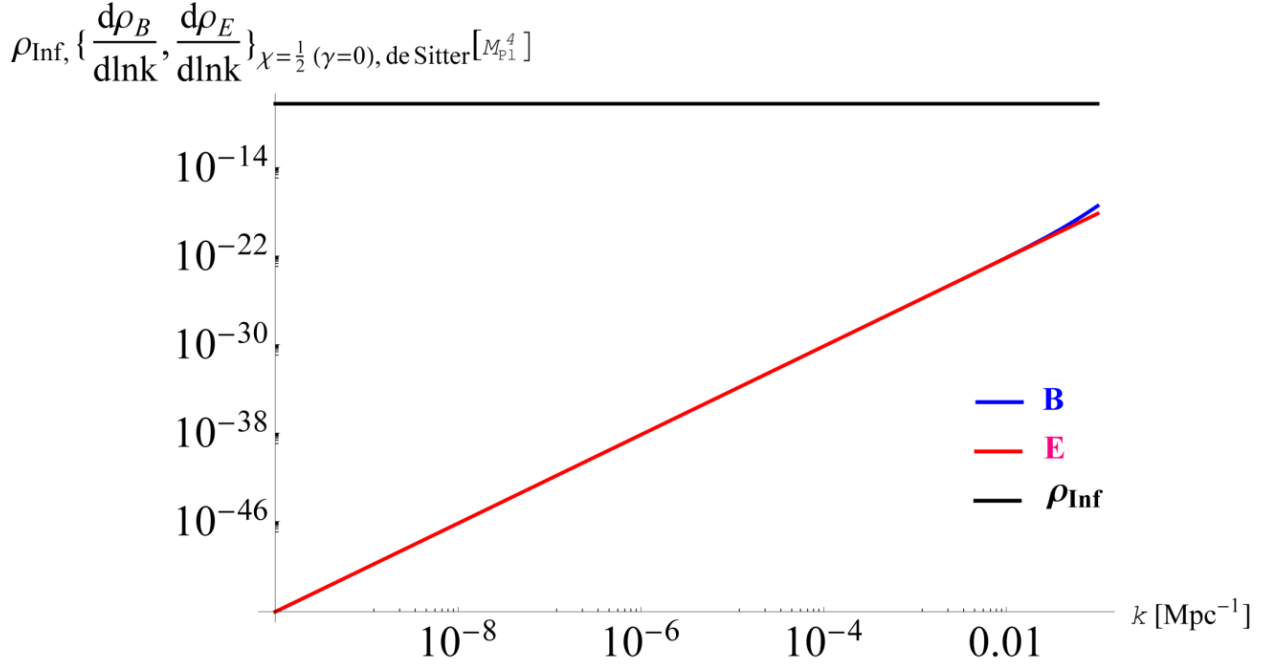

$$\rho_{\mathrm{Inf}}, \{\frac{\mathrm{d}\rho_B}{\mathrm{dln}k}, \frac{\mathrm{d}\rho_E}{\mathrm{dln}k}\}_{\chi = \frac{1}{2} \, (\gamma = 0), \, \mathrm{de\,Sitter}} [M_{\mathrm{Pl}}^4]$$

**Fig.7.2**. The electromagnetic spectra and the inflationary energy density, $\rho_{\mathrm{Inf}}$, in NI at $\chi = 1/2$ ($\gamma = 0$, in exponential potential), $\eta = -20$, $\sigma \approx M_{\mathrm{Pl}} = 1$, $\Lambda = 10^{-3}$, $H_i = 3.6 \times 10^{-5} M_{\mathrm{Pl}}$, $\alpha = 2$, and $D = 1$. The spectrum of electric field is of the same order of magnitude as the spectrum of the magnetic field for $k \ll 1$. At relatively high $k$, they start diverging from each other. However, the energy density of inflation generated by NI, $\rho_{\mathrm{Inf}}$ is much larger than the electromagnetic energy density.

On the other hand, if the BICEP2 limit is relaxed and the scale invariance condition is enforced, $\chi_1 = 5/2$ (compared with $\gamma = +3, -2$, in exponential potential), then $c_2$ becomes

$$c_2 = \sqrt{\frac{1}{27} \frac{\alpha \, \Lambda^4}{H_i^2 M_{\mathrm{Pl}}^2}} \,. \tag{7.26}$$

The coupling function can be written as,

$$f(\eta) = D \, \sin\left[\arctan\left(\sqrt{\frac{1}{27} \frac{\alpha \, \Lambda^4}{H_i^2 M_{\mathrm{Pl}}^2}} \, \eta^{-\frac{\Lambda^4}{3H_i^2 f^2}}\right)\right]^{\frac{2f^2\alpha}{3M_{\mathrm{Pl}}^2}}. \tag{7.27}$$

The electric and magnetic spectra in this case can be seen in Fig.7.3. We can see that at extremely long wavelength ( $k \ll 1$ ), the electric field spectrum far exceeds that of the magnetic field and the energy density of the inflation which is generated by NI and the energy of inflation $\rho_{\mathrm{Inf}}$. It may cause the backreaction problem. However for $k_{\min} \gtrsim 8.0 \times 10^{-7} \, \mathrm{Mpc}^{-1}$, the electromagnetic energy can go below that of inflation. Most of the observable scale is above $k_{\min}$. That range of $k$ includes most of the observable scales according to Planck, 2015. For example, it includes the standard pivot scale, $k_* = 0.05 \mathrm{Mpc}^{-1}$. Further, it includes some of the cut-off scale, $\ln(k_c / \mathrm{Mpc}^{-1}) \in [-12, -3]$, chosen by Planck, 2015 [97]. Therefore, the backreaction problem might be avoided in generating PMF by the $f^2 FF$ model in NI, under de Sitter expansion for $k_{\min}\eta < k\eta \ll 1$.

Likewise, plotting the electromagnetic spectra as a function of the Hubble parameter $H_i$, shows that the electric field is always greater than the magnetic field and can exceed the energy of inflation for $H_{\min} \gtrsim 1.25 \times 10^{-3} M_{\mathrm{Pl}}$.(See Fig.7.4). This value is well above the upper limit of the Hubble parameter, obtain by Planck, 2015, $H_i \lesssim 3.6 \times 10^{-5} M_{\mathrm{Pl}}$. Hence, this model can be free from the backreaction problem.

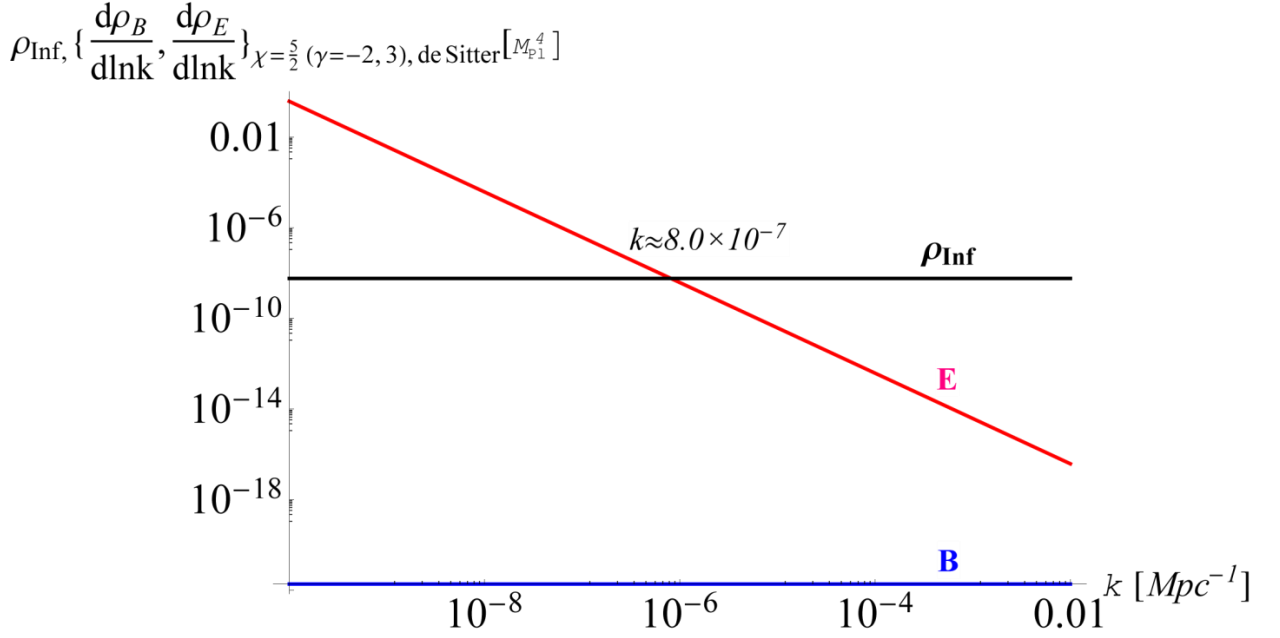

**Fig.7.3**. The electromagnetic spectra and the inflationary energy density, $\rho_{\text{Inf}}$, in NI model at $\chi = 5/2$ ( $\gamma = = -2, 3$, in exponential potential), $\eta = -20$, $\sigma \approx M_{\text{Pl}} = 1$, $\Lambda = 10^{-3}$, $H_i = 3.6 \times 10^{-5} M_{\text{Pl}}$, $\alpha = 2$, and $D = 1$. The spectrum of electric field is much greater than the spectrum of magnetic field for extremely low, $k \ll 1$. For $k_{\min} \gtrsim 8.0 \times 10^{-7} \text{Mpc}^{-1}$, the electromagnetic energy density can go below that of inflation. Hence, the backreaction problem might be avoided for $k_{\min} \eta < k\eta \ll 1$.

Similarly, plotting energy density of inflation and the electromagnetic spectra as a function of $\zeta$ clearly shows that the backreaction problem can be avoided for the possible values (See Fig.7.5). The value $\zeta$ plays an effective role in the characteristics of the natural inflation and its implications. For example, in the case of $\zeta \gg 1$, the natural inflation behaves like quadratic inflation, see Fig.7.1 [61].

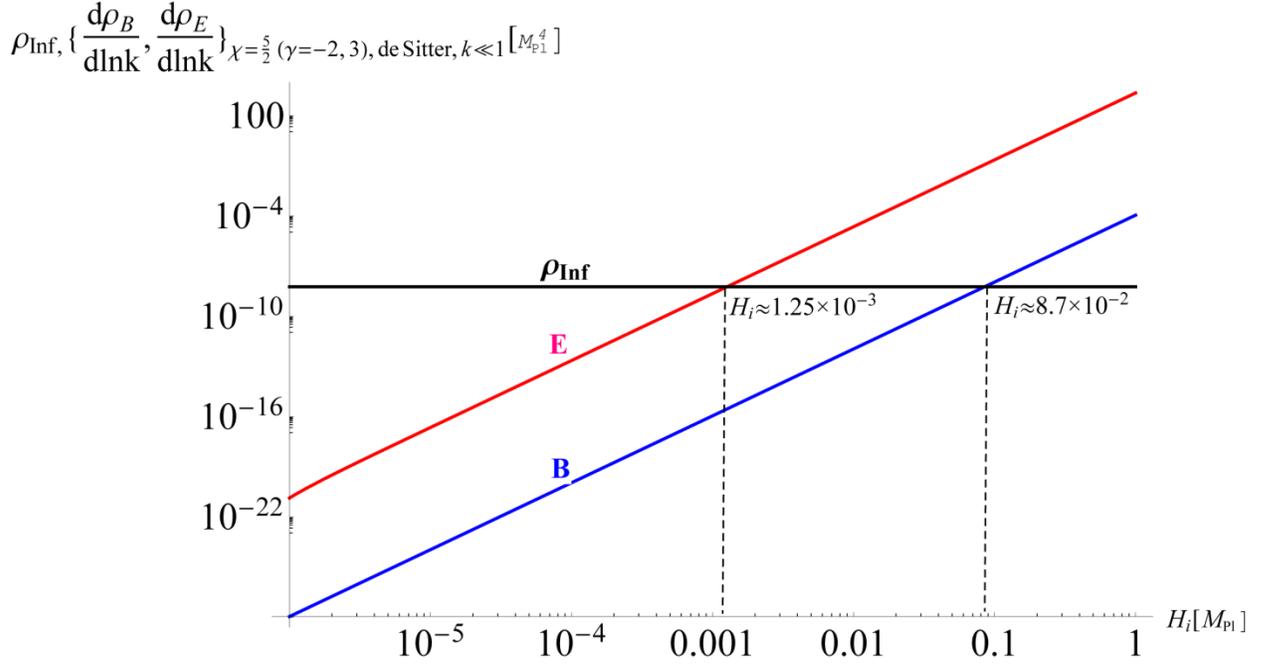

**Fig.7.4**. The electromagnetic spectra and the inflationary energy density, $\rho_{\text{Inf}}$ in NI model function of $H_i$ at $\chi = 5/2$ , ($\gamma = -2, 3$, in exponential potential) $\eta = -20$, $\sigma \approx M_{\text{Pl}} = 1$ , $\Lambda = 10^{-3}$, $k_i = 10^{-3}\text{Mpc}^{-1}$, $\alpha = 2$ and $D=1$. The spectrum of electric field is always much greater than the spectrum of magnetic field and can exceed the energy of inflation for $H_{\text{min}} \gtrsim 1.25 \times 10^{-3} M_{\text{Pl}}$. This value is well above the upper limit of the Hubble parameter, obtained by Planck, 2015, $H_i \lesssim 3.6 \times 10^{-5} M_{\text{Pl}}$. Below $H_{\text{min}}$, all electromagnetic spectra will be less than $\rho_{\text{Inf}}$. Hence, that might avoid the backreaction problem.

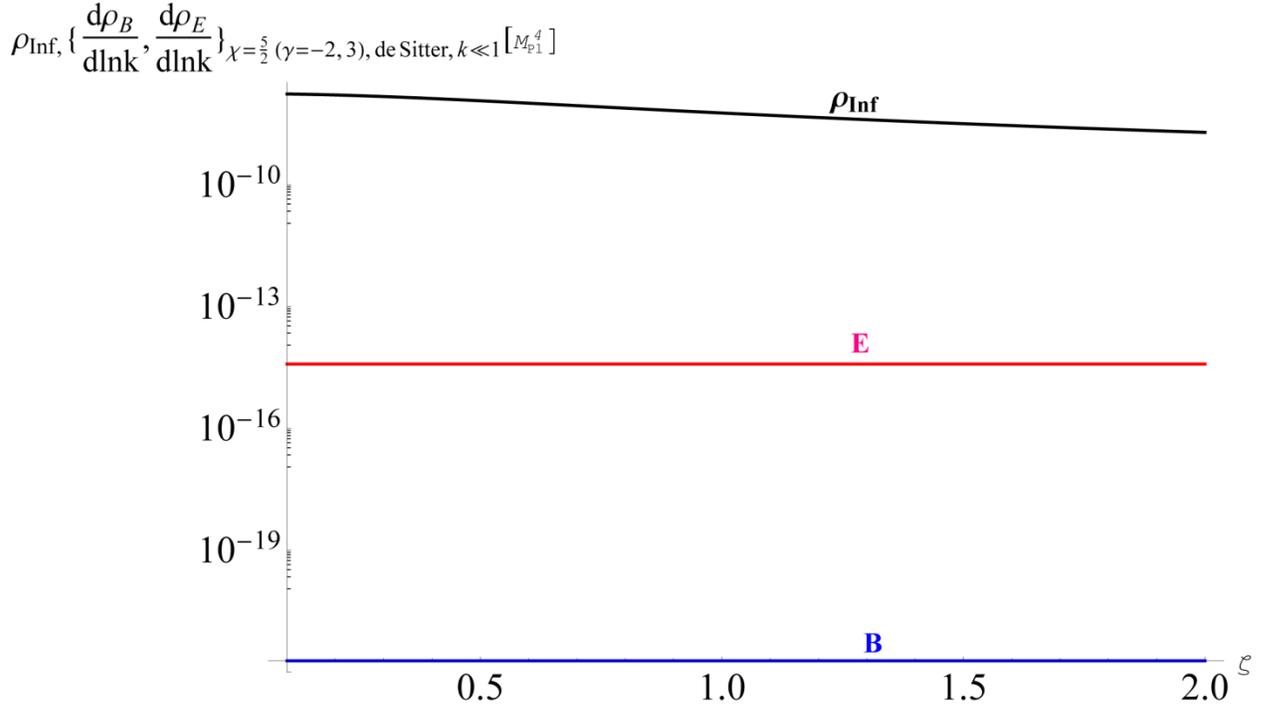

**Fig.7.5.** The electromagnetic spectra and the inflationary energy density, $\rho_{\text{Inf}}$, in NI model as a function of $\zeta = \sigma / M_{\text{Pl}}$ at $\chi = 5/2$, ($\gamma = -2$, 3, in exponential potential), $\eta = -10^2$, $\Lambda = 10^{-3}$, $\alpha = 2$, $H_i = 3.6 \times 10^{-5} M_{\text{Pl}}$, and $D = 1$. The spectrum of electric field is always much greater than the spectrum of magnetic field for $k\eta \ll 1$. However, both electric and magnetic spectra are much less than $\rho_{\text{Inf}}$. The value $\zeta$ plays an effective role in the characteristics of the natural inflation and its implications.

Finally, one can analyze the shape of the electromagnetic spectra as a function of $\Lambda$. As seen in Fig.7.6, there is a narrow range of $\Lambda$ ($\sim 0.00874 M_{\text{Pl}}$), at which the electric fields can even fall below the magnetic field. In order to decide the range of $k$ for which the electric field energy is less than the magnetic fields, one can plot the electromagnetic spectra as a function of $k$, as in Fig.7.7. The range is $k \gtrsim 2.53 \times 10^{-3} \text{Mpc}^{-1}$, for $k \ll 1$, around $k_{\text{min}} \sim 0.0173 \text{Mpc}^{-1}$. As we choose $M_{\text{Pl}} = 1 (\approx 1 \times 10^{19} \text{GeV})$, the appropriate values of $\Lambda$ is in the order of $M_{\text{GUT}} \sim 10^{16} \text{GeV}$ that fits with the results of Ref.[61].

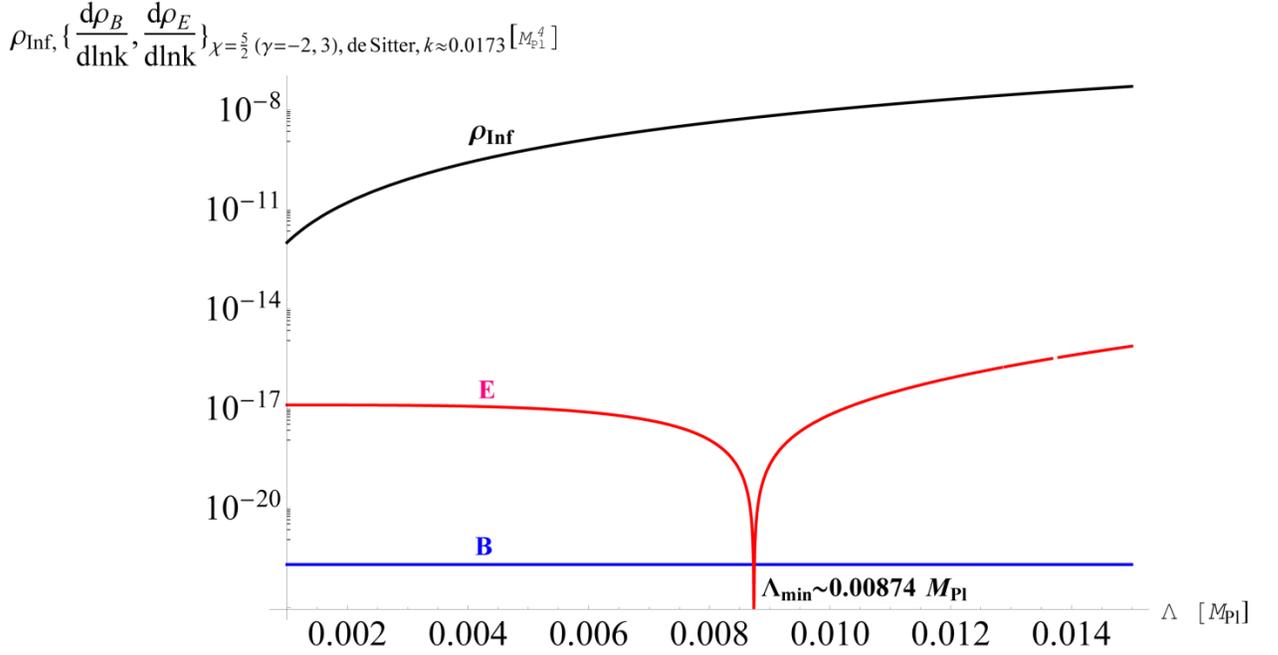

**Fig.7.6**. The electromagnetic spectra and the inflationary energy density, $\rho_{\text{Inf}}$, in NI, as a function of $\Lambda$ at $\chi = 5/2$, ($\gamma = -2, 3$, in exponential potential), $\eta = -20$, $\sigma \approx M_{\text{Pl}} = 1$, $\alpha = 2$, $D = 1$, and $k = 10^{-3}\text{Mpc}^{-1}$. The spectrum of electric field falls below the spectrum of magnetic field around $\Lambda = 0.00874 M_{\text{Pl}}$. However, both of them are much less than $\rho_{\text{Inf}}$, which may avoid the backreaction problem.

On the other hand, the range of $k$ includes most of the observable scales according to Planck, 2015. For example, it includes the standard pivot scale, $k_* = 0.05\text{Mpc}^{-1}$. Further, it includes some of the cut-off scale, $\ln(k_c/\text{Mpc}^{-1}) \in [-12, -3]$, chosen by Planck, 2015 [97]. However, the relatively narrow range of $k$ at which magnetic spectrum is higher than electric spectrum, may cause serious challenge to this model. This is so because after sufficient number of e-foldings the wave number may go below $k < 10^{-3}\text{Mpc}^{-1}$.

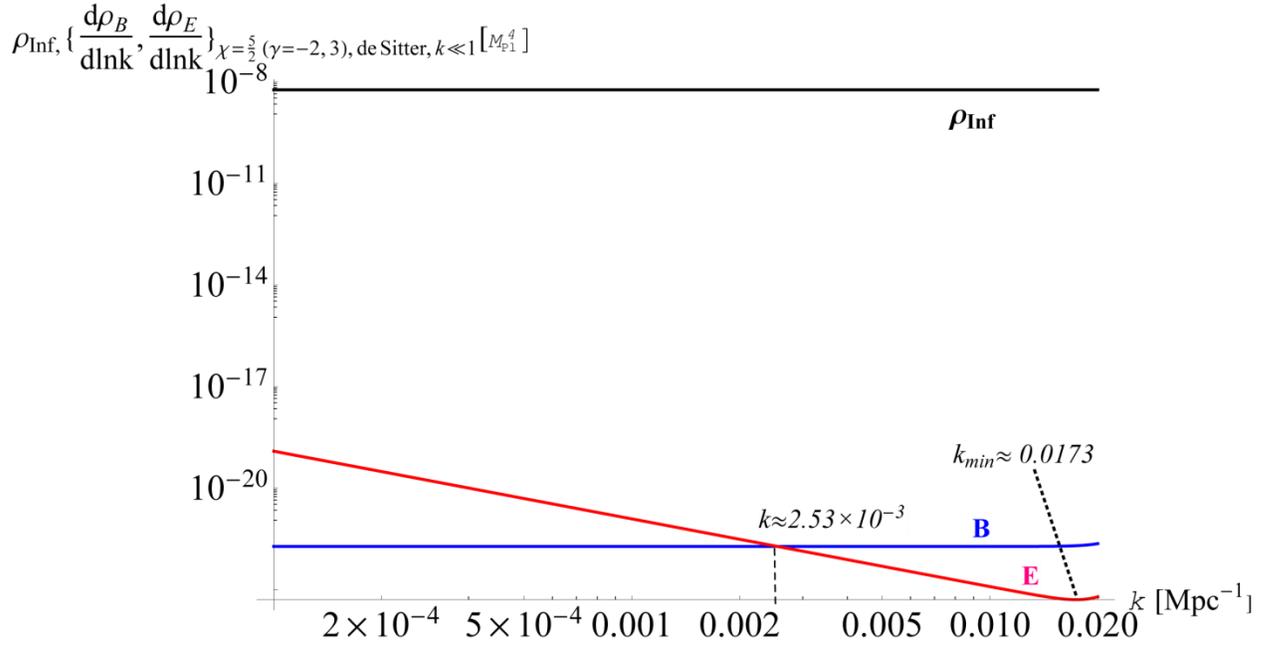

**Fig.7.7.** The electromagnetic spectra and the inflationary energy density, $\rho_{\text{Inf}}$, in NI as a function of $k$ at at $\chi = 5/2$ , ($\gamma = = -2, 3$, in exponential potential), $\eta = -20$, $\sigma \approx M_{\text{pl}} = 1$, $\alpha = 2$, $D = 1$, and $\Lambda = 0.00874 M_{\text{pl}}$. The spectrum of electric field falls below the spectrum of magnetic field on the range of, $k \gtrsim 2.53 \times 10^{-3} \text{Mpc}^{-1}$, at which the backreaction problem can be avoided.

## 7.4 Summary and Discussion of the PMF Generated in NI

In this chapter, we used the same method of [44, 92] to investigate the PMF in natural inflation, NI. We first presented the slow roll analysis of the NI, and derived the, $r - n_s$ relation. Unlike the large field inflation [92], for sufficiently large e-folding number, $N \geq 50$, the power law expansion can lead to the same result as the simple de Sitter model of expansion in NI.

We find that PMFs can in principle be generated in the NI model for all values of $\zeta = \sigma / M_{Pl}$. Under $V(0) \approx 0$ model of inflation, the scale invariant PMF is unlikely to be generated in the context of NI. That is similar to the case in the context of LFI [92]. However, if this constraint is relaxed, a scale invariant PMF can be achieved in NI. In this case, the magnitude of the PMF spectrum, at extremely low $k \ll 1$ is much smaller than the spectrum of the associated electric field. In comparison with the inflationary energy density, $\rho_{Inf}$, in NI and the upper bound of the energy density of inflation derived from WMAP7, $(\rho_{end})_{CMB} \lesssim 2.789 \times 10^{-10} M_{Pl}^4$ [109], the energy of the electric field may exceed the energy scale of inflation at $k \lesssim 8.0 \times 10^{-7} \text{Mpc}^{-1}$ and $H_i \gtrsim 1.25 \times 10^{-3} M_{Pl}$. That may prevent inflation from occurring at all. This is the problem of backreaction. One can conclude that for small enough value of $k$, this problem cannot be avoided in the $f^2FF$ model under natural inflation.

On the other hand for $k > 8.0 \times 10^{-7} \text{Mpc}^{-1}$ and $H_i \lesssim 1.25 \times 10^{-3} M_{Pl}$, both electric and magnetic energy densities can fall below the inflationary energy density. In this case, one can consider these values as, respectively, a lower bound of $k$ and an upper bound of $H_i$ for a backreaction-free model of PMF. Moreover, these scales include most of the observable ranges of $k$ and $H_i$.

Furthermore, there is a range of $\Lambda_{\min}(\sim 0.00874 M_{\text{Pl}})$, and $k \gtrsim 2.53 \times 10^{-3} \text{Mpc}^{-1}$, at which the energy density of the electric field can even fall below the energy density of the magnetic field. Again these values lie on the observable range of $k$ and the anticipated scale of $\Lambda$. Therefore, the problem of backreaction can be avoided easily in these ranges of values. Also, in this range, one does need the huge amount of conductivity needed to suppress the electric field in reheating era, which comes after the inflation. However, the relatively short range of $k$, presents a serious challenge to the viability of this model. One way to extend this research is to include the effect of reheating era and then to calculate the present value of PMF generated in NI as we do in the context of $R^2$-inflation [93].

# CHAPTER EIGHT: THE PMF GENERATED BY $f^2FF$ IN $R^2$-INFLATION AFTER PLACNK 2015

Very recently, the joint analysis of BICEP2/Keck Array and Planck (BKP) data was released on Feb 2015. The joint data of three probes eliminates the effect of dust contamination and shows that the upper limit of tensor to scalar ration, $r_{0.05} < 0.12$ at 95% CL, and the gravitational lensing B-modes (not the primordial tensor) are detected in $7\sigma$ [94]. Similarly, the inflationary models are constrained based on the new analysis of data. The scalar spectral index was constrained by Planck, 2015 to be $n_s = 0.9682 \pm 0.0062$ [96]. As a result, the more standard inflationary models, like $R^2$-inflation, which result low value of $r$, are the most favored ones. However, the chaotic inflationary models like large field inflation (LFI) and natural inflation (NI) are disfavored. These results ruled out the first results of BICEP2, 2014 [54].

In this chapter, the simple model, $f^2FF$, of PMF will be investigated in the context of $R^2$-inflation, in the same way we did for NI and LFI, in the last two chapters [92-93]. Further, we will constrain the reheating parameters under the same inflationary by using the reported upper limit of PMF by Planck, 2015 [95]. Also, the present PMF will be constrained based on the scale invariant magnetic field generated during inflationary era [97].

$R^2$-inflation was first proposed by Starobinsky in 1980 [2]. It is called $R^2$-inflation because, its action in Jordan frame, can be written as,

$$S = -\frac{M_{Pl}^2}{2} \int \sqrt{-g} \left( R - \frac{R^2}{6\mu^2} \right) + S_{matter},$$

(8.1)

where the parameter, $\mu$, is fixed by the normalization of the amplitude of scalar perturbation. However, the above action can be transformed into the Einstein frame by using the conformal symmetry of the metric, $g_{\mu\nu} \to e^{\sqrt{2/3}\phi/M_{Pl}} g_{\mu\nu}$. The new action in the Einstein frame becomes,

$$S = -\frac{M_{Pl}^2}{2} \int \sqrt{-g} \left( R - \frac{1}{2} \partial_\mu \partial^\mu \phi + V(\phi) \right) + S'_{matter} \ . \tag{8.2}$$

The potential $V(\phi)$ can be written [98-99] as,

$$V(\phi) = M^4 \left[ 1 - \exp\left(-\sqrt{2/3}\phi/M_{Pl}\right) \right]^2 , \tag{8.3}$$

where, $M$ is the amplitude of the potential and it can be determined by the amplitude of CMB anisotropies. It can be constrained as $M \sim 4.0 \times 10^{-5} M_{Pl}$ [81]. In Eq.(8.3) and throughout this chapter, we assume $\phi \gg M_{Pl}$.

The order of this chapter will be as follows, in the next section 8.1, the slow roll inflation formulation is presented for both de Sitter and power law expansion in the context of $R^2$-inflation. In section 8.2, the PMF and associated electric fields are computed in the same model of inflation. The constraining of reheating parameters in $R^2$-inflation and by using the reported upper limit of PMF is discussed in section 8.3. In section 8.4, the present PMF is constrained based on the magnetogensis, computed in the second section. Finally in 8.5, we summarize and discuss the results of the chapter.

## 8.1 Slow Roll Analysis of $R^2$-Inflation

The slow roll parameters, see Eq.(7.2)-(7.3), of inflation in terms of the potential, (8.3), of the $R^2$-inflation, can be written as [55, 81],

$$\epsilon_{1V}(\phi) = \frac{4}{3}\left[-1 + \exp\left(\sqrt{\frac{2}{3}}\frac{\phi}{M_{\text{Pl}}}\right)\right]^2, \tag{8.4}$$

$$\epsilon_{2V}(\phi) = \frac{2}{3}\left[\sinh\left(\frac{\phi}{\sqrt{6}M_{\text{Pl}}}\right)\right]^{-2}, \tag{8.5}$$

$$\epsilon_{3V}(\phi) \simeq \frac{2M_{\text{Pl}}^4}{\epsilon_{2V}(\phi)}\left[\frac{V_{\phi\phi\phi}V_\phi}{V^2} - 3\frac{V_{\phi\phi}}{V}\left(\frac{V_\phi}{V}\right)^2 + 2\left(\frac{V_\phi}{V}\right)^4\right] = \frac{2}{3}\left(\coth\left[\frac{\phi}{\sqrt{6}M_{\text{Pl}}}\right] - 1\right)\coth\left[\frac{\phi}{\sqrt{6}M_{\text{Pl}}}\right]. \tag{8.6}$$

As emphasized before, all parameters are assumed to be very small during the slow roll inflation, $(\epsilon_{1V}, \epsilon_{2V}, \epsilon_{3V}) \ll 1$. Also, inflation ends when, their values reach unity. In that case we have, $\phi_{\text{end}} \simeq 0.94M_{\text{Pl}}$, $1.83M_{\text{Pl}}$, $1.51M_{\text{Pl}}$, for $\epsilon_{1V}$, $\epsilon_{2V}$, and $\epsilon_{3V}$ at the end of inflation. Therefore, the first value will be used to avoid violating the slow roll condition.

Similarly, the scalar spectral index, $n_s$, and tensor-to-scalar ratio, $r$, can be written as follows [81],

$$n_s = 1 - 6\epsilon_{1V} + 2\epsilon_{2V} = 1 - 8\left(e^{\sqrt{\frac{2}{3}}\frac{\phi}{M_{\text{Pl}}}} - 1\right)^{-2} + \frac{4}{3}\text{csch}\left[\frac{\phi}{\sqrt{6}M_{\text{Pl}}}\right]^2 \approx 1 + \frac{4}{3}\text{csch}\left[\frac{\phi}{\sqrt{6}M_{\text{Pl}}}\right]^2, \tag{8.7}$$

$$r = \frac{A_t}{A_s} = 16\epsilon_{1V} = \frac{64}{3}\left(e^{\sqrt{\frac{2}{3}}\frac{\phi}{M_{\text{Pl}}}} - 1\right)^{-2}. \tag{8.8}$$

One can find the relation between $r$ and $n_s$ which depends on the number of e-folds of inflation, $N$. The first order of approximation for $N$ can be written as,

$$N = \ln\left(\frac{a(\eta)}{a(\eta_i)}\right) \simeq -\sqrt{\frac{1}{2M_{\text{Pl}}^2}}\int_\phi^{\phi_{\text{end}}}\frac{1}{\sqrt{\epsilon_1}}d\phi \simeq \frac{3}{4}\exp\left(\sqrt{\frac{2}{3}}\frac{\phi}{M_{\text{Pl}}}\right), \tag{8.9}$$

where, $\phi$, is the initial field and $\phi_{end}$ is the field at the end of inflation. Solving for $\phi$ from (8.9),

$$\phi \simeq \sqrt{\frac{3}{2}} M_{Pl} \ln\left(\frac{4(N_{end} - N)}{3}\right), \qquad (8.10)$$

where, $N_{end}$ is the number of e-folds at the end of inflation, we have assumed $\phi \gg \phi_{end}$. For simplicity, we will denote $(N_{end} - N) \equiv N$. Hence, plugging (8.10) into (8.7)-(8.8) yields,

$$r \simeq \frac{192}{(3 - 4N)^2 n_s - 64N}. \qquad (8.11)$$

The relation (8.11) is drawn for some interesting values of $N$ in Fig.8.1. The values of $r$, are well below the limit of BKP, 2015 ($r < 0.12$). Eq.(8.11) has no singularity for interesting values of $N$ and the observed value of $n_s$.

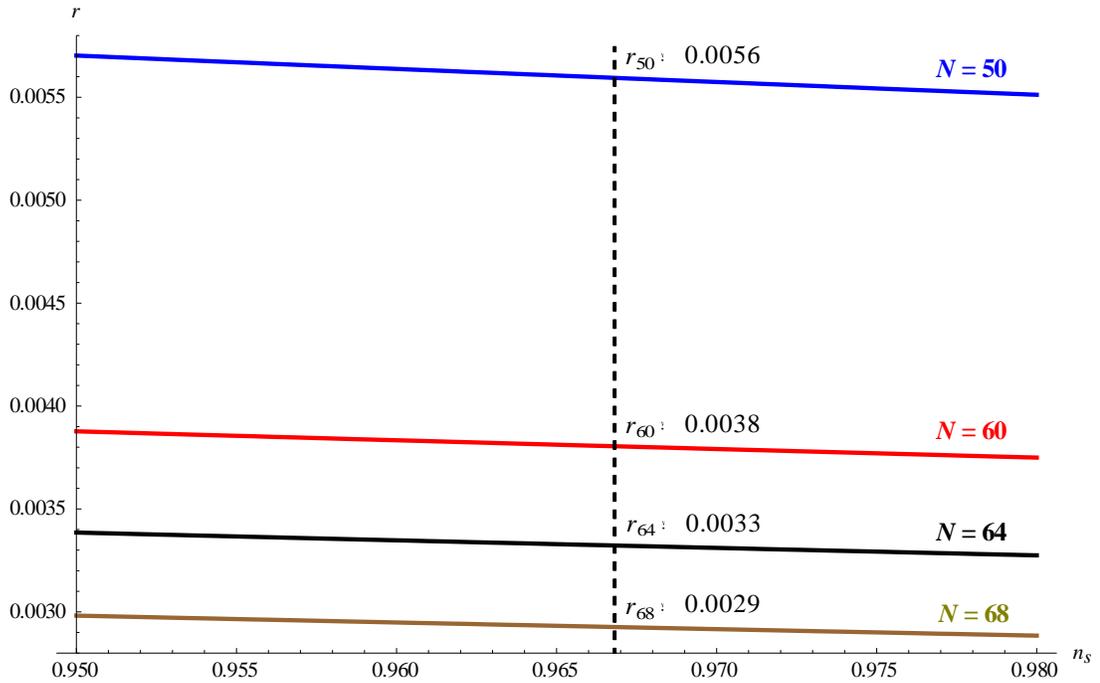

**Fig.8.1.** The $r - n_s$ relation in $R^2$–inflation, for $N = 50, 60, 64$, and 68. All values of $r_N$ at $n_s \simeq 0.968$, are well below the upper limit of $r$ ($< 0.12$) by BKP, 2015.

## 8.2 The Electromagnetic Spectra Generated in $R^2$-Inflation

As assumed in the last two chapters, the relation between the coupling function and the scale factor is of the power law form [44], $f(\eta) \propto a^\alpha$, where, $\alpha$ is free index to be determined later from the shape of the electromagnetic spectra. Also, $\alpha$ has different physical implications for different inflationary potentials. So, there is no unique relation between $\alpha$ and power index of expansion, $\beta$, in all models of inflation. Substituting (8.10) into (8.3) and then into (5.56) yields the coupling function as a function of $N$ for $R^2$-Inflation,

$$f(N) = D\left[\left(\frac{4N}{3}\right)^{\frac{\alpha}{4}} e^{-\frac{N\alpha}{3}}\right],\tag{8.12}$$

where, $D$ is a coupling constant. Substituting (8.12) into (5.16) gives,

$$\mathcal{A}''(\eta,k) + \left(k^2 - Y(\eta)\right)\mathcal{A}(\eta,k) = 0,\tag{8.13}$$

where the function $Y(\eta) = \dfrac{f''}{f}$. Hence, we need to write the derivative in terms of e-folds number, $N$, which has different forms in the case of de Sitter and power law. In the next two sections, the electromagnetic spectra are calculated for both de Sitter and power law expansion.

### 8.2.1 Inflationary Electromagnetic Spectra in de Sitter Expansion

For the zeroth approximation and after the first few e-foldings, we can consider $H$ as a constant ratio. The $i$ subscript is omitted in $H$ in this chapter to avoid confusing it with the reheating Hubble parameters in the next section. The de Sitter expansion is exactly exponential expansion as described by (6.9). In fact, de Sitter model does not have graceful exit from inflation [91]. But it can be assumed as a valid approximation on most parts of the inflation.

Assuming, $a(\eta_i) = \text{const}$, and substituting of $a(\eta) = -1/H\eta$ into (8.9) and differentiating both sides twice, yields that, $\partial_{\eta\eta}f(\eta) = \eta^{-2}\partial_{NN}f(N)$. Therefore, the function $Y(\eta)$ can be written as,

$$Y_{dS}(\eta) = \frac{\alpha\left[(3-4N)^2\alpha - 36\right]}{144N^2\eta^2}.$$ (8.14)

Also, the logarithmic relation between $N$ and $\eta$ makes the variation in $N$ very small compared with the variation in $\eta$. Hence, we assume that $N$ is quasi-constant and it may be considered as an independent function of $\eta$ explicitly in Eq.(8.13), $N \ncong N(\eta)$. So, plugging (8.14) into (8.13), yields,

$$\mathcal{A}(\eta,k) = (k\eta)^{1/2}\left[C_1(k)\mathbf{J}_\chi(k\eta) + C_2(k)\mathbf{J}_{-\chi}(k\eta)\right],$$ (8.15)

where, $\mathbf{J}_\chi(k\eta)$, is the Bessel function of the first kind, and the argument of the function,

$$\chi_{dS}(\alpha,N) = \frac{\sqrt{36N^2 - 36\alpha + (3-4N)^2\alpha^2}}{12N}.$$ (8.16)

The relevant PMF is obtained in the long wavelength regime, $k \ll 1$ (outside Hubble radius). In this limit, Eq.(8.15) can be written as [44],

$$\mathcal{A}_{k\ll 1}(\eta,k) = (k)^{-1/2}\left[D_1(\chi)(k\eta)^{\chi+\frac{1}{2}} + D_2(\chi)(k\eta)^{\frac{1}{2}-\chi}\right].$$ (8.17)

The constants, $D_1(\chi)$ and $D_2(\chi)$, can be fixed by using the normalization of $\mathcal{A}(\eta,k)$ and the other limit, $\mathcal{A}_{k\gg 1}(\eta,k) \to e^{-k\eta}/\sqrt{2k}$. Thus, they can be written as

$$D_1(\chi) = \frac{\sqrt{\pi}}{2^{\chi}} \frac{e^{-i\pi(\chi+\frac{1}{2})/2}}{\Gamma(\chi+1)\cos\left(\pi(\chi+\frac{1}{2})\right)}, \quad D_2(\chi) = \frac{\sqrt{\pi}}{2^{1-\chi}} \frac{e^{-i\pi(\chi+\frac{3}{2})/2}}{\Gamma(1-\chi)\cos\left(\pi(\chi+\frac{1}{2})\right)}. \quad (8.18)$$

By substituting (8.17) into (5.22)-(5.23), we get the spectra of the magnetic and electric fields. If the PMF is scale invariant, then the magnetic spectrum should be constant, which could be achieved at $\chi = 5/2$. This value can only be obtained around $\alpha \approx \{-7.44, \ 7.44\}$ in the range of interesting values of $N$, see Fig.8.2.

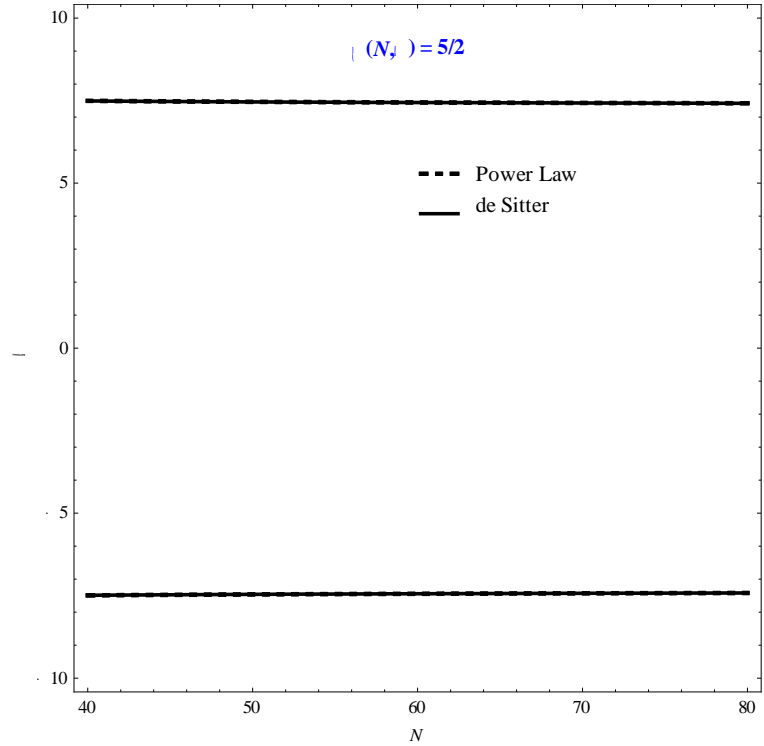

**Fig.8.2.** The $\alpha - N$ relation, at $\chi(N,\alpha) = 5/2$, the case at which we can generate the scale invariant PMF, $\alpha \approx 7.44$.

Although these values of $\alpha$ are too high, which may exceed dynamo limit in the case of exponential inflationary model [44], we adopt the positive value, $\alpha \approx 7.44$. In the last section of this chapter, we will discuss this point. So, let us assume it is valid at this point and use it to investigate the electromagnetic spectra generated by $f^2 FF$ for long wavelength approximation ($k \ll 1$). Hence, one can plot the electromagnetic spectra for different values of $H$ and $\alpha$, see Fig.3, for $H = 10^{-5} M_{Pl}$, which is around the Planck, 2015 upper limit of pivot Hubble parameter, ($H_* \lesssim 3.6 \times 10^{-5} M_{Pl}$) [96]. The pivot moment is the time when the commoving scale ($k_* = a_* H_* \simeq 0.05 Mpc^{-1}$ [96]) exits the Hubble radius. As a result, the conformal time can be taken as $\eta \approx -20$. In this case, the scale invariant PMF can be generated and the backreaction problem may be avoided at $\alpha \simeq 7.4359$. However, for $\alpha = 2$, the backreaction problem can be avoided easily, but the scale invariant PMF is not generated, see Fig.8.3.

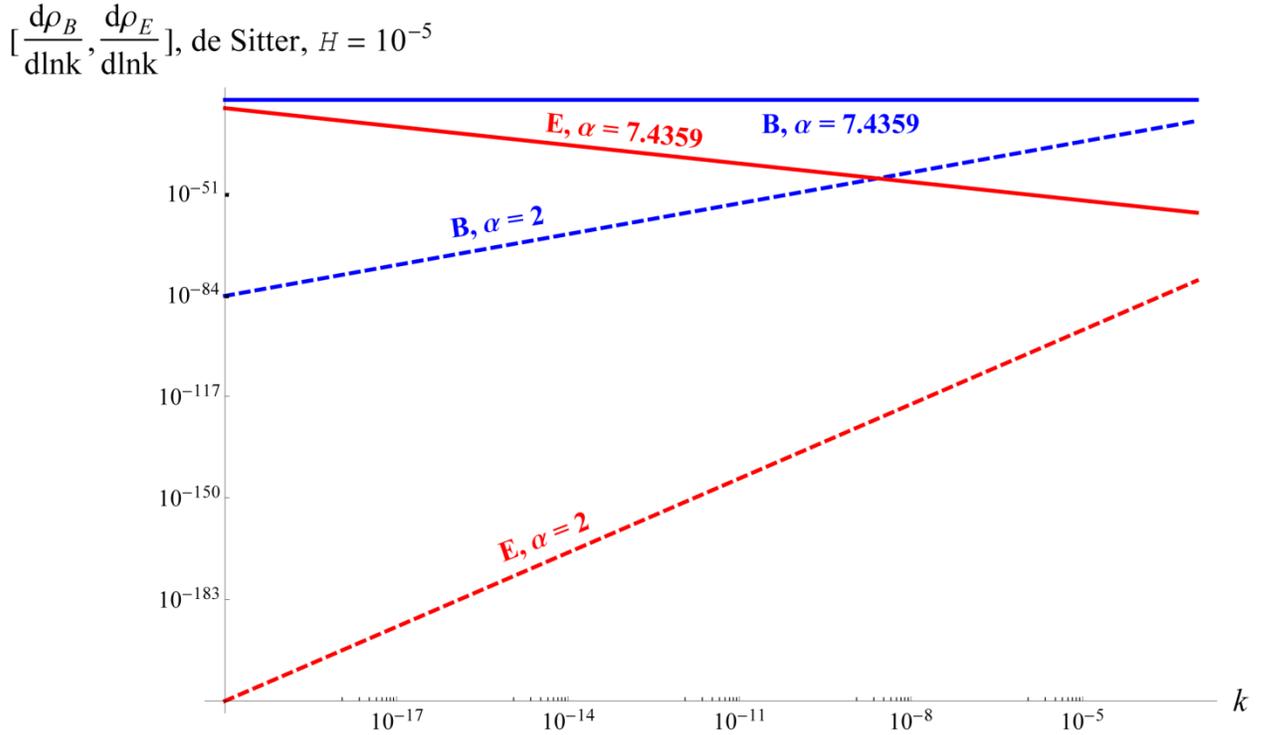

**Fig.8.3**. The EM spectra for $\alpha = 2$, 7.4359, $N = 64$, $\eta = -20$ and the expansion rate of inflation, $H = 10^{-5} M_{Pl}$. For $\alpha = 2$, the scale invariant PMF cannot be achieved although the backreaction problem is easily to be avoided. However, for $\chi(N, \alpha) = 5/2$ ($\alpha = 7.4359$), the scale invariant PMF can be generated and the magnetic field energy is more than the electric field.

On the other hand, in the large expansion rate case, $H = 0.2 M_{\text{Pl}}$, the scale invariant PMF can be generated at $\alpha \simeq 7.4359$ but the problem of backreaction may not be avoided in the low, $k\eta \ll 1$, see Fig.8.4. As the energy density at the end of inflation ($\epsilon_{1V} \approx 1$, $\rho_{\text{end}} = \frac{3}{2} V_{\text{end}}$) by adopting the value of $M \sim 4.0 \times 10^{-5} M_{\text{Pl}}$, is $\left(\rho_{\text{end}}\right)_{\text{R}^2\text{-inflation}} \simeq 1.1 \times 10^{-18} M_{\text{Pl}}^4$. It is clear from Fig.8.4 that the energy of electromagnetic spectra is much higher than $\rho_{\text{end}}$. In fact, this value of $H$ is way above the constraint upper limit [96]. Again, for $\alpha = 2$, the backreaction problem can be avoided easily, but the scale invariant PMF is not maintained throughout the inflationary era.

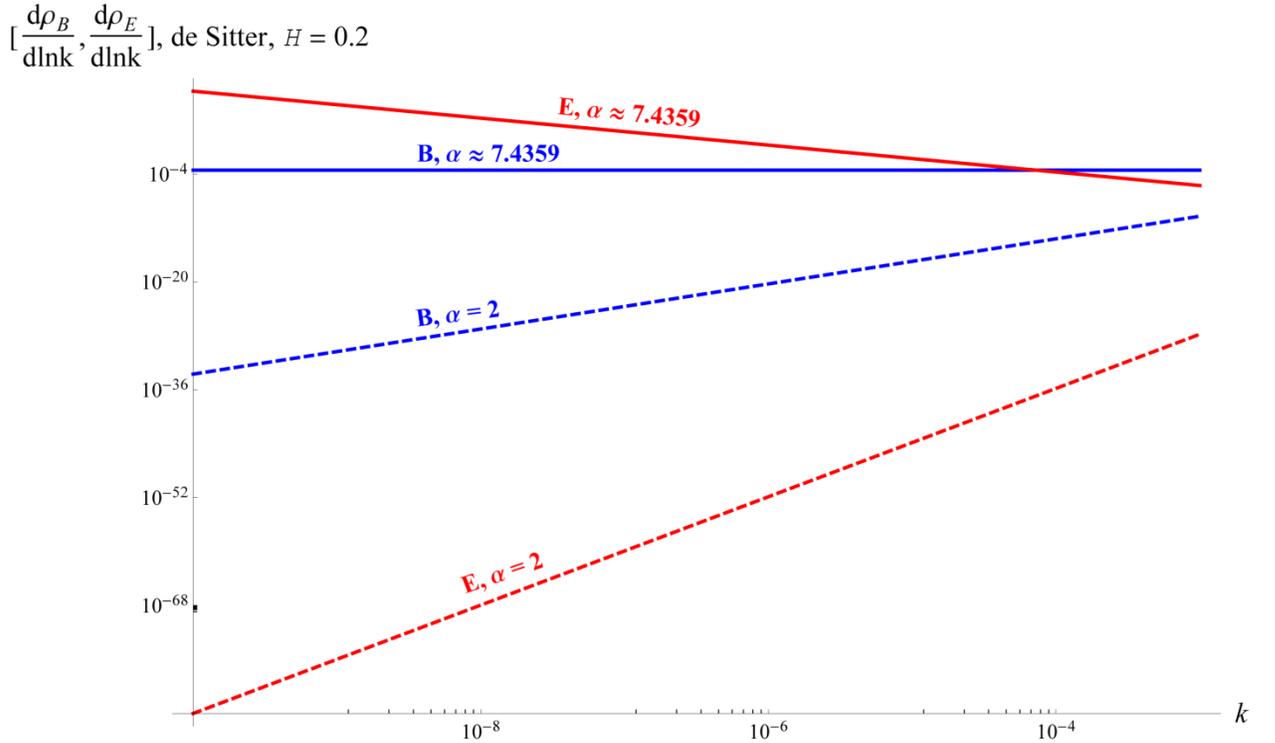

**Fig.8.4.** The EM spectra for $\alpha = 2$, $7.4359$, $N = 64$, $\eta = -20$ and the expansion rate of inflation, $H = 0.2 M_{\text{Pl}}$. For $\alpha = 2$, the scale invariant PMF cannot be achieved although the backreaction problem is easily to be avoided. However, for $\chi(N, \alpha) = 5/2$ ($\alpha = 7.4359$), the scale invariant PMF can be generated but the magnetic field energy is less than the electric field for low wavenumber.

The value of the index, $\alpha$, at which the scale invariant PMF can be generated, varies as the value of $N$ changes. There is a slit different between this relation in de Sitter and power law expansion, see Fig.8.5. These values are around, $\alpha \sim 7.44$, for an interesting values of e-folds, ($50 < N < 70$). The validity of $\alpha \sim 7.44$ will be discussed in the section 8.5, but from now onward, we adopt this value at $N = 64$ to study the scale invariant PMF. In general, the whole results of de Sitter are very close to the power law results.

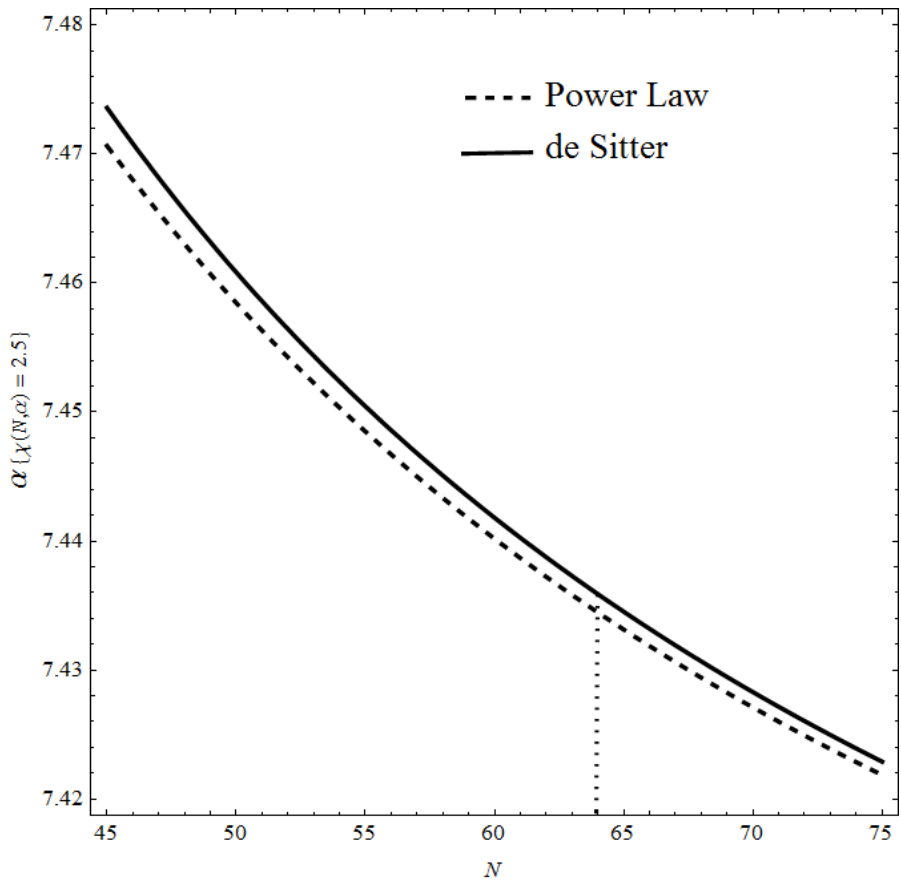

**Fig.8.5.** The relation between $\alpha$ and $N$ at which $\chi(N, \alpha) = 5/2$, for both de Sitter and power law expansion. For the interesting range of $N$ ($50 < N < 70$), the value of $\alpha$ is around, $\alpha \sim 7.44$.

First, let us draw the relation between electromagnetic spectra, Hubble parameter, $H$, and e-folding number, $N$, see Fig.8.6. It is clear, that, for the values of $H \lesssim 0.2 M_{\text{Pl}}$, the magnetic spectra is more dominant than the electric ones. In this case, we still have $\dfrac{d\rho_B}{d\ln k} \gg \rho_{\text{Inf}}$.

However, we can have $\dfrac{d\rho_B}{d\ln k} < \rho_{\text{Inf}}$ below the upper constraint of the Hubble parameter,

$H \lesssim 3.6 \times 10^{-5} M_{\text{Pl}}$. The current investigation will be bounded in $10^{-5} M_{\text{Pl}} \lesssim H \lesssim 0.2 M_{\text{Pl}}$.

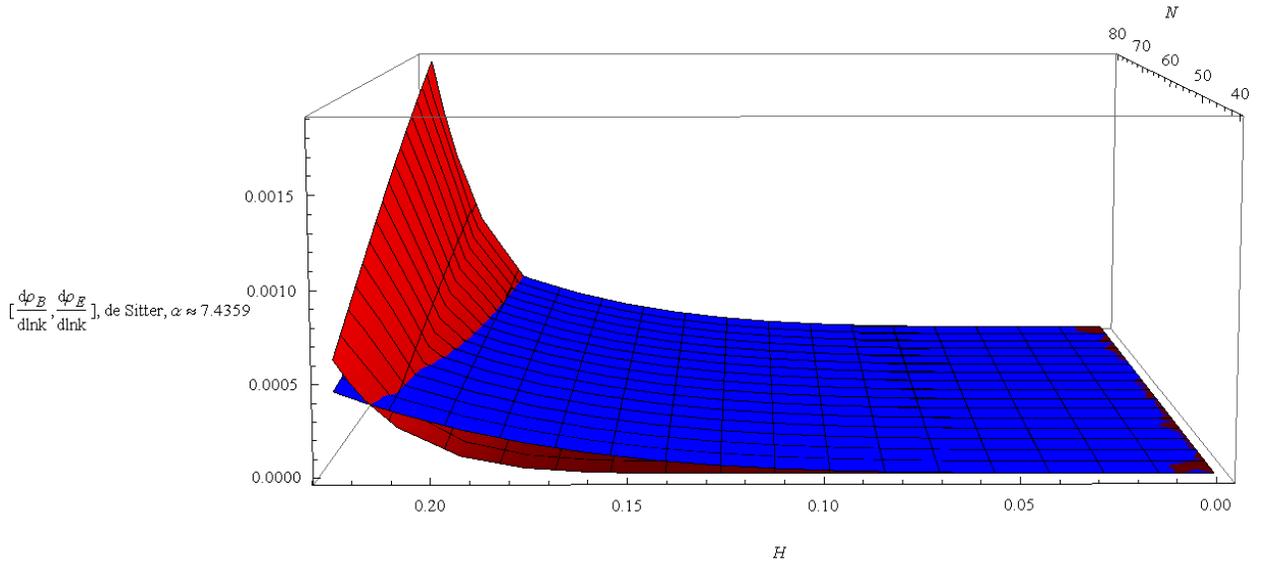

**Fig.8.6.** The relation between the EM spectra, Hubble parameter $H$, and e-folds number, $N$, for de Sitter expansion. Around $H \sim 0.2 M_{\text{Pl}}$, electric field can exceed the magnetic field in all interesting range of $N$ $(50 < N < 70)$.

## 8.2.2 Electromagnetic Spectra in Power Law Expansion

If the scalar field of inflation falls below a certain value, it starts to oscillate and then converts into particles in the reheating era, right after inflation. Hence, during inflation, the scale factor follows the power law, (6.7). In the slow roll limit, we can approximate $\epsilon_{1V} \simeq \epsilon_{1H}$. Hence, substituting (8.4) into (6.7) yields,

$$a(\eta) = l \; \eta^{\,-1-\frac{4}{3\left(-1+\frac{4N}{3}\right)^2}} \;,$$
(8.19)

where, $l$ is the integration constant. For relatively high ($N \approx 50$), the exponent in (8.19) is $\sim -1.0003$, so we can approximate $l \simeq H^{-1}$. Therefore, one cannot expect significant differences between the two cases. Nevertheless, we will investigate this case to have more precise results.

Again, as we did in the last section, we will assume that $N$ is quasi-constant and we will not write it as a function of conformal time explicitly. Hence, in the power law expansion, we substitute (8.19) into (8.9) and differentiate both sides, after doing few simplifications, we get,

$$Y_{PL}(\eta) = \frac{\left(3-4N\right)^2 \left(63-4N\left(63+4N\left(-9+4N\right)\right)\right)^2 \alpha \left(-36+\left(3-4N\right)^2 \alpha\right)}{144 N^2 \left(21+8N\left(-3+2N\right)\right)^4 \eta^2} \;,$$
(8.20)

$$\chi_{PL}(\alpha,N) = \frac{\sqrt{\left(\begin{array}{l}36N^2\left(21+8N\left(-3+2N\right)\right)^4 - 36\left(3-4N\right)^2\left(63-4N\left(63+4N\left(-9+4N\right)\right)\right)^2 \alpha \\ +\left(3-4N\right)^4\left(63-4N\left(63+4N\left(-9+4N\right)\right)\right)^2 \alpha^2\end{array}\right)}}{12N\left(21+8N\left(-3+2N\right)\right)^2} .$$
(8.21)

Plugging (8.20) into (8.13) yields the same solution as (8.15). In this case, $\chi_{PL}(\alpha,N) = 5/2$, at $\alpha \simeq \{-7.4339,\ 7.4345\}$, see Fig.8.2 for all interesting values of $N$. Therefore, the scale invariant PMF can be obtained in the long wavelength regime, $k\eta \ll 1$ (outside Hubble radius) around these values of $\alpha$. Also, the electromagnetic spectra can be obtained in the same way as done in last section. The results are very close to those obtained by de Sitter case, see Fig.8.7-8.8.

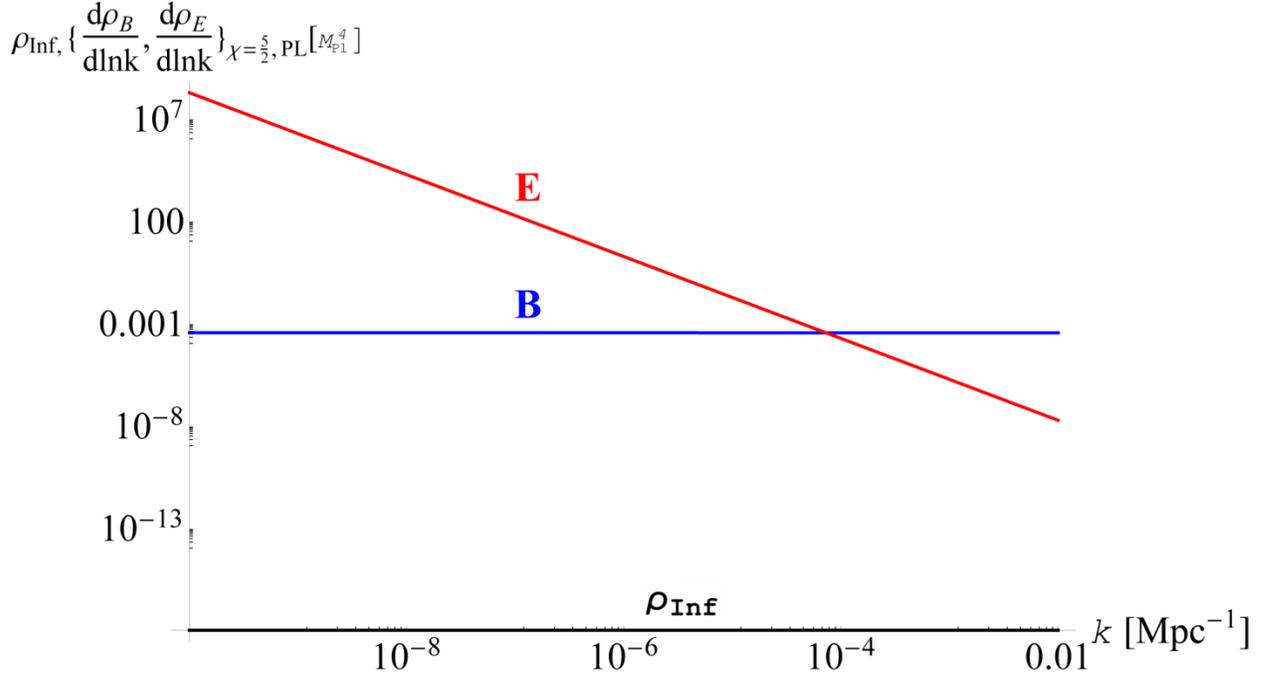

**Fig.8.7.** The EM spectra for $\alpha \approx 7.43$, $N = 64$, $\eta = -20$ and the expansion rate of inflation (PL), $H \approx l^{-1} = 0.2 M_{\text{Pl}}$. The PMF exceeds the electric fields at $k \sim 7 \times 10^{-5} \text{Mpc}^{-1}$.

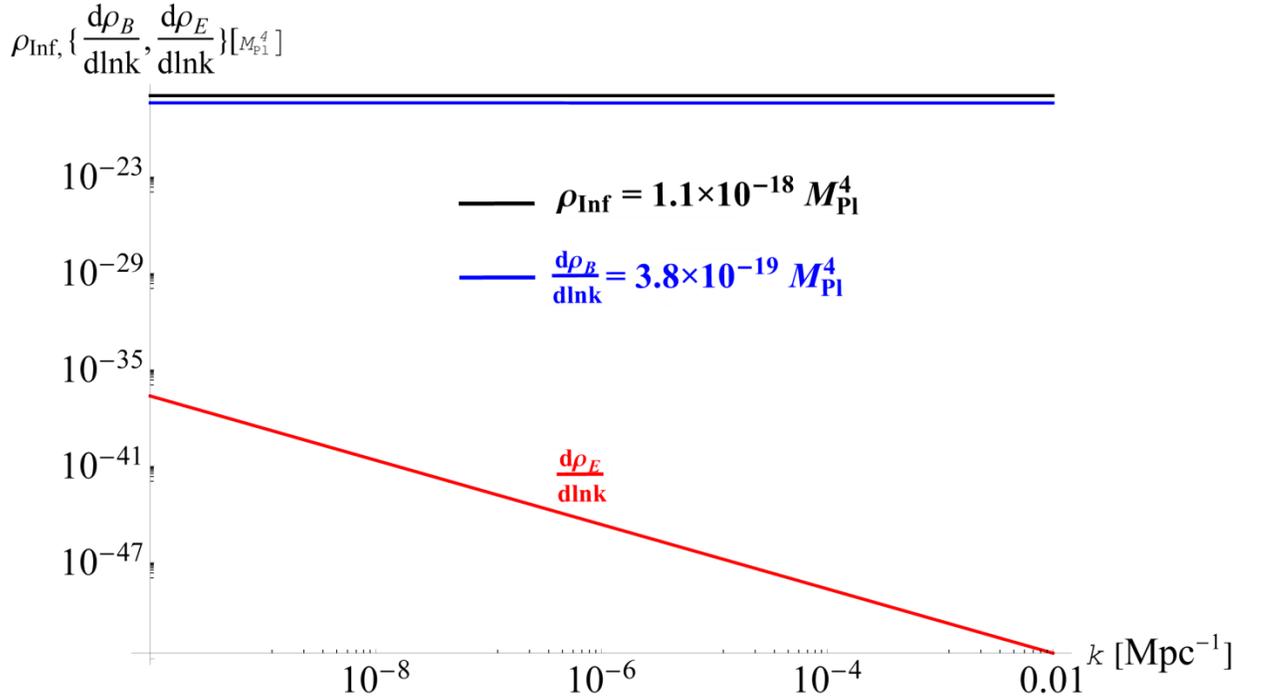

**Fig.8.8.** The EM spectra for $\alpha \approx 7.43$, $N = 64$, $\eta = -20$ and the expansion rate (PL) at the pivot scale, $H \approx l^{-1} = 3.6 \times 10^{-5} M_{\text{Pl}}$. The relation, $\rho_{\text{Inf}} > \dfrac{d\rho_B}{d\ln k} \gg \dfrac{d\rho_E}{d\ln k}$ stays valid, throughout inflation at $k\eta \ll 1$.

It is clear from Fig.8.8 that, $\rho_{Inf} > \frac{d\rho_B}{d\ln k} \gg \frac{d\rho_E}{d\ln k}$ at $H \simeq H_* = 3.6 \times 10^{-5} M_{Pl}$. In fact that

value of the Hubble parameter is the upper value during inflation or at the time of pivot scale, at

which the space time exits the Hubble radius [96]. Therefore, the problem of the backreaction

caused by the divergence of the electric field at $k \ll 1$, is avoided. In order to make sure, the

above relation stays valid throughout the inflationary era, we can plot the EM spectra versus the

conformal time, $\eta$, see Fig.8.9. We usually take $\eta$ as a constant, but we consider it here as free

parameter to make sure that the backreaction problem can be avoided during all relevant

(conformal) time. Similarly, plotting an electromagnetic spectra as a function of e-foldings, $N$,

shows the same relation, see Fig.8.10.

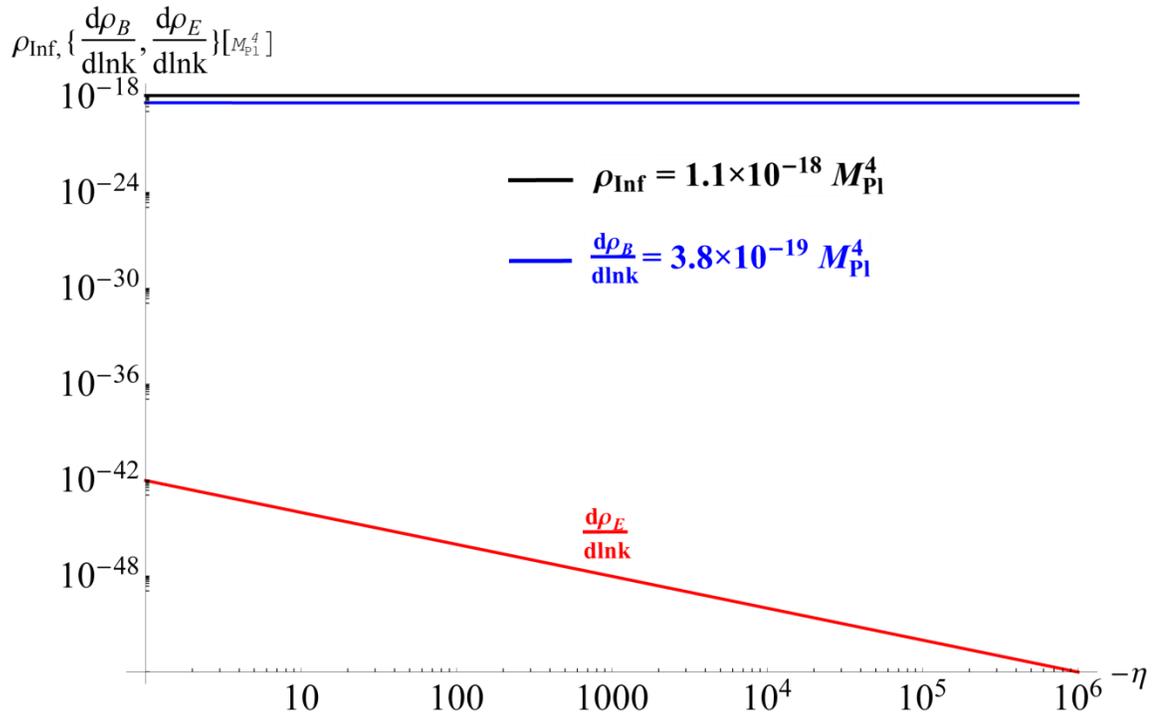

**Fig.8.9.** The EM spectra as a function of $\eta$, for $\alpha \approx 7.43$, $N = 64$, $k = 10^{-3} \mathrm{Mpc}^{-1}$ and the upper limit of Hubble

parameter during inflation, $H \approx l^{-1} = 3.6 \times 10^{-5} M_{Pl}$. The relation, $\rho_{Inf} > \frac{d\rho_B}{d\ln k} \gg \frac{d\rho_E}{d\ln k}$ stays valid, throughout the period, $\eta \ll -1$.

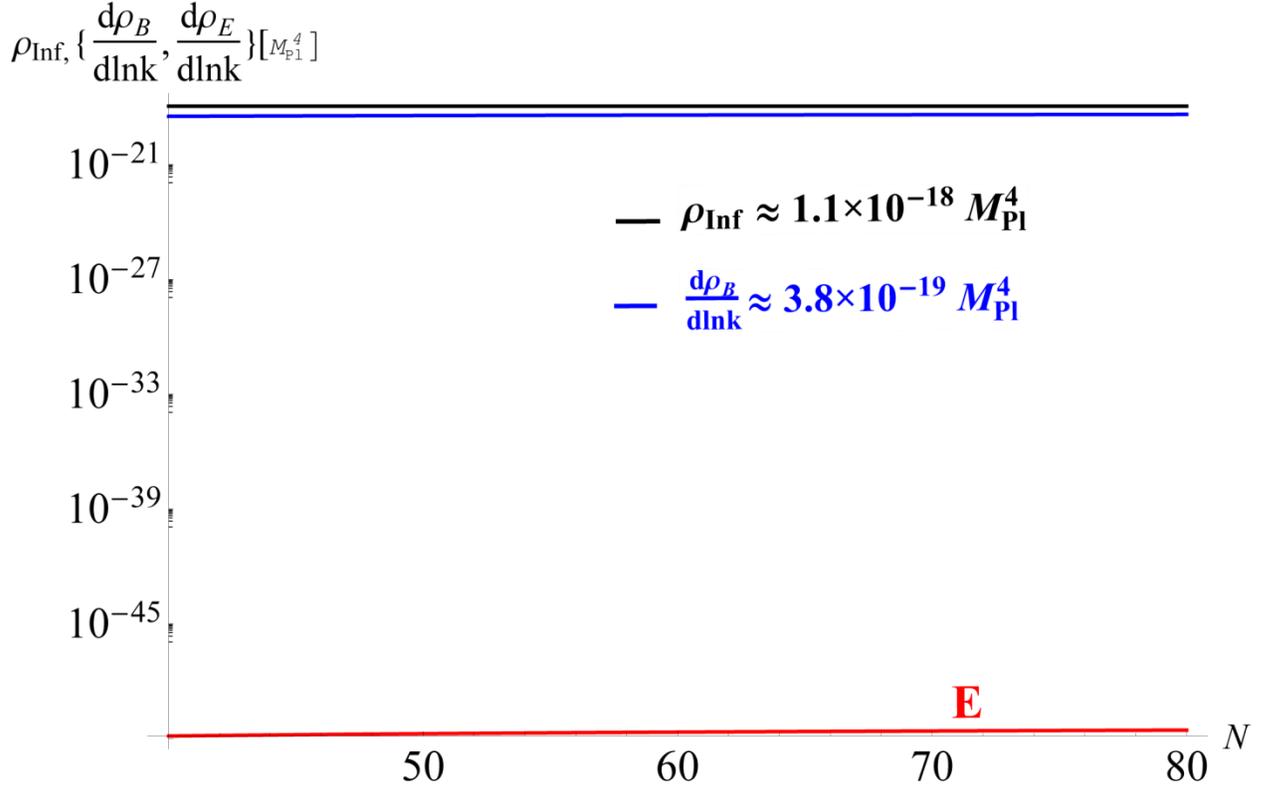

**Fig.8.10**. The EM spectra as a function of $N$, for $\alpha \approx 7.43$, $k = 10^{-3}\,\mathrm{Mpc}^{-1}$ and the upper limit of Hubble parameter during inflation, $H \approx l^{-1} = 3.6 \times 10^{-5} M_{\mathrm{Pl}}$. The relation, $\rho_{\mathrm{Inf}} > \dfrac{d\rho_B}{d\ln k} \gg \dfrac{d\rho_E}{d\ln k}$ stays valid, throughout the period, $40 \leq N \leq 80$.

On the other hand, plotting the electromagnetic spectra as a function of Hubble parameter shows that the upper limit, $H \lesssim 4.6 \times 10^{-5} M_{\mathrm{Pl}}$, can be considered as an upper limit for this model to avoid the backreaction problem. It is slightly higher than $H_*$, see Fig 8.11. Furthermore, plotting the electromagnetic spectra as a function of the free index, $\alpha$, shows that the positive upper limit, $\alpha \lesssim 7.76$, can be considered as an upper limit for this model to avoid the backreaction problem. It is also slightly higher than, $\alpha \approx 7.43$, at which one can generate a scale invariant PMF, see Fig 8.12.

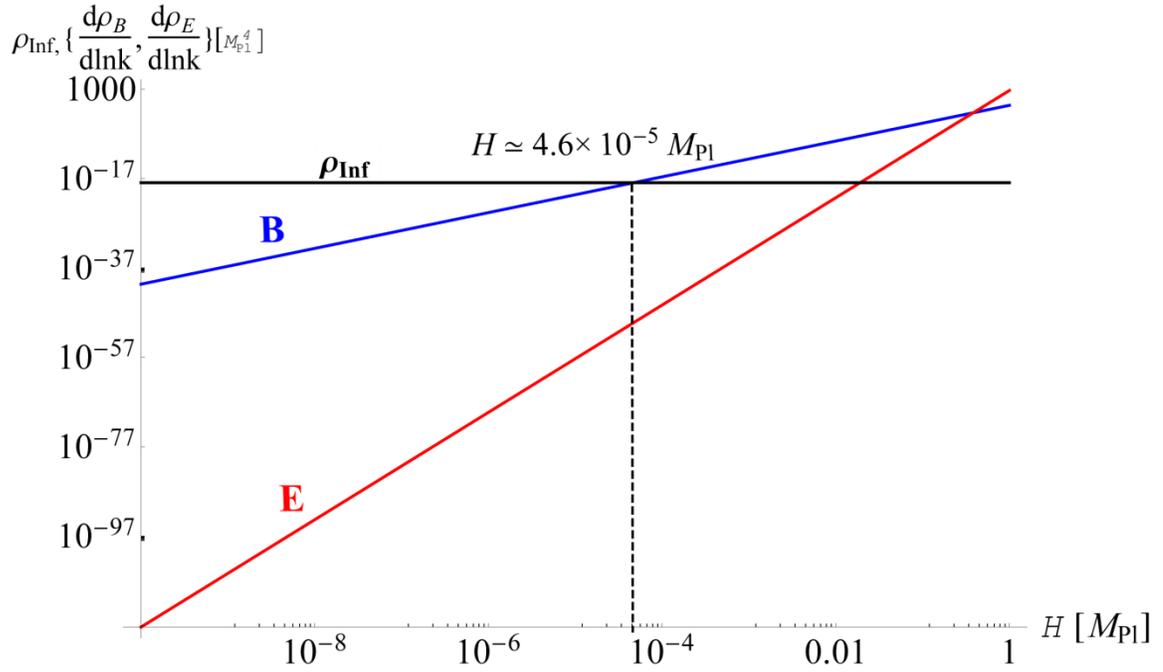

**Fig.8.11**. The EM spectra as a function of $H$, for $\alpha \approx 7.43$, $k = 10^{-3}\,\text{Mpc}^{-1}$ and $N = 64$. One can consider, $H \simeq 4.6 \times 10^{-5}\,M_{\text{Pl}}$, as an upper limit of the Hubble parameter during inflation to avoid the backreaction problem.

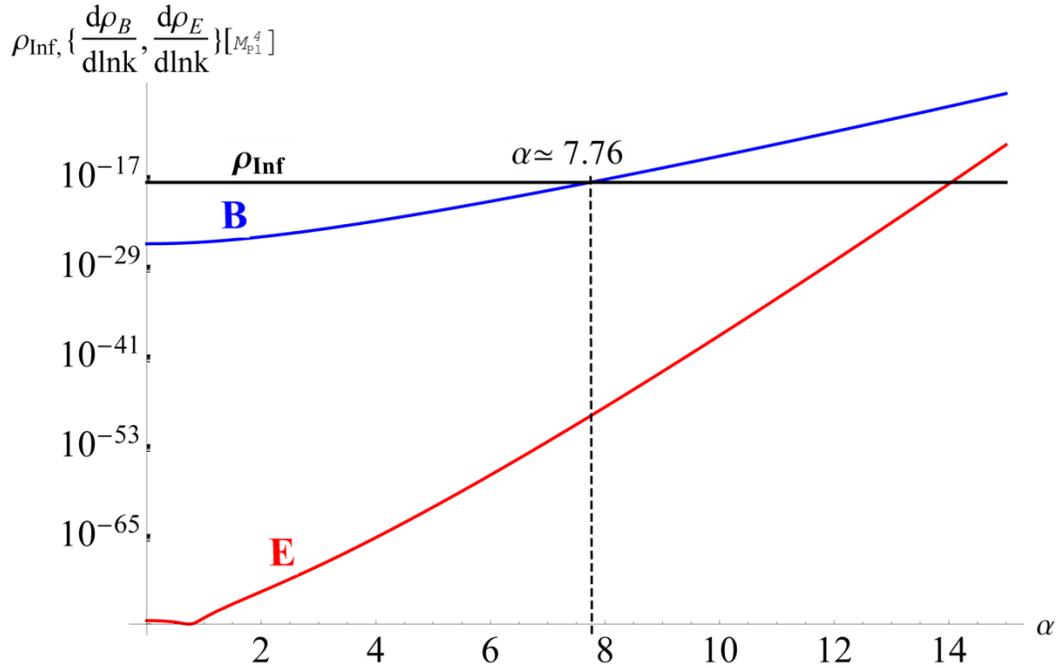

**Fig.8.12**. The EM spectra as a function of $\alpha$, for $H \approx l^{-1} = 3.6 \times 10^{-5}\,M_{\text{Pl}}$, $k = 10^{-3}\,\text{Mpc}^{-1}$ and $N = 64$. One can consider, $\alpha \simeq 7.76$, as an upper limit of the free index during inflation to avoid the backreaction problem.

So far, we make sure that $\rho_{Inf} > \dfrac{d\rho_B}{d\ln k} \gg \dfrac{d\rho_E}{d\ln k}$ during the inflation to avoid the backreaction problem. However, to test the contribution of the magnetic field in the inflationary energy, one has to consider the constraints of post-inflation eras. That point will be discussed in the next section.

## 8.3 Constraining Reheating Parameters by PMF

It is widely believed, that there was a phase of *pre-heating* or *reheating* at the end of inflation and before the radiation dominated era [100-102]. In this phase, as the temperature of inflation falls to certain value, $T_{reh}$, the scalar field of inflation starts oscillates around some value and decays into standard matter, which populates the Universe later. As the temperature continues to fall, the Big Bang Nucleosynthesis (BBN) starts taking place at $T_{BBN} \sim 1\,\text{MeV}$. These produced particles are perturbatively decayed into radiation in the radiation era.

Initially, this process was thought it only occurs in a complicated inflationary model that has more than one field [103-105]. However, later on it was shown that it could be occur even in a single field model but at sub-Hubble scale of perturbations [106-108]. In order to constrain this phase, one has to define the reheating parameters based on both the inflationary model and the observations, like the Cosmic Microwaves Background (CMB) [109], the Large Scale Structure (LSS) [110], the BBN [111], and magnetogenesis [112].

There are two main difficulties in investigating this era. First, no direct cosmological observations can constrains the reheating parameters. Second, the physics of this period is highly uncertain. Therefore, indirect constraints of the reheating parameters are usually calculated from other cosmological observations. Also, several models have been proposed to explain this era. In this section, the effect of the scale invariant PMF on the parameters of reheating is investigated. That is basically similar to Ref [112] but by considering $R^2$-inflationary model.

We adopt the new upper limits of present PMF which was constrained by Planck, 2015 [95] and the instantaneous transition to the reheating and the epochs come after. For this reason, we start by discussing the reheating parameters. Next, we discuss how to constrain them by $R^2$-inflation and the present upper limit of PMF.

### 8.3.1 Reheating Parameters

The reheating era can be specified mainly by three parameters, the reheating parameter, $R_{rad}$, the reheating temperature, $T_{reh}$, and the equation of state parameter, $\omega_{reh}$ [55, 113-114]. The first one is defined as,

$$R_{rad} \equiv \frac{a_{end}}{a_{reh}} \left( \frac{\rho_{end}}{\rho_{reh}} \right)^{1/4},$$
(8.22)

where, $\rho$ is the energy density, and "end" and "reh" stand respectively to the end of inflation and the end of reheating era. From the conservation of energy during the reheating era, $\rho = \rho_\phi + \rho_\gamma$ and $P = P_\phi + \rho_\gamma / 3$, where $\rho_\phi$ is the energy density of the scalar field of inflation and $\rho_\gamma$ is the energy density of the radiation. Assuming these are the main constituent of the reheating era, one can write,

$$\rho_{reh} = \rho_{end} \exp\left\{ -3 \int_{N_{end}}^{N_{reh}} \frac{[1+\omega(a)]}{a} da \right\} = \rho_{end} \exp[-3\Delta N(1+\bar{\omega}_{reh})],$$
(8.23)

where, $\bar{\omega}_{reh}$ is mean value of the equation of state parameter ($P / \rho = \omega(a)$) defined by,

$$\bar{\omega}_{\text{reh}} \equiv \frac{1}{N_{\text{reh}} - N_{\text{end}}} \int_{N_{\text{end}}}^{N_{\text{reh}}} \frac{\omega(a)}{a} da \ . \tag{8.24}$$

Taking the logarithm of (8.22) and making use of (8.23) yields

$$\ln R_{\text{rad}} = \frac{(N_{\text{reh}} - N_{\text{end}})}{4} \left( -1 + 3\bar{\omega}_{\text{reh}} \right) = \frac{1 - 3\bar{\omega}_{\text{reh}}}{12(1 + \bar{\omega}_{\text{reh}})} \ln \left( \frac{\rho_{\text{reh}}}{\rho_{\text{end}}} \right) \ . \tag{8.25}$$

In terms of the pivot quantities [81],

$$\ln R_{\text{rad}} = N_* + \ln \left( \frac{k / a_0}{\rho_\gamma^{1/4}} \right) + \frac{1}{4} \ln \left( \frac{3V_{\text{end}}}{\varepsilon_{1V*} V_*} \frac{3 - \varepsilon_{1V*}}{3 - \varepsilon_{1V\text{end}}} \right) - \frac{1}{4} \ln \left( \frac{H_*^2}{M_{\text{pl}}^2 \varepsilon_{1V*}} \right). \tag{8.26}$$

On the other hand, the relation between the reheating temperature and energy density can be written as,

$$\rho_{\text{reh}} = \frac{\pi^2}{30} g_{\text{reh}} T_{\text{reh}}^4 \ , \tag{8.27}$$

where, $g_{\text{reh}}$ is the number of relativistic degree of freedom at the end of reheating. One also can relate the reheating temperature to today temperature of CMB, $T_0$, [115] as

$$T_{\text{reh}} = T_0 \left( \frac{a_0}{a_{\text{reh}}} \right) \left( \frac{43}{11 g_{\text{reh}}} \right)^{1/3} = T_0 \left( \frac{a_0}{a_{\text{eq}}} \right) e^{N_{\text{eq}}} \left( \frac{43}{11 g_{\text{reh}}} \right)^{1/3} \ . \tag{8.28}$$

where, $a_{\text{eq}} / a_{\text{reh}} = e^{N_{\text{eq}}}$, during radiation era, and "eq" stands for the period of equality between radiation and matter dominant phases.

In terms of the pivot scale, $k = a_k H_k$, at which the commoving scale crosses the Hubble radius (horizon size) during the inflation, one can write down the relation between the scale factors and Hubble parameters [55] for different epochs of the Universe as

$$\frac{k}{a_0 H_0} = \frac{a_k H_k}{a_0 H_0} = \frac{a_k}{a_{\text{end}}} \frac{a_{\text{end}}}{a_{\text{reh}}} \frac{a_{\text{reh}}}{a_{\text{eq}}} \frac{a_{\text{eq}}}{a_0} \frac{H_k}{H_0} .$$  (8.29)

From (8.29), one can write the ratio,

$$\frac{a_0}{a_{\text{eq}}} = \frac{a_0 H_k}{k} e^{-N_k} e^{-N_{\text{reh}}} e^{-N_{\text{eq}}} ,$$  (8.30)

In (8.28) and (8.30), we assume the exponential expansion during reheating and radiation epoch too. Substituting of (8.30) into (8.28) gives

$$T_{\text{reh}} = \left( \frac{T_0 a_0}{k} \right) \left( \frac{43}{11 g_{\text{reh}}} \right)^{1/3} H_k e^{-N_k} e^{-N_{\text{reh}}} .$$  (8.31)

If the equation of state is assumed to be constant during reheating then, $\rho \propto a^{-3(1+\omega)}$. Also, the relation between the energy density and potential [115] can be written as,

$$\rho = \left( 1 + \frac{1}{3/\epsilon_{1V} - 1} \right) V$$  (8.32)

Making use of these relations, at which $\epsilon_{1V} \approx 1$, and substituting of (8.31) into (8.27) and taking natural logarithm yields the equations of reheating e-folds numbers and temperature,

$$N_{\text{reh}} = \frac{4}{1 - 3\omega_{\text{reh}}} \left[ \begin{array}{l} -N_k - \ln\left( \dfrac{k}{a_0 T_0} \right) - \dfrac{1}{4}\ln\left( \dfrac{30}{g_{\text{reh}}\pi^2} \right) - \dfrac{1}{3}\ln\left( \dfrac{11 g_{\text{reh}}}{43} \right) \\[2ex] -\dfrac{1}{4}\ln(V_{\text{end}}) - \dfrac{1}{4}\ln\left( 1 + \dfrac{1}{3/\epsilon_{1V} - 1} \right) + \dfrac{1}{2}\ln\left( \dfrac{\pi^2 r A_s}{2} \right) \end{array} \right] ,$$  (8.33)

$$T_{\text{reh}} = \left[ \left( \frac{T_0 a_0}{k} \right) \left( \frac{43}{11 g_{\text{reh}}} \right)^{1/3} H_k e^{-N_k} \left[ \left( 1 + \frac{1}{3/\epsilon_{1V} - 1} \right) \frac{V_{\text{end}}}{\pi^2 g_{\text{reh}}} \right]^{-\frac{1}{3(1+\omega_{\text{reh}})}} \right]^{\frac{3(1+\omega_{\text{reh}})}{3\omega_{\text{reh}} - 1}}, \qquad (8.34)$$

where, $V_{\text{end}}$ is the potential at the end of inflation and $A_s$ is the scalar power spectrum magnitude, obtained by (6.11).

By adopting Planck, 2015 results [96, 115, 116], we have $\dfrac{k}{a_0} = 0.05 \text{Mpc}^{-1}$, $n_s = 0.9682 \pm 0.0062$, $A_s = 2.196 \times 10^{-9}$, and $H_k < 3.6 \times 10^{-5} M_{\text{Pl}}$, also by using $g_{\text{reh}} \approx 100$, Eqs.(8.33)-(8.34) become

$$N_{\text{reh}} = \frac{4}{1 - 3\omega_{\text{reh}}} \left[ 61.1 - N_k - \ln \left( \frac{V_{\text{end}}^{1/4}}{H_k} \right) \right], \qquad (8.35)$$

$$T_{\text{reh}} = \exp \left[ -\frac{3}{4} (1 + \omega_{\text{reh}}) N_{\text{reh}} \right] \left[ \left( \frac{3}{10\pi^2} \right) \left( 1 + \frac{1}{3/\varepsilon_{1V} - 1} \right) V_{\text{end}} \right]^{1/4}. \qquad (8.36)$$

The equation of state parameter at reheating era, $\omega_{\text{reh}}$, for general inflationary potential is usually taken in the interval, $-1/3 < \omega_{\text{reh}} < 1$ [117]. But for Starobinsky inflation ($R^2$-inflation), the interval that fits well with Planck, 2015 results is $0 < \omega_{\text{reh}} < 1/3$ [115].

### 8.3.2 Constraining Reheating Parameters by the Present PMF in $R^2$-inflation

The next step is to relate the reheating parameters to the present PMF which is constrained by the results of Planck, 2015 [95]. As the conformal invariance of electromagnetic field is restored after inflation, the present super Hubble magnetic field $B_0$ is redshifted since the end of inflation [112] as,

$$B_0 = \frac{B_{\text{end}}}{(1 + z_{\text{end}})^2} \, , \qquad (8.37)$$

where, $B_{\text{end}}$ is the magnetic field at the end of inflation and $z_{\text{end}}$ is the redshift at the end of inflation. Hence, at the end of inflation we can write,

$$1 + z_{\text{end}} = \frac{a_0}{a_{\text{end}}} = \frac{a_0}{a_{\text{eq}}} \frac{a_{\text{eq}}}{a_{\text{reh}}} \frac{a_{\text{reh}}}{a_{\text{end}}} = \frac{1}{R_{\text{rad}}} \left( \frac{\rho_{\text{end}}}{\rho_\gamma} \right)^{1/4} \, , \qquad (8.38)$$

where, $\rho_\gamma \sim (5.7 \times 10^{-125} M_{Pl}^4)$ is energy density of radiation today.

We substitute (8.38) into (8.37) and make sure there is no backreaction problem at the end of inflation, $\rho_{B_{\text{end}}} < \rho_{\text{end}}$. Since, $\rho_B = B^2 / 2$, then combining (8.37) and (8.38) yields the constraint in reheating parameters from the present PMF [112],

$$R_{\text{rad}} \gg \frac{B_0^{1/2}}{(2\rho_\gamma)^{1/4}} \, . \qquad (8.39)$$

The upper limit of present PMF calculated by Planck, 2015 is $\sim 10^{-9}\,\text{G}$. Therefore, the lower limit of reheating parameter is, $R_{\text{rad}} \gtrsim 1.761 \times 10^{-2}$. But this limit is independent of the inflationary model. Hence, for a more accurate limit of $R_{\text{rad}}$ associated with $R^2$-inflation, one can use (8.26) and adopting the Planck, 2015 results, at which the middle value, $N_* \approx 58.5$, if $\omega_{\text{reh}}$ is not constant [96]. It implies that, $R_{\text{rad}} \gtrsim 6.888$, which is three orders of magnitude more than the previous one. Therefore, the reheating in this case is more constraint by the inflationary model bound.

On the other hand, the reheating energy density scale, $\rho_{\text{reh}}$, e-folds number, $N_{\text{reh}}$, and temperature, $T_{\text{reh}}$, are not model independent. Hence, we need to constrain these values in the context of ($R^2$-inflation). In order to constrain the reheating energy scale, one can substitute (8.39) into (8.25) to obtain,

$$\rho_{\text{reh}} > \rho_{\text{end}} \left[ \frac{B_0}{(2\rho_\gamma)^{1/2}} \right]^{\frac{6(1+\bar{\omega}_{\text{reh}})}{1-3\bar{\omega}_{\text{reh}}}}. \tag{8.40}$$

Therefore, one has to find $\rho_{\text{end}}$ for the model of inflation. Also, the lower limit of reheating energy density should be greater than BBN energy, $\rho_{\text{reh}} > \rho_{\text{nuc}}$, where $\rho_{\text{nuc}}$ is in the order of ($\sim 10\,\text{MeV}$). We can use the upper bound of the energy density of inflation derived from WMAP7, $(\rho_{\text{end}})_{\text{CMB}} < 2.789 \times 10^{-10} M_{\text{Pl}}^4$ [109] and the lower limit of $R_{\text{rad}}(\gtrsim 6.888)$ and substitute them into (8.38), one can find the upper bound of, $1 + z_{\text{end}} < 6.828 \times 10^{27}$. If we substitute the redshift into (8.37), we can find the upper limit of magnetic field at the end of inflation, $B_{\text{end}} \lesssim 4.662 \times 10^{46}$ G ($2.3541 \times 10^{-5} M_{\text{Pl}}^2$), where we have used, $1\,\text{G} \approx 3.3 \times 10^{-57} M_{\text{Pl}}^2$. Hence, the energy density of the magnetic field is $(\rho_{B_{\text{end}}})_{\text{CMB}} \lesssim 1.184 \times 10^{-20} M_{\text{Pl}}^4$. It is ten orders of magnitude less than the upper limit of $\rho_{\text{end}}$ found from CMB [109]. Therefore, the backreaction problem can be avoided easily.

On the other hand, we can use Eq.(8.32) to find the energy density at the end of inflation ($\varepsilon_{1V} \approx 1$, $\rho_{\text{end}} = \frac{3}{2} V_{\text{end}}$). By adopting the value of $M \sim 4.0 \times 10^{-5} M_{\text{Pl}}$, calculated from the amplitude of the CMB anisotropies [81], we find $(\rho_{\text{end}})_{R^2-\text{inflation}} \simeq 1.1 \times 10^{-18} M_{\text{Pl}}^4$. Similarly, substituting of this value with the limit, $R_{\text{rad}}(> 6.888)$ into (8.37)-(8.38), implies that, $1 + z_{\text{end}} < 5.414 \times 10^{25}$ and $\left( \rho_{B_{\text{end}}} \right)_{R^2-\text{inflation}} < 4.6788 \times 10^{-29} M_{\text{Pl}}^4$, which is free more from backreaction problem. Therefore, $\left( \rho_{B_{\text{end}}} \right)_{R^2-\text{inflation}}$ calculated by inflationary model and the present limits of PMF found by Planck, 2015 puts more constraints on the $\rho_{B_{\text{end}}}$ than the constraints found by $\left( \rho_{B_{\text{end}}} \right)_{\text{CMB}}$ with present limits of PMF.

Similarly, one can plot both $N_{\text{reh}}$ and $T_{\text{reh}}$ as a function of $n_s$ by using Eq.(8.36), see Fig.8.13-8.14. As shown in Fig.8.14, all curves of possible $\omega_{\text{reh}}$ intersect at $T_{\text{reh}} \approx 4.32 \times 10^{13} \text{GeV}$ and $n_s \approx 0.9674$, the value of spectral index fits well with the range of Planck, 2015 for $R^2$-inflation. This temperature is much more than the range of reheating temperature obtained from CMB in the context of LFI, see Eq. (54) of Ref.[109]. As we use the upper limit of $H_*$ in Eq.(8.35), we can consider the above value as the upper limit of $T_{\text{reh}}$.

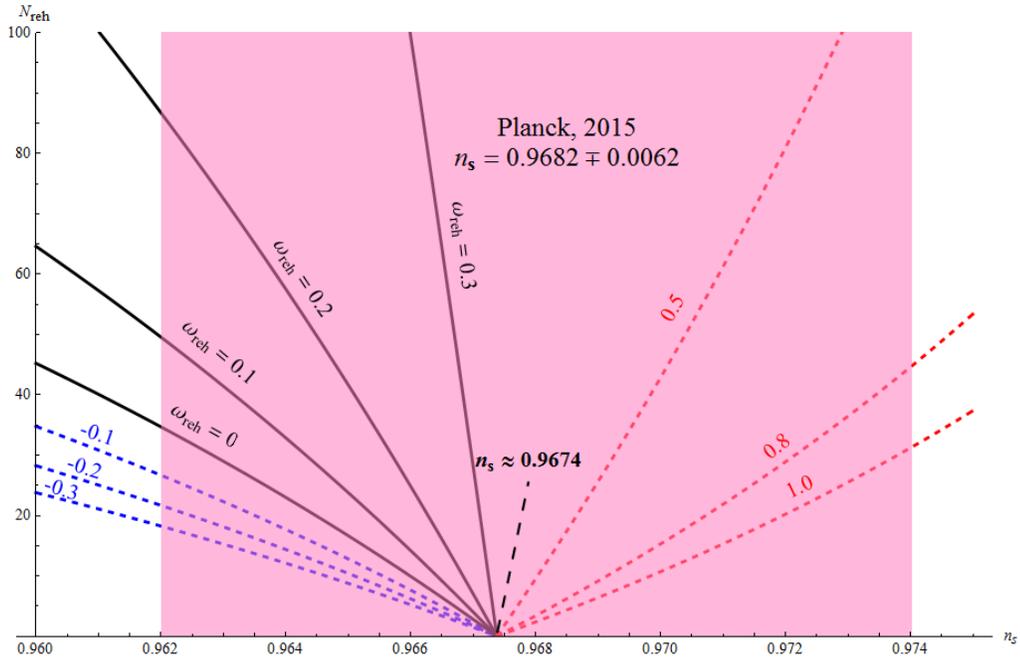

**Fig.8.13**. The plot of reheating e-folds, $N_{reh}$, versus spectral index, $n_s$, at the end of $R^2$-inflation, for some values in $-0.3 < \omega_{reh} < 1$. They all intersect at $n_s \approx 0.9674$, which lies well in the Planck, 2015 range, $n_s = 0.9682 \pm 0.0062$.

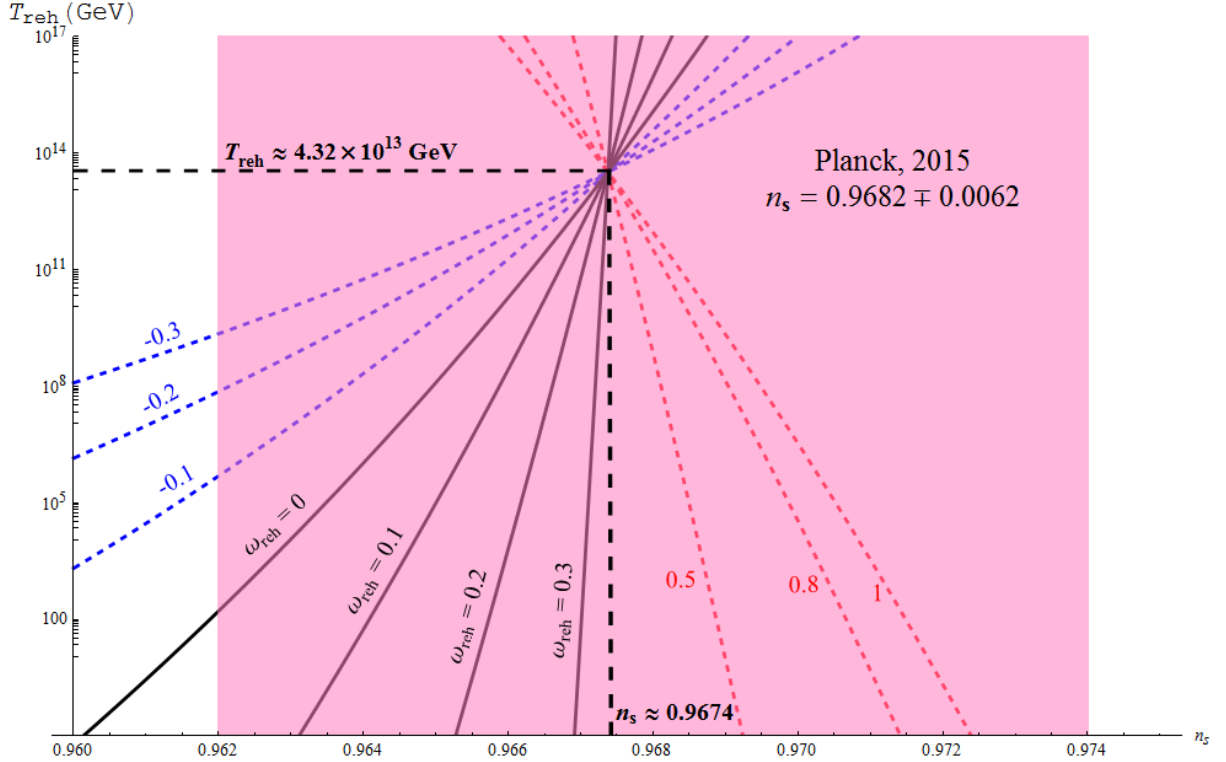

**Fig.8.14**. The plot of reheating temperature, $T_{reh}$, versus spectral index, $n_s$, at the end of $R^2$-inflation, for some values in $-0.3 < \omega_{reh} < 1$. They all intersect into $T_{reh} \sim 4.32 \times 10^{13}$ GeV at $n_s \approx 0.9674$. The value of the temperature is below $(\rho_{end})^{1/4}$ and the spectral index lie in the Planck, 2015 range, $n_s = 0.9682 \pm 0.0062$.

Thus, adopting the this temperature for all $\omega_{reh}$ models of reheating, enables us to constrain the $N_{reh}$ on the range $1 < N_{reh} < 8.3$, for all possible values of , $\omega_{reh}$ see Fig.8.15. The average value of $N_{reh}$ is relatively low, $N_{reh} \simeq 4.7$. Also, from (8.27), one can find $\rho_{reh} \approx 3.259 \times 10^{-18} M_{Pl}^4$, which is in the same order of magnitude of the upper limit of $(\rho_{end})_{R^2-inflation}$ and two orders of magnitude more than the upper limit of $(\rho_{B_{end}})_{CMB}$. Interestingly enough this value of $\rho_{reh}^{1/4}(1.035 \times 10^{14} GeV)$ is one order of magnitude less than the range of $\rho_{end}$

, obtained from CMB for the large field inflation [109], see Eq.(8.22) of the same reference. Therefore, the instantaneous reheating at which, $\rho_{end} \approx \rho_{reh}$, can be manifested by this result.

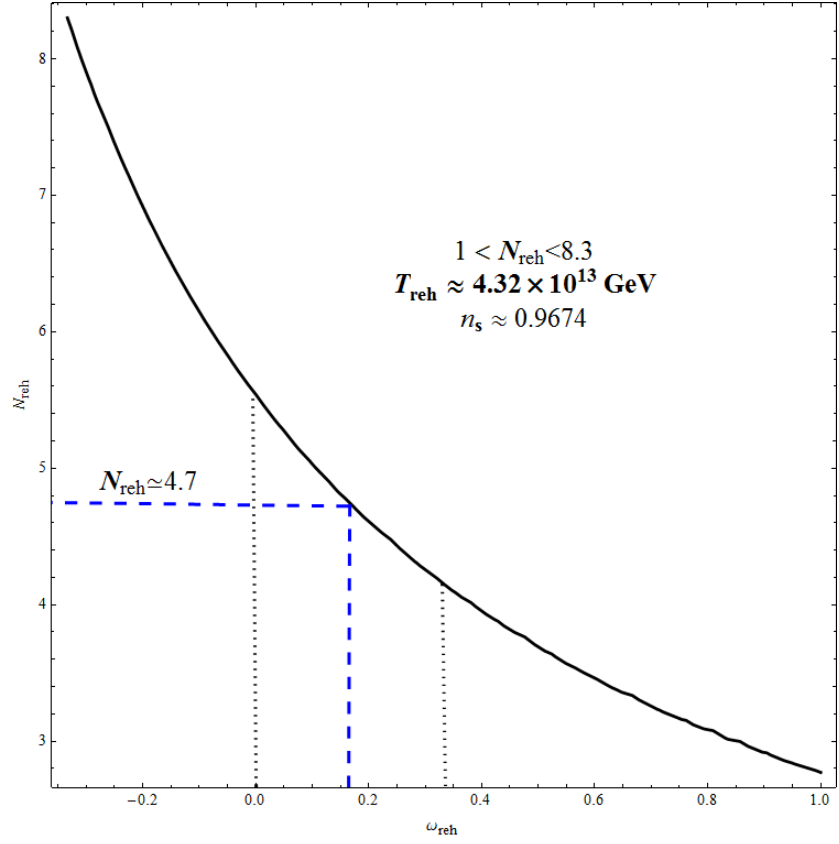

**Fig.8.15.** The number of e-folds, $N_{reh}$, during reheating, for all possible values of the equation of state parameter, $-1/3 < \omega_{reh} < 1$, at $n_s \approx 0.9637$. The range of e-folds is, $1 < N_{reh} < 8.3$. In $R^2$-inflation, the values of state parameter lies onto $0 < \omega_{reh} < 1/3$, hence the middle value is $N_{reh} \approx 4.7$.

At the end of this section, we constrain the upper limit of reheating energy density, $\rho_{reh}$, based on the lower limits of both $N_{reh}$ and reheating parameter, $R_{rad}$. The last one was obtained from the present upper limit of PMF by Planck, 2015. Hence, Eq.(8.22) can be written as,

$$R_{rad} = e^{-N_{reh}} \left( \frac{\rho_{end}}{\rho_{reh}} \right)^{1/4},$$ 

(8.41)

If we use $\left(\rho_{\text{end}}\right)_{\text{CMB}}$, the upper limit, $\left(\rho_{\text{reh}}\right)_{\text{CMB}} < 8.480 \times 10^{-22} M_{\text{Pl}}^4$, which is much more than the lower limit derived from WMAP7 for both large and small field inflation [109]. However, if $\left(\rho_{\text{end}}\right)_{R^2-\text{inflation}}$ is used, the upper limit is $\left(\rho_{\text{reh}}\right)_{R^2-\text{inflation}} < 3.344 \times 10^{-30} M_{\text{Pl}}^4$. It is still much more than the lower limit derived from WMAP7 for both large and small field inflation.

## 8.4 Constraining the Present PMF from the Magnetogensis

In this section, we will constrain the value of present PMF, based on both the predicted scale invariant magnetic fields generated in inflationary era by $f^2FF$ model in $R^2$-inflation, which are calculated in section 8.2, and by using the constraint values of reheating parameters, which are calculated in section 8.3. In order to do that, we need to impose some necessary assumptions. First as shown in [44], the magnetic field energy density, $\rho_B$, scales as $1/a^4$ independently of the dominant constituent in the reheating era and the eras come after. That is basically the implication of Eq.(8.37). Second, and to insure that, this model does not suffer from the backreaction problem, we generalize the validity of Eq.(102) in [44] to the end of inflation.

$$\frac{d\rho_E}{d\ln k}\bigg|_{\text{end}} + \frac{d\rho_B}{d\ln k}\bigg|_{\text{end}} < \rho_{\text{end}} \ , \tag{8.42}$$

where, the EM spectra in (8.42) are calculated by (5.22)-(5.23). The third assumption is that, the reheating is not going to affect the magnitude and the shape of EM spectra.

As a result of the above assumptions, one can neglect the electric field in Fig.8.9 without specifying the constraints of electric conductivity of the reheating ($\sigma_c \gg H$) which may lead to zero electric field and constant magnetic field. Also, as $N_{\text{reh}}$, is relatively small we may assume that, $\left.\dfrac{d\rho_B}{d\ln k}\right|_{\text{end}} \simeq \left.\dfrac{d\rho_B}{d\ln k}\right|_{\text{reh}}$ . Since, we use Planck units in computation $(M_{\text{Pl}}=1)$, then the scale of

the spectra is in $(M_{\text{Pl}}^4)$. Also, by using the upper limit of Hubble parameter during inflation,

$H=3.6\times10^{-5}M_{\text{Pl}}$, the magnitude of magnetic spectra, which are obtained from Fig.8.9-8.10, is

$\left.\dfrac{d\rho_B}{d\ln k}\right|_{\text{end}} \simeq 3.842\times10^{-19}M_{\text{Pl}}^4$. This value is well below both the upper limit of $(\rho_{\text{end}})_{\text{CMB}}$ and

$(\rho_{\text{end}})_{\text{R}^2-\text{inflation}}$. However, it is one order of magnitude more than the upper limit of $(\rho_{B_{\text{end}}})_{\text{CMB}}$ and much more than the upper limit of $(\rho_{B_{\text{end}}})_{\text{R}^2-\text{inflation}}$. Also, $\left.\dfrac{d\rho_B}{d\ln k}\right|_{\text{end}} \gg (\rho_{\text{reh}})_{\text{R}^2-\text{inflation}}, (\rho_{\text{reh}})_{\text{CMB}}$.

The last result shows that the inflationary magnetogensis may play significant role during the reheating era.

By taking the magnitude of the magnetic spectra as equivalent to $\rho_{B_{\text{end}}}$, then the upper bound of magnetic field at the end of inflation, $B_{\text{end}} < 3.76\times10^{47}\,\text{G}$. Therefore, if we use the upper bound of redshift derived from $(\rho_{\text{end}})_{\text{CMB}}$, $1+z_{\text{end}} < 6.828\times10^{27}$, the upper limit of the present PMF is, $B_0 < 8.058\times10^{-9}\,\text{G}$. It is in the same order of magnitude of the upper bound of

PMF obtained by Planck, 2015. However, if we use the upper bound of redshift derived from

$(\rho_{\text{end}})_{\text{R}^2-\text{inflation}}$, $1+z_{\text{end}} < 5.414\times10^{25}$, the upper limit of the present PMF will be

$B_0 < 1.282\times10^{-4}\,\text{G}$. That is even higher than the galactic magnetic field which is in the order of

$10^{-6}\,\text{G}$. Therefore, the second limit is too weak.

## 8.5 Summary and Discussion of the PMF Generated in $R^2$-Inflation

In this chapter, we have shown that the scale invariant PMF can be generated by the simple model $f^2FF$ in $R^2$-inflationary model, which is mostly favored by the latest result of Planck, 2015 [96]. Similar to generating PMF in NI [92] and LFI [93], we can avoid the problem of backreaction in generating it in $R^2$-inflation. Further, in $R^2$-inflation, one can hold the relation

$$\rho_{\text{Inf}} > \frac{d\rho_B}{d\ln k} \gg \frac{d\rho_E}{d\ln k}$$ true for scales of $k$, $H$, $N$ that fit with observations. So, avoiding the

backreaction problem in this model is easier than the other models. It is easily to avoid this problem as long as, the rate of inflationary expansion, $H$, is in the order of or less than, $H \lesssim 4.6 \times 10^{-5} M_{\text{Pl}}$ which is slightly higher than the upper bound reported by Planck ( $\lesssim 3.6 \times 10^{-5} M_{\text{Pl}}$) [96]. Also, the positive upper limit, $\alpha \lesssim 7.76$, can be considered as an upper limit for this model to avoid the backreaction problem. It is also slightly higher than, $\alpha \approx 7.43$, at which one can generate a scale invariant PMF in this model. We do this investigation for both simple exponential (de Sitter) and power law expansion. At sufficiently high e-folds number, $N$, there is no significant differences in their results.

The second main result is constraining the reheating parameters from the upper limits of PMF reported by Planck, 2015 [95]. In the context of $R^2$-inflation, we calculate the lower limits of the reheating parameter, $R_{\text{rad}} > 6.888$. Also, we find the other reheating parameters based on the upper limit of energy density at the end of inflation calculated from CBM data, $(\rho_{\text{end}})_{\text{CMB}} < 2.789 \times 10^{-10} M_{\text{Pl}}^4$, and from the inflationary model, $(\rho_{\text{end}})_{R^2-\text{inflation}} \simeq 1.1 \times 10^{-18} M_{\text{Pl}}^4$. As a result, we find that the magnetic field energy density at the end of inflation as $(\rho_{B_{\text{end}}})_{\text{CMB}} < 1.184 \times 10^{-20} M_{\text{Pl}}^4$ and $(\rho_{B_{\text{end}}})_{R^2-\text{inflation}} < 4.6788 \times 10^{-29} M_{\text{Pl}}^4$. Similarly, we find the upper limit at the end of reheating, $(\rho_{\text{reh}})_{\text{CMB}} < 8.480 \times 10^{-22} M_{\text{Pl}}^4$ and $(\rho_{\text{reh}})_{R^2-\text{inflation}} < 3.344 \times 10^{-30} M_{\text{Pl}}^4$. All of foregoing results are more than the lower limit derived from WMAP7 for both large and small field inflation [109]. These results show the significance of PMF during reheating era.

On the other hand, we constrain the reheating parameters by using the Planck inflationary constraints, 2015 [96] in the context of $R^2$-inflation. The upper limit of reheating temperature and energy density for all possible values of , $\omega_{\text{reh}}$ are respectively constrained as, $T_{\text{reh}} < 4.32 \times 10^{13} \text{GeV}$ and $\rho_{\text{reh}} < 3.259 \times 10^{-18} M_{\text{Pl}}^4$ at $n_s \approx 0.9674$. This value of $n_s$ spectral index is well consistent with Planck, 2015 results. Adopting $T_{\text{reh}}$ for all $\omega_{\text{reh}}$ models of reheating, enables us to constrain the $N_{\text{reh}}$ on the range $1 < N_{\text{reh}} < 8.3$, for all possible values of , $\omega_{\text{reh}}$.

At the end, we constrain $B_0$, from the scale invariant PMF, generated by $f^2 FF$ in $R^2$-inflation in section.8.3. From the PMF spectra, Fig.8.9-8.10, we find that the upper limit of magnetic field in the end of inflation is $B_{\text{end}} < 3.76 \times 10^{47} \text{G}$. Therefore, if we use the upper bound of redshift derived from $(\rho_{\text{end}})_{\text{CMB}}$, then $B_0 < 8.058 \times 10^{-9} \text{G}$. It is in the same order of PMF obtained by Planck, 2015. However, if we use the upper bound of redshift derived from $(\rho_{\text{end}})_{R^2-\text{inflation}}$, then $B_0 < 1.282 \times 10^{-4} \text{G}$. That is even higher than the interplanetary or galactic magnetic field which is of the order of $10^{-6} \text{G}$. Therefore, the second limit is too weak.

In order to achieve the scale invariant PMF by this model, the free index of the coupling function has a relatively high values, $\alpha \approx \{-7.44, 7.44\}$. However, at $\alpha = 2$, which is the typical value, we cannot generate scale invariant magnetic field. The main problem with this model is the value, $|\alpha| \approx 7.44$, which is out of the dynamo constraints imposed by CMB, BBN, and Faraday rotation, RM, see Eq.(94) and Fig.3 of Ref.[44]. In fact those limits are derived mainly for exponential and large field inflation models. Therefore, this subject needs more investigation on the context of $R^2$-inflation.

## CHAPTER NINE: DISCUSSION OF THE RESULTS AND CONCLUSIONS

The PMF can be generated by the simple inflation model $f^2FF$ in the exponential model of inflation, and requires a breaking of the conformal symmetry of the electromagnetism. In this research, we use the same method of [44] to investigate the PMF in large field inflation, LFI, and natural inflation, NI, and $R^2$-inflation. LFI and NI gained more attention after BICEP2 [54], because they fit more with its results.

We investigate the PMF generated in the context of LFI and NI by both the de Sitter expansion, and the power law expansion. The simple de Sitter expansion is only zeroth order approximation which does not have graceful exit from inflation [91]. But it can be assumed valid on most of inflationary era.

The slow roll analysis for both models shows that for sufficiently large e-folding number, $N \geq 50$, the power law inflation can lead to the same general results as the simple de Sitter model of expansion. On both cases, we find that, under the constraints (limits) of BICEP2, $V\left(\phi\left(\eta_f\right)\right) \ll 1$, the PMF can be generated in principle in all parameters of the two models. However, the scale invariant PMF cannot be generated in these limits. In this case the electric and magnetic fields generated are of the same order of magnitude.

However, by releasing BICEP2 limits and enforce the scale invariance condition, $\chi = 5/2$, one can generate the scale invariant PMF in both models and both ways of expansion. A scale invariance property explains why PMF is detected nearly in all scales of the universe. We investigate the PMF under the scale invariance condition in LFI and NI outside Hubble radius, ($k\eta \ll 1$) in both ways of expansion. A scale invariant PMF can be generated in both cases, but the associated electric field has energy scale increasing excessively and becomes greater than the energy of inflation at extremely low $k$, that is basically the problem of backreaction.

On the other hand, for some observable scales of wave numbers, $k$, the electromagnetic spectra can fall below $\rho_{\text{Inf}}$ in LFI. Thus, the backreaction problem can be avoided in this model. For example, the backreaction problem can be avoided for $k \gtrsim 8 \times 10^{-7} \text{Mpc}^{-1}$ and $k \gtrsim 4 \times 10^{-6} \text{Mpc}^{-1}$ in de Sitter and power law expansion respectively in the context of LFI model.

Similarly, computing the electromagnetic spectra as a function of $p$, $H_i$, $N$, $l_0$, $M$, and $c_2$, shows that the electromagnetic spectra can fall below $\rho_{\text{Inf}}$ at certain ranges, in the context of LFI. Under de Sitter expansion, the backreaction problem can be avoided on the ranges, $H_i \lesssim 1.3 \times 10^{-3} M_{\text{Pl}}$, and $M \gtrsim 8.5 \times 10^{-5} M_{\text{Pl}}$. However, under the power law of expansion, it can be avoided on the ranges, $N \gtrsim 51$, $p < 1.66$, $p > 2.03$, $l_0 \gtrsim 3 \times 10^5 M_{\text{Pl}}^{-1} (H_i \lesssim 3.3 \times 10^{-6} M_{\text{Pl}})$, $M \gtrsim 2.8 \times 10^{-3} M_{\text{Pl}}$, and $c_2 \gtrsim 1$. Interestingly enough, all of the above ranges fit with the observational constraints. Beyond these ranges, the backreaction problem is more likely to occur. In these cases, the results of this research provide more arguments against the simple gauge invariant coupling $f^2 FF$, as way of generating PMF in LFI.

On the other hand in the context of NI, for $k \gtrsim 8.0 \times 10^{-7} \text{Mpc}^{-1}$ and $H_i \lesssim 1.25 \times 10^{-3} M_{\text{Pl}}$, both electric and magnetic energy densities can fall below the inflationary energy density, $\rho_{\text{Inf}}$. In this case, one can consider these values as, respectively, a lower bound of $k$ and an upper bound of $H_i$ for a backreaction-free model of PMF. Moreover, these scales include most of the observable ranges of $k$ and $H_i$.

Furthermore, there is a range of $\Lambda_{min}(\sim 0.00874 M_{Pl})$, and $k \gtrsim 2.53 \times 10^{-3} \text{Mpc}^{-1}$, at which the energy density of the electric field can even fall below the energy density of the magnetic field. Again these values lie on the observable range of $k$ and the anticipated scale of $\Lambda$. Therefore, the problem of backreaction can be avoided in these ranges of values. Also, the role of electric field may be neglected in the post-inflation eras. However, the relatively short range of $k$, presents a serious challenge to the viability of this model.

One of the main part of this research is the investigation of magnetogensis in the context of the $R^2$-inflation, which is favored most, by Planck, 2015. Further, in $R^2$-inflation, one can hold the relation $\rho_{\text{Inf}} > \dfrac{d\rho_B}{d\ln k} \gg \dfrac{d\rho_E}{d\ln k}$ true for scales of $k$, $H$, $N$ that fit with observations. So, avoiding the backreaction problem in this model is easier than the other models. It is easily to avoid this problem as long as, the rate of inflationary expansion, $H$, is in the order of or less than, $H \lesssim 4.6 \times 10^{-5} M_{Pl}$ which is slightly higher than the upper bound reported by Planck ( $\lesssim 3.6 \times 10^{-5} M_{Pl}$) [96]. Also, the positive upper limit, $\alpha \lesssim 7.76$, can be considered as an upper limit for this model to avoid the backreaction problem. It is also slightly higher than, $\alpha \approx 7.43$, at which one can generate a scale invariant PMF in this model.

In the same context, we investigate the post-inflation era which may affect the evolution of PMF. As a result, we calculate the lower limits of the reheating parameter, $R_{\text{rad}} > 6.888$. Also, we find the other reheating parameters based on the upper limit of energy density at the end of inflation calculated from CBM data, $(\rho_{\text{end}})_{\text{CMB}} < 2.789 \times 10^{-10} M_{Pl}^4$, and from the inflationary model, $(\rho_{\text{end}})_{R^2-\text{inflation}} \simeq 1.1 \times 10^{-18} M_{Pl}^4$. We find that the magnetic field energy density at the end of inflation as $(\rho_{B_{\text{end}}})_{\text{CMB}} < 1.184 \times 10^{-20} M_{Pl}^4$ and $(\rho_{B_{\text{end}}})_{R^2-\text{inflation}} < 4.6788 \times 10^{-29} M_{Pl}^4$. Similarly, we find the upper limit at the end of reheating, $\left(\rho_{\mathrm{reh}}\right)_{\mathrm{CMB}} < 8.480 \times 10^{-22} M_{\mathrm{Pl}}^4$ and $\left(\rho_{\mathrm{reh}}\right)_{R^2-\mathrm{inflation}} < 3.344 \times 10^{-30} M_{\mathrm{Pl}}^4$. All of foregoing results are more than the lower limit derived from WMAP7 for both large and small field inflation [109]. These results show the significance of PMF role during reheating era, which might be needed to consider in investigating the reheating era.

In the same way, we constrain the reheating parameters by using the Planck inflationary constraints, 2015 [96] in the context of $R^2$-inflation. The upper limit of reheating temperature and energy density for all possible values of $, \omega_{\mathrm{reh}}$ are respectively constrained as, $T_{\mathrm{reh}} \lesssim 4.32 \times 10^{13} \mathrm{GeV}$ and $\rho_{\mathrm{reh}} \lesssim 3.259 \times 10^{-18} M_{\mathrm{Pl}}^4$ at $n_{\mathrm{s}} \approx 0.9674$. This value of $n_{\mathrm{s}}$ spectral index is well consistent with Planck, 2015 results. Adopting $T_{\mathrm{reh}}$ for all $\omega_{\mathrm{reh}}$ models of reheating, enables us to constrain the $N_{\mathrm{reh}}$ on the range $1 \lesssim N_{\mathrm{reh}} \lesssim 8.3$, for all possible values of $, \omega_{\mathrm{reh}}$, see Fig.8.13.

The final result is the most important one in this research. It is the constraining of the present PMF, $B_0$, based on the scale invariant PMF, generated by $f^2 FF$ in $R^2$-inflation. By referring to the spectra, Fig.8.9-8.10, we find that the upper limit of magnetic field in the end of inflation is $B_{\mathrm{end}} < 3.76 \times 10^{47} \mathrm{G}$. Therefore, if we use the upper bound of redshift derived from $\left(\rho_{\mathrm{end}}\right)_{\mathrm{CMB}}$, then $B_0 < 8.058 \times 10^{-9} \mathrm{G}$. This result is in the same order of PMF obtained by Planck, 2015. However, if we use the upper bound of redshift derived from $\left(\rho_{\mathrm{end}}\right)_{R^2-\mathrm{inflation}}$, then $B_0 < 1.282 \times 10^{-4} \mathrm{G}$. That is even higher than the interplanetary or galactic magnetic field which is of the order of $10^{-6} \mathrm{G}$. Therefore, the second limit is too weak.

In fact, Planck, 2015 results do not include the inflationary initial conditions of PMF. They only include the passive and compensated modes which occur way after the inflation era. The passive modes are generated if PMF contributes in CMB before neutrino decoupling ( $T_\nu \gg \text{MeV}$ ), and the compensated modes are generated if PMF contributes in CMB after neutrino decoupling ( $T_\nu \sim \text{MeV}, t \sim s$ ).

On the other hand, there are some problems relating to this model need to be investigated in order to have more robust model. The first one is the high value of the free index of the coupling function, $\alpha \approx \{-7.44, 7.44\}$, at which we can generate a scale invariant PMF. However, at $\alpha = 2$, which is the typical value, we cannot generate scale invariant magnetic field in this model. This high value is out of the dynamo constraints imposed by CMB, BBN, and Faraday rotation, RM, see Eq.(94) and Fig.3 of Ref.[44]. In fact those limits are derived mainly for exponential and large field inflation models. Therefore, this subject needs more investigation on the context of $R^2$-inflation.

The second problem is the approximation of quasi-constant e-folding number, $N$, in comparing with the conformal time. This assumption was necessary to have a Bessel solution to Eq.(8.13). That enables us to decide the scale invariance conditions and to calculate the magnitude of PMF at the end of inflation. This approximation may need more investigation under different inflationary quantities.

By now, it is clear that the results of BICEP2 [54] have been disapproved since the new results of PKB has been announced [94]. Also, the $f^2FF$ model is apparently incompatible with both LFI and NI and the BICEP2 results in the same time. Furthermore, if the above problems are not resolved and $R^2$-inflation shows more consistency with cosmological observations, the last results of this research may go against the same model as way of generating PMF. These difficulties add constraints to this model, in addition to those found by other researches, such as new stringent upper limits on the PMF, derived from analyzing the expected imprint of PMF on the CMB power spectra [56], bi-spectra [82], tri-spectra [83], anisotropies and B-modes [84], and the curvature perturbation and scale of inflation [84, 85].

However, if there is a way to justify the relatively high $\alpha$ and the approximation of quasi-constant $N$, in the framework of the standard cosmological model, $\Lambda CDM$, the results of chapter 8 may be considered as a possible avenue to solve the problem of backreaction and the $f^2 FF$ model may be viable model in the context of $R^2$-inflation. It also might be a contribution in solving the open question about the generation of PMF. Also, the agreement between the result of this model and the upper limit of present PMF found by Planck, 2015 casts some credits to this model.

In addition to investigating the possibility of the high value of $\alpha$, one can extend this research by investigating this model in quantum field realm. The problem of strong coupling can be addressed in this context. If this problem is resolved too, the model will become more robust and more self-consistent.

## REFERENCES

[1]  A. H. Guth, Phys. Rev. D 23, 347  (1981).

[2]  A. A. Starobinsky, Phys. Lett., B91, 99 (1980).

[3]  V. F. Mukhanov, G.V. Chibisov, ZhETF Pisma Redaktsiiu 33, 549 (1981).

[4]  K. Sato, MNRAS 195, 467 (1981).

[5]  A. Albrecht, P.J. Steinhardt, Phys. Rev. Lett. 48, 1220 (1982).

[6]  A. H. Guth, S.-Y. Pi, Phys. Rev. Lett. 49, 1110 (1982).

[7]  S. W. Hawking, Phys. Lett. B 115, 295 (1982).

[8]  A. D. Linde, Phys. Lett. B 108, 389  (1982).

[10]  A. A. Starobinsky, Phys. Lett. B 117, 175 (1982).

[11]  J. M. Bardeen, P. J. Steinhardt, and M.S. Turner, Phys. Rev. D 28, 679 (1983).

[12]  H. K. Eriksen, F. K. Hansen, A. J. Banday, K. M. Gorski and P. B. Lilje, Astrophys. J. 605, 14 (2004), arXiv:astro-ph/0307507.

[13]  A. de Oliveira-Costa, M. Tegmark, M. Zaldar-riaga and A. Hamilton, Phys. Rev. D 69,063516 (2004), arXiv:astro-ph/0307282; K. Land and J. Magueijo, Phys. Rev. Lett. 95, 071301 (2005), arXiv:astro-ph/0502237.

[14]  C. Copi, D. Huterer, D. Schwarz and G. Starkman, Phys. Rev. D 75, 023507 (2007), arXiv:astro-ph/0605135.

[15]  J. Hoftuft, H. K. Eriksen, A. J. Banday, K. M. Gorski, F. K. Hansen and P. B. Lilje, Astrophys. J. 699, 985 (2009), arXiv:0903.1229.

[16]  P. K. Samal, R. Saha, P. Jain and J. P. Ralston, Mon. Not. Roy.Astron. Soc. 396, 511 (2009), arXiv:0811.1639.

[17]  E. Komatsu et al, WMAP Collaboration, Astrophys. J. Suppl. 192, 18 (2011), arXiv:1001.4538; G. Hinshaw et al. , WMAP Collaboration, Astrophys. J. Suppl. 208, 19 (2013), arXiv:1212.5226.

[18]  R. M. Wald, Phys. Rev. D 28, 2118 (1983).

[19]  L. H. Ford, Phys. Rev. D 40, 967 (1989).

[20] C. Armendariz-Picon, JCAP 0407, 007 (2004), arXiv:astro-ph/0405267.

[21] K. Dimopoulos, Phys. Rev. D 74, 083502 (2006), arXiv:hep-ph/0607229.

[22] T. Koivisto and D. F. Mota, Astrophys. J. 679, 1 (2008), arXiv:0707.0279.

[23] A. Golovnev, V. Mukhanov and V. Vanchurin, JCAP 0806, 009 (2008), arXiv:0802.2068.

[24] K. Dimopoulos and M. Karciauskas, JHEP 0807, 119 (2008), arXiv:0803.3041.

[25] S. Kanno, M. Kimura, J. Soda and S. Yokoyama, JCAP 0808, 034 (2008), arXiv:0806.2422.

[26] K. Dimopoulos, M. Karciauskas, D. H. Lyth and Y. Rodriguez, JCAP 0905, 013 (2009), arXiv:0809.1055.

[27] Linde A D, Phys. Lett. B 129 177 (1983).

[28] C. Armend´ariz-Pic´on, JCAP 407, 007 (2004), arXiv:astro-ph/0405267v2.

[29] Jose Beltr´an Jim´enez and Antonio L. Maroto, arXiv:0807.2528v1.

[30] Tomi Koivisto and David F Mota, JCAP 808, 21 (2008), arXiv:0805.4229v3.

[31] B. Himmetoglu, C. R. Contaldi and M. Peloso, Phys. Rev. Lett 102, 111301 (2009), arXiv:0809.2779; B. Himmetoglu, C. R. Contaldi and M. Peloso, Phys. Rev. D **79**, 63517 (2009), arXiv:0812.1231; B. Himmetoglu, C. R. Contaldi and M. Peloso, Phys. Rev. D 80, 123530 (2009), arXiv:0909.3524.

[32] Burak Himmetoglu, PhD Thesis, "*Vector Fields During Cosmic Inflation: Stability Analysis and Phenomenological Signatures*", (2010).

[33] M. A. Watanabe, S. Kanno and J. Soda, Phys. Rev. Lett. 102, 191302 (2009), arXiv:0902.2833.

[34] S. Kanno, J. Soda and M. a. Watanabe, JCAP 0912, 009 (2009), arXiv:0908.3509.

[35] Emanuela Dimastrogiovanni, PhD Thesis, "*Cosmological Correlation Functions in Scalar and Vector Inflationary Models*", (2010), arXiv:1004.1829v1.

[36] T. R. Dulaney and M. I. Gresham, Phys. Rev. D81, 103532 (2010), arXiv:1001.2301.

[38] M. Watanabe, S. Kanno, and J. Soda, Prog. Theor. Phys. 123, 1041 (2010), arXiv:1003.0056v2.

[39] A. Golovnev, V. Mukhanov and V. Vanchurin, JCAP 0811, 018 (2008), arXiv:0810.4304.

[40] V. Demozzi, V. Mukhanov, and H. Rubinstein, JCAP 908, 25 (2009), arXiv:0907.1030v1; R. J. Z. Ferreira, R. K. Jain, and M. S. Sloth, JCAP 10, 004 (2013), arXiv:1305.7151v3.

[41] R. Kumar Jain, R. Durrer and L. Hollenstein, J. Phys. Conf. Ser. 484, 012062 (2014), arXiv:1204.2409v1.

[42] M. S. Turner and L. M. Widrow, Phys. Rev. D **37** 2743 (1988).

[43] B. Ratra, Astrophys. J. 391 L1 (1992).

[44] J. Martin and J. Yokoyama, JCAP 801, 25 (2008), arXiv:0711.4307v1.

[45] K. Subramanian, Astron. Nachr. 331, 110 (2010), arXiv:0911.4771v2.

[46] G. F. R. Ellis, "*Cargese Lectures in Physics*", Vol. 6, Ed. E. Schatzman, 1. (1973).

[47] S. Kanno, J. Soda and M. a. Watanabe, JCAP 912, 009 (2009), arXiv:0908.3509.

[48] C. G. Tsagas, et al, Phys. Lett. B 561, 17 (2003).

[49] C. G. Tsagas, Phys. Rev. D81, 043501 (2010), arXiv:0912.2749v3.

[50] C. Caprini and R. Durrer, Phys. Rev. D65, 023517 (2001), arXiv:astro-ph/0106244v2.

[51] L. Pogosian. et al. arXiv:1210.0308v1.

[52] A. Neronov, and I. Vovk, Sci, 328, 73 (2010); T. Fujita, and S. Mukohyama, JCAP 2012, 34 (2012), arXiv:1205.5031v3; F. Tavecchio et al, MNRAS 406, L70 (2010), arXiv:1004.1329v2.

[53] L. M. Widrow. et. al, Space Sci Rev 166, 37 (2012), arXiv:1109.4052v1.

[54] P. A. R. Ade et al, BICEP2 Collaboration, Phys. Rev. Lett. 112, 241101 (2014), arXiv:1403.3985.

[55] A. R. Liddle and D. H. Lyth. "*Cosmological Inflation and Large-Scale Structure*" (2000); P. Peter and J. Uzan, "*Primordial Cosmology*" (2005); A. R. Liddle, P. Parsons and J. D. Barrow, Phys. Rev D50, 7222 (1994), arXiv:astro-ph/9408015v1; J. Martin, and D. Schwarz, Phys. Rev D62, 103520 (2000), arXiv:astro-ph/9911225v2.

[56] P. A. R. Ade et al, Planck 2013 results. XXII. Constraints on Inflation, Astron. Astrophys. 571, A22 (2014), arXiv:1303.5082v2 ; A. Linde, arXiv:1402.0526v2.

[57] J. Martin, C. R. R. Trotta, and V. Vennin, JCAP 03, 039 (2014).

[58] J. Martin et al, Phys. Rev. D90, 063501 (2014), arXiv:1405.7272v2.

[59] R. Kallosh, A. Linde, and A. Westphal, Phys. Rev. D 90, 023534 (2014), arXiv:1405.0270v1; R. Kallosh, A. Linde and D. Roest, JHEP 08, 052, (2014), arXiv:1405.3646v1; T. Kobayashi, O. Seto, Phys. Rev. D 89, 103524 (2014), arXiv:1403.5055v2.

[60] T. Chiba and K. Kohri, Prog. Theor. Exp. Phys. 093, E01 (2014), arXiv:1406.6117v1.

[61] K. Freese, J. A. Frieman, and A. V. Olinto, Phys. Rev. Lett. 65, 3233 (1990); K. Freese, and W H. Kinney, JCAP 2015, 044 (2015), arXiv:1403.5277v3.

[62] L. Boubekeur and D. Lyth, JCAP 0507,010 (2005), hep-ph/0502047.

[63] P. A. R. Ade et al, Planck intermediate results. XXXIII, arXiv:1411.2271v1.

[64] J. V. Jose and E. J. Saletan. *"Classical Dynamics"* (1998).

[65] Ch. Quigg. *"Gauge Theories of the Strong, Weak, and Electromagnetic Interactions"* (1983).

[66] L. M. Widrow, Rev. Mod. Phys. 74, 775 (2002), arXiv:astro-ph/0207240v1.

[67] D. G. Yamazaki. et. al, Physics Reports 517, 141 (2012), arXiv:1204.3669v1.

[68] M. J. Reid & E. M. Siverstein, Astrophys. J. 361, 483 (1990).

[69] G. Rybicki, A. Lightman *"Radiative Processes in Astrophysics"* (2004).

[70] K. T. Chy˙zy and R. Beck, Astron. & Astrophys. 417, 541 (2004), arXiv:astro-ph/0401157.

[71] A. Kandus. et. al, Phys. Repr. 505, 1 (2011), arXiv:1007.3891v2.

[72] C. Vogt and T. A. Enßlin, Astron. & Astrophys. 412, 373 (2003).

[73] K. T. Kim, et al, Astrophys. J. 355, 29 (1990).

[74] P. P. Kronberg et al, Astrophys. J. 659, 267 (2007), arXiv:0704.3288.

[75] R. Plaga, Nature, 374, 430 (1995).

[76] F. Tavecchio et al, MNRAS 414, 3566 (2011), arXiv:1009.1048.

[77] A. M. Beck et al., MNRAS 429, L60 (2012).

[78] P. A. R. Ade et al, Planck 2013 results. XVI. Cosmological parameters, Astron & Astrophy. 571, A16 (2014), arXiv:1303.5076v1.

[79] F. Bowman, "*Introduction to Bessel Functions*" (1958).

[80] F. Lucchin and S. Matarrese, Phys. Rev. D 32, 1316 (1985).

[81] J. Martin, C. Ringeval, and V. Vennin, Phys. Dark. Univ. 5, 75 (2014), arXiv:1303.3787v3.

[82] P. Trivedi, K. Subramanian and T. R. Seshadri, Phys. Rev. D **82**, 123006 (2010).

[83] P. Trivedi, K. Subramanian and T. R. Seshadri, Phys. Rev. D 89, 043523 (2014), arXiv:1312.5308v1.

[84] R. Z. Ferreira, R. K. Jain, and M. S. Sloth, JCAP 1406, 053 (2014), arXiv:1403.5516v2.

[85] T. Fujitaa, and Sh. Yokoyama, JCAP 1403, 013 (2014), arXiv:1402.0596v3.

[86] C. Bonvin, R. Durrer, and R. Maartens, Phys. Rev. Lett. 112, 191303 (2014), arXiv:1403.6768v1.

[87] D. Larson et al., Astrophys. J. Supp. 192, 16 (2011).

[88] C. L. Bennett et al, The Astrophys. J. Suppl. 208, 2, arXiv:1212.5225v1.

[89] P. A. R. Ade et al, Planck intermediate results. XXX. The angular power spectrum of polarized dust, Astron & Astrophy (2014), arXiv:1409.5738v1.

[90] M. J. Mortonson and U. Seljak, JCAP 10, 035, (2014), arXiv:1405.5857; M. Cortes, A. R. Liddle, and D. Parkinson, Phys. Rev. D 92, 063511 (2015), arXiv:1409.6530v1.

[91] V. Mukhanov, "*Physical Foundation of Cosmology*" (2005).

[92] A. AlMuhammad, R. Lopez-Mobila, "Primordial magnetic fields in the $f^2FF$ model in large field inflation under de Sitter and power law expansion" (sent to publication).

[93] A. AlMuhammad, R. Lopez-Mobilia, General Relativity and Gravitation 47, 134 (2015), arXiv:1505.04668v1.

[94] P. A. R. Ade et al, A Joint Analysis of BICEP2/Keck Array and Planck Data 2015, Phys. Rev. Lett. 114, 101301 (2015), arXiv:1502.00612v1.

[95] P. A. R. Ade et al, Planck 2015 results. XIX. Constraints on primordial magnetic fields, arXiv:1502.01594v1.

[96] P. A. R. Ade et al, Planck 2015 results. XX. Constraints on inflation, Astron & Astrophy 571, A22 (2014), arXiv:1502.02114v1.

[97] A. S. AlMuhammad, "Inflationary Magnetogenesis in $R^2$-Inflation after Planck, 2015", (to be sent to publication), arXiv:1505.05204v1.

[98] A. A. Starobinsky, Phys. Lett. B 91, 99 (1980).

[99] F. Bezrukov and M. Shaposhnikov, Phys. Lett. B659, 703 (2008), arXiv:0710.3755v2.

[100] M. S. Turner, Phys. Rev. D28, 1243 (1983).

[101] L. Kofman, A. D. Linde, and A. A. Starobinsky, Phys. Rev. D56, 3258 (1997), hep-ph/9704452.

[102] B. A. Bassett, S. Tsujikawa, and D. Wands, Rev. Mod. Phys. 78, 537 (2006), astro-ph/0507632.

[103] F. Finelli and R. H. Brandenberger, Phys. Rev. Lett. 82, 1362 (1999), arXiv:hep-ph/9809490v2.

[104] B. A. Bassett, D. I. Kaiser, and R. Maartens, Phys. Lett. B455, 84 (1999), arXiv:hep-ph/9808404.

[105] F. Finelli and R. H. Brandenberger, Phys. Rev. D62, 083502 (2000), arXiv:hep-ph/0003172.

[106] K. Jedamzik, M. Lemoine, and J. Martin, JCAP 1009, 034, (2010), arXiv:1002.3039.

[107] K. Jedamzik, M. Lemoine, and J. Martin, JCAP 1004, 021, (2010), arXiv:1002.3278.

[108] R. Easther, R. Flauger, and J. B. Gilmore, arXiv:1003.3011.

[109] J. Martin and C. Ringeval, Phys. Rev. D82, 023511 (2010), arXiv:1004.5525v2.

[110] M. Bastero-Gil, V. Di Clemente, and S. F. King, Phys. Rev. D67, 103516 (2003), arXiv:hep-ph/0211011.

[111] Mazumdar and J. Rocher, Phys. Rept. 497, 85 (2011), arXiv:1001.0993.

[112] V. Demozzi and C. Ringeval, JCAP 1205, 009 (2012), arXiv:1202.3022v2.

[113] A. R. Liddle and S. M. Leach, Phys. Rev. D68, 103503 (2003), arXiv:astro-ph/0305263.

[114] J. Martin and C. Ringeval, JCAP 0608, 009 (2006), arXiv:astro-ph/0605367v2.

[115] J. L. Cook, E. Dimastrogiovanni, D. A. Easson, L. M. Krauss, arXiv:1502.04673v1.

[116] J. B. Munoz and M. Kamionkowski, Phys. Rev. D 91, 043521 (2015), arXiv:1412.0656.

[117] D. Grasso and H. Rubinstein, Phys. Rept. 348, 163 (2001), arXiv:astro-ph/0009061v2.

[118] J. Martin and C. Ringeval, JCAP 08, 009 (2006), astro-ph/0605367.

[119] W. Esseya, S. Andob, and A. Kusenko, Astroparticle Physics 35, 135 (2011), arXiv:1012.5313.

[120] C. Ringeval, T. Suyama, and J. Yokoyama, JCAP 2013, 020 (2013), arXiv:1302.6013v1.

[121] A. G. Riess et al, Astron. J. 116, 1009 (1998), arXiv:astro-ph/9805201v1.

**VITA**

Anwar Saleh AlMuhammad was born in AlAwjam, Saudi Arabia, in 1973. He got his B.S in physics from King Fahd University of Petroleum and Minerals (KFUPM) in 1996. Also, he got his M.S in Physics from the same university in 2003. His M.S thesis was about the effect of gravitational mass on sub-millisecond pulsars.

Anwar started his Ph.D program in physics in the Department of Physics and Astronomy at University of Texas at San Antonio (UTSA) in 2011 and he is expected to finish the degree in 2015. His Ph.D dissertation is in inflationary cosmology field. It is about the generation of the primordial magnetic fields (PMF) in inflation by $(f^2 FF)$ model in large field inflation (LFI), natural inflation (NI) and $R^2$-inflation models.

As the subject investigated by the dissertation is a very active subject, Anwar intends to continue his researches in the same area of cosmology.